\documentclass{aa}
\usepackage{graphicx}
\usepackage{color}

\def\mb#1{\setbox0=\hbox{$#1$}\kern-.025em\copy0\kern-\wd0
\kern-0.05em\copy0\kern-\wd0\kern-.025em\raise.0233em\box0}

\begin{document}

   \title{Relativistic stars with a linear equation of state:\\ analogy
   with classical isothermal spheres and black holes}

\titlerunning{Relativistic stars with a linear equation of state}

   \author{P.H. Chavanis}

\institute{ Laboratoire de Physique Th\'eorique (UMR 5152 du CNRS), Universit\'e Paul
Sabatier, 118 route de Narbonne 31062 Toulouse, France\\
\email{chavanis@irsamc.ups-tlse.fr}}

   \date{\today}

\abstract{We complete our previous investigations 
concerning the structure and the stability of ``isothermal'' spheres
in general relativity. This concerns objects that are described by a
linear equation of state, $P=q\epsilon$, so that the pressure is
proportional to the energy density. In the Newtonian limit
$q\rightarrow 0$, this returns the classical isothermal equation of
state. We specifically consider a self-gravitating radiation
($q=1/3$), the core of neutron stars ($q=1/3$), and a gas of baryons
interacting through a vector meson field ($q=1$). Inspired by recent
works, we study how the thermodynamical parameters (entropy,
temperature, baryon number, mass-energy, etc) scale with the size of
the object and find unusual behaviours due to the non-extensivity of
the system. We compare these scaling laws with the area scaling of the
black hole entropy. We also determine the domain of validity of these
scaling laws by calculating the critical radius (for a given central
density) above which relativistic stars described by a linear equation
of state become dynamically unstable. For photon stars
(self-gravitating radiation), we show that the criteria of dynamical
and thermodynamical stability coincide. Considering finite spheres, we
find that the mass and entropy present damped oscillations as a
function of the central density. We obtain an upper bound for the
entropy $S$ and the mass-energy $M$ above which there is no
equilibrium state. We give the critical value of the central density
corresponding to the first mass peak, above which the series of
equilibria becomes unstable. We also determine the deviation from the
Stefan-Boltzmann law due to self-gravity and plot the corresponding
caloric curve. It presents a striking spiraling behaviour like the
caloric curve of isothermal spheres in Newtonian gravity.  We
extend our results to $d$-dimensional spheres and show that the
oscillations of mass-versus-central density disappear above a critical
dimension $d_{crit}(q)$. For Newtonian isothermal stars ($q\rightarrow
0$), we recover the critical dimension $d_{crit}=10$. For the stiffest
stars ($q=1$), we find $d_{crit}=9$ and for a self-gravitating
radiation ($q=1/d$) we find $d_{crit}=9.96404372...$ very close to
$10$. Finally, we give simple analytical solutions of relativistic
isothermal spheres in two-dimensional gravity. Interestingly,
unbounded configurations exist for a unique mass
$M_{c}=c^{2}/(8G)$. }

\maketitle

\section{Introduction}
\label{sec_introduction}

Self-gravitating systems have a very strange thermodynamics due to the
attractive long-range, unshielded nature of the gravitational
potential (Padmanabhan 1990, Chavanis 2006c). In particular, they can
display negative specific heats, inequivalence of statistical
ensembles, and phase transitions associated with gravitational
collapse. Furthermore, since these systems are spatially inhomogeneous
and their energy non-additive, the usual thermodynamic limit
($N\rightarrow +\infty$ with $N/V$ fixed) is clearly irrelevant and
must be reconsidered.  As a result, the thermodynamical parameters
have unusual scalings with respect to the number of particles or
system size (see Sect. 7.1 of Chavanis 2006c). For classical
self-gravitating isothermal spheres, the mass scales like the size of
the object $M\sim R$, which is the exact scaling of the singular
isothermal sphere. As a result, the classical thermodynamic limit
(CTL) for self-gravitating systems corresponds to $N\rightarrow
+\infty$ in such a way that $N/V^{1/3}$ is fixed (de Vega \& Sanchez
2002).  It is then found that the temperature, the energy, and the
entropy scale like $T\sim 1$, $E\sim N$, and $S\sim N$ (Chavanis \&
Rieutord 2003). When quantum effects are taken into account, the mass
scales with the radius as $M\sim R^{-3}$, which is the exact
mass-radius relation for nonrelativistic white dwarf stars. As a
result, the quantum thermodynamic limit (QTL) for self-gravitating
fermions corresponds to $N\rightarrow +\infty$ in such a way that $N
V$ is fixed. It is then found that the temperature, the energy, and
the entropy scale as $T\sim N^{4/3}$, $E\sim N^{7/3}$, and $S\sim N$
(Hertel \& Thirring 1971, Chavanis 2002d).  These exotic scalings
(with respect to conventional thermodynamics) come from the long-range
action of gravity, which makes the system spatially inhomogeneous and
renders the energy and the entropy non additive.

On the other hand, in the context of black hole physics, Bekenstein
(1973) and Hawking (1975) have shown that the entropy of a black hole
scales as its area $S\sim R^2$ and that its temperature (measured
from an observer at infinity) scales as $T\sim 1/R$. These results
largely remain mysterious and are usually considered to
reflect a fundamental description of spacetime at the quantum
level. In order to explain the area scaling of the entropy,
researchers have invoked holography (Bousso 2002), quantum gravity,
string theory, entanglement entropy (Srednicki 1993), and brick-wall
models ('t Hooft 1985).

In recent works, Banks et al. (2002) and Pesci (2007) considered
perfect fluids in general relativity with a linear equation of state
$P=q\epsilon$ and found that for the stiffest case compatible with
causality ($q=1$), the entropy scales as the area $S\sim R^2$ and
the temperature as $T\sim 1/R$. Here, these unusual scalings are a
consequence of the non-extensivity of the system due to the action of
gravity. As mentioned above, exotic scalings are also encountered in
Newtonian gravity. However, that purely classical systems
can exhibit scaling laws analogous to black holes is intriguing and
deserves further investigation. The study of these systems should
evidence what, in the area scaling of the black hole entropy, is the
reflect of a fundamental theory of quantum gravity and what mainly
stems from the non-extensive nature of the system.

In a preceding paper (Chavanis 2002b), denoted Paper I, we studied the
structure and the stability of spherically symmetric relativistic
stars with a linear equation of state $P=q\epsilon$. These systems are
sometimes called ``isothermal'' (by an abuse of language) because in
the Newtonian limit $q\rightarrow 0$, they reduce to classical
isothermal spheres described by the Emden equation (Chandrasekhar
1972). Just as for their Newtonian counterparts, they must be enclosed
within a box (of radius $R$) so as to prevent their evaporation and
make their total mass finite. For a given volume, we showed the
existence of a critical mass-energy $M_c$ above which there is no
equilibrium state. In that case the system is expected to collapse and
form a black hole. Furthermore, considering the series of equilibria,
we showed that the mass vs central density relation $M(\epsilon_0)$
presents a series of damped oscillations. The configurations of
hydrostatic equilibrium become unstable above a certain value of the
central density $\epsilon_{0,c}$ corresponding to the maximum mass
(first peak). Secondary extrema in the series of equilibria correspond
to new modes of instability.  We also numerically observed that the
baryon number vs central density $N(\epsilon_0)$ presents extrema at
the same locations as the mass $M(\epsilon_0)$, but we were not able to
explain this observation.

A first motivation of the present paper is to clarify this result.  In
Sects. \ref{sec_ege}-\ref{sec_gre} and in Appendix \ref{sec_f}, applying
the general argument of Weinberg (1972) to a system described by a
linear equation of state, we show that the Oppenheimer-Volkoff
equation of hydrostatic equilibrium in general relativity can be
obtained by extremizing the baryon number $N\lbrack
\epsilon\rbrack$ at fixed mass-energy $M\lbrack
\epsilon\rbrack$. From the condition $\delta N=\mu\delta M$ (where
$\mu$ is a Lagrange multiplier), it becomes clear that extrema of
$M(\epsilon_{0})$ in the series of equilibria correspond to extrema of
$N(\epsilon_{0})$, as we observed numerically. We also argue in
Sect. \ref{sec_ege} and Appendix \ref{sec_ff} that only {\it maxima} of
baryon number $N\lbrack
\epsilon\rbrack$ at fixed mass $M\lbrack \epsilon\rbrack$ are
dynamically stable. More precisely, the maximization of $N\lbrack
\epsilon\rbrack$ at fixed $M\lbrack
\epsilon\rbrack$  forms a criterion of formal nonlinear dynamical
stability for a perfect fluid with respect to the Einstein
equations. In Paper I, we considered the linear dynamical stability
problem and showed that the system becomes unstable after the first
mass peak in the series of equilibria $M(\epsilon_0)$. In Appendix
\ref{sec_ff} of the present paper, we show that the first mass peak
also corresponds to the point at which the hydrostatic configurations
cease to be maxima of the baryon number $N\lbrack \epsilon\rbrack$ (at
fixed mass) and become saddle points. Therefore, the conditions of
linear and nonlinear dynamical stability coincide. Similar results are
obtained for barotropic spheres described by the Euler-Poisson system
in Newtonian gravity (see Chavanis 2006a).

Another motivation of our paper, inspired by the works of Banks et
al. (2002) and Pesci (2007), is to investigate the scaling behaviour
of the thermodynamical parameters (mass-energy $M$, baryon number $N$,
entropy $S$, temperature $T$, etc) with the system size $R$.  In
particular, for $R\rightarrow +\infty$, it is found in
Sects. \ref{sec_ss} and \ref{sec_abhe} that $M\sim R$, $N\sim S\sim
R^{\frac{3q+1}{q+1}}$, and $T\sim R^{-\frac{2q}{q+1}}$. These scaling
laws were implicit in our preceding paper, but they deserve to be
emphasised. We complete previous studies in two respects. In
Secs. \ref{sec_me} and \ref{sec_bn}, we use the stability criteria
obtained in Paper I and Appendix
\ref{sec_dsa} to determine  the domain of validity of these
scaling laws precisely. We show that, for a fixed central density
$\epsilon_{0}$, the system becomes unstable above a critical radius
$R_{c}$ so that the scaling laws only hold approximately close to this
maximum radius. Pure scaling-law profiles, corresponding to singular
spheres with infinite central energy or to regular spheres with very
large radii, are dynamically unstable.  Secondly, we consider general
relativistic systems of astrophysical interest described by a linear
equation of state for which all the parameters entering into the scaling
laws (including the multiplicative factor) can be calculated
explicitly. In Sect. \ref{sec_sgr}, we consider a self-gravitating
radiation (photon star) corresponding to $q=1/3$.  This problem was
first studied by Sorkin et al. (1981) in a seminal paper.  For this
system, we note that the entropy is proportional to the particle
number ($S=\lambda N$) and the energy is proportional to the mass
($E=Mc^2$). Therefore, the maximization of $N$ at fixed $M$ (dynamical
stability) is equivalent to the maximization of $S$ at fixed $E$
(thermodynamical stability). We immediately conclude that the
conditions of dynamical and thermodynamical stability coincide.  We
also note that the dynamical variational problem $\delta N=\mu\delta
M$ can be rewritten in the form of a thermodynamical variational
problem $\delta S=T^{-1}\delta E$ where $T=c^2/(\lambda\mu)$ is to be
interpreted as a temperature (we shall see that $T$ corresponds to the
temperature at infinity $T_{0}$ given by the Tolman relation). We can
therefore apply the results of Paper I to that system. For a fixed box
radius, we find that there exists a maximum mass $M_c=0.2465...
M_{P}R/L_{P}$ (where $M_{P}$ and $L_{P}$ are the Planck mass and the
Planck length) and a maximum entropy
$S_c=0.6217... k_{B}(R/L_{P})^{3/2}$ above which there is no
equilibrium state.  Moreover, the series of equilibria $M(\epsilon_0)$
and $S(\epsilon_0)$ present damped oscillations and become unstable
(dynamically and thermodynamically) above a critical central density
$\epsilon_{0,c}=0.439...c^{4}/(GR^{2})$ corresponding to the first
mass peak. New modes of instability appear at each secondary
peaks. These stability results, which have been proved {\it
analytically} in Paper I and which are developed in the present paper,
complete and simplify the early analysis of Sorkin et al. (1981).  For
a fixed central density, the photon star becomes unstable above a
critical radius $R_c=0.663...(c^{4}/G\epsilon_{0})^{1/2}$. For large
radii, the entropy scales as $S\sim R^{3/2}$. This differs from the
black hole scaling, but this is consistent with the Bekenstein
inequality (Bekenstein 1981). We also determine the deviation from the
Stefan-Boltzmann law due to self-gravity and plot the corresponding
caloric curve. In Sect. \ref{sec_ns}, we find similar
results for a gas of completely degenerate ultra-relativistic fermions
at $T=0$ modelling the core of neutron stars. In particular, the baryon
number scales as $N\sim R^{3/2}$ for $R\rightarrow +\infty$. We also
consider, in Sect. \ref{sec_bvm}, a gas of baryons interacting through a
vector meson field. This model, introduced by Zel'dovich (1962), is
described by a linear equation of state with $q=1$. For this system
the baryon number scales as the area $N\sim R^2$, analogously to the
black hole entropy.

A last motivation of our paper is to extend our results to a space of
dimension $d$. Such generalization is quite common in black hole
physics and quantum gravity, where is it advocated that
extra-dimensions can appear at the micro-scales, an idea stemming from
the Kaluza-Klein theory. In Newtonian gravity, the influence of the
dimension of space on the laws of physics has only been considered
recently (Sire \& Chavanis 2002, Chavanis \& Sire 2004, Chavanis 2004,
2006a, 2006b, 2007a). We have shown that the structure of the system
is highly dependent on the dimensionality of space and that the
problem is very rich because it involves several critical dimensions.
Therefore, it is natural to complete this type of investigations. In
continuity with our study of classical isothermal spheres (Sire \&
Chavanis 2002), we show in Sect. \ref{sec_dimd} that the oscillations
in the mass-central density profile disappear above a critical
dimension $d_{crit}(q)$ depending on the index $q$. Above this
dimension, the configurations are stable for {\it any} central
density, contrary to the case $d<d_{crit}$. This implies that the pure
scaling law profiles corresponding to the singular solution are now
stable. For Newtonian isothermal stars ($q\rightarrow 0$) we recover
the critical dimension $d_{crit}=10$ (Sire \& Chavanis 2002). For the
stiffest stars ($q=1$), we find $d_{crit}=9$ and for a
self-gravitating radiation ($q=1/d$), we find
$d_{crit}=9.96404372...$, very close to $10$. The oscillations exist
for any $q\in [0,1]$ when $d<9$, and they cease to exist for any $q\in
[0,1]$ when $d\ge 10$. We also note that the dimension $d=2$ is
critical so that the results obtained for $d>2$ do not pass to the
limit $d\rightarrow 2$. In two-dimensional gravity, we obtain in
Sect. \ref{sec_tdg} analytical expressions for the density profile of
relativistic isothermal spheres for any $q$. Unexpectedly, they have a
finite radius for $q\neq 1$ (and the density profile decreases like a
Gaussian for $q=1$) contrary to their Newtonian analogue where the
density decreases as $r^{-4}$ (see, e.g., Sire \& Chavanis
2002). Furthermore, they exist at a unique mass
$M_{c}=c^{2}/8G$. Similarly, classical isothermal spheres in $d=2$
exist at a unique mass $M_{c}=4k_{B}T/Gm$ for a given temperature, or
equivalently at a unique temperature $k_{B}T_{c}=GMm/4$ for a given
mass (see, e.g., Chavanis 2007b). Because of that, we may expect that
these structures are only marginally stable.

\section{Relativistic stars}
\label{sec_rs}

In this section, we complete the results obtained in Paper I
concerning the structure and the stability of relativistic stars
with a linear equation of state.

\subsection{The equations governing equilibrium}
\label{sec_ege}

The condition of hydrostatic equilibrium for a spherically symmetric
perfect fluid in general relativity is described by the
Oppenheimer-Volkoff (1939) equation
\begin{eqnarray}
\left\lbrace 1-\frac{2GM(r)}{c^{2}r}\right\rbrace \frac{dP}{dr}=
-\frac{\epsilon+P}{c^{2}}\left\lbrace \frac{GM(r)}{r^{2}}+\frac{4\pi G}{c^{2}}Pr\right\rbrace,\nonumber\\
\label{ege1}
\end{eqnarray}
where $P(r)$ is the pressure, $\epsilon(r)$ is the energy density,
and
\begin{equation}
M(r)=\frac{1}{c^{2}}\int_{0}^{r}\epsilon(r) 4\pi r^{2}\; dr
\label{ege2}
\end{equation}
is the mass-energy contained within the sphere of radius $r$. If
$R$ denotes the radius of the configuration, the total mass-energy
is $M=M(R)$. We also need the baryon number $N$. It is
obtained by multiplying the baryon number density $n(r)$ by the
proper volume element $e^{\lambda(r)/2}4\pi r^{2}dr$ and integrating
over the whole configuration. This  leads to an expression of the
form 
\begin{equation}
N=\int_{0}^{R}n(r) \left\lbrack
1-\frac{2GM(r)}{rc^{2}}\right\rbrack^{-1/2}4\pi r^{2}\; dr.
\label{ege3}
\end{equation}

If we restrict ourselves to spherically symmetric configurations and
isentropic perturbations, it
can be shown that the maximization problem
\begin{equation}
{\rm Max}\quad \lbrace N[\epsilon,n]\quad |\quad M[\epsilon]=M \quad
{\rm fixed}\rbrace, \label{ege4}
\end{equation}
determines stationary solutions of the Einstein equations that are
dynamically stable (Weinberg 1972). The critical points  of the baryon
number $N$ at fixed mass $M$ for isentropic perturbations solve the
variational problem
\begin{equation}
\delta N-\mu \delta M=0, \label{ege5}
\end{equation}
where $\mu$ is a Lagrange multiplier enforcing the conservation of
mass. This variational principle leads to the Oppenheimer-Volkoff
equation (Weinberg 1972); therefore, it determines steady states of
the Einstein equations. However, these first-order variations tell
nothing about the stability of the system. Only {\it maxima} of
$N$ at fixed mass $M$ are dynamically stable, so we must consider the
sign of the second-order variations of $N$ to settle the stability of
the system. In Appendix \ref{sec_dsa}, we consider the case of a
linear equation of state $P=q\epsilon$ (see below) so that the baryon
number is a functional $N[\epsilon]$ of the energy density alone. In
that case, we can evaluate the second-order variations of $N$ and
study the stability of the system.

The maximization problem (\ref{ege4}) or (\ref{dsa1}) is similar to
the minimization of the energy functional ${\cal W}[\rho]$ at fixed
mass $M[\rho]=M$ for a self-gravitating barotropic gas in Newtonian
gravity. It is known that this minimization problem provides a
criterion of formal {\it nonlinear} dynamical stability for the Euler-Poisson
system (see Chavanis 2006a). Similarly, we expect that the maximization
problem (\ref{ege4}) or (\ref{dsa1}) provides a criterion of formal
nonlinear dynamical stability for the Einstein equations.

\subsection{The equation of state}
\label{sec_es}

To close the system of Eqs. (\ref{ege1})-(\ref{ege2}), we need to
specify an equation of state relating the pressure $P$ to the energy
density $\epsilon$.  The first law of thermodynamics can be expressed
as
\begin{equation}
d\left (\frac{\epsilon}{n}\right )=-P d\left (\frac{1}{n}\right )+T
d\left (\frac{s}{n}\right ), \label{es1}
\end{equation}
where $n$ is the baryon number density and $s$ the entropy density
in the rest frame. We assume in the following that the term
$Td(s/n)$ can be neglected. In that case, the first law of
thermodynamics reduces to
\begin{equation}
d\epsilon=\frac{P+\epsilon}{n}dn.
\label{es2}
\end{equation}
We now assume a ``gamma law'' equation of state of the form
\begin{equation}
P=q\epsilon \quad {\rm with}\quad q=\gamma-1.
\label{es3}
\end{equation}
In that case, Eq. (\ref{es2}) can be integrated at once and we find the polytropic relation 
\begin{equation}
P=K n^{\gamma},
\label{es4}
\end{equation}
where $K$ is a constant. Combining Eqs. (\ref{es3}) and (\ref{es4}), we find that the baryon  density  is related to the energy density by 
\begin{equation}
\label{es5} n=\left(\frac{q}{K}\right )^{1/\gamma}
\epsilon^{1/\gamma}.
\end{equation}
The velocity of sound is given by $(dP/d\epsilon)^{1/2} c=q^{1/2}c$ so that
the principle of causality requires $q\le 1$. In the following, we
 consider $0\le q\le 1$ (i.e. $1\le \gamma\le 2$).

There are two situations where the term $Td(s/n)$ can be neglected.
The first situation is when $T=0$. This is the situation that prevails
in the core of neutron stars where the thermal energy is much smaller
than the Fermi energy so that the neutrons are completely
degenerate. The second situation is when the entropy per baryon
$s/n=\lambda$ is a constant. This is the case in supermassive stars
where convection keeps the star stirred up and produces a uniform
entropy distribution. This is also the case for a gas of
self-gravitating photons where the pressure is entirely due to
radiation (see Sect. \ref{sec_sgr}). For a self-gravitating radiation,
the Gibbs-Duhem relation
\begin{equation}
\epsilon=-P+Ts+\mu n,
\label{es6}
\end{equation}
simplifies itself since the chemical potential for photons vanishes
($\mu=0$). In that case, we obtain
\begin{equation}
\epsilon=-P+Ts.
\label{es7}
\end{equation}
Combining the foregoing relations, we find that
\begin{equation}
n(r)=\frac{s(r)}{\lambda}=\left (\frac{\lambda q}{1+q}\frac{T(r)}{K}\right )^{\frac{1}{\gamma-1}},
\label{es8}
\end{equation}
\begin{equation}
P(r)=q\epsilon(r)={K} \left (\frac{\lambda
q}{1+q}\frac{T(r)}{K}\right )^{\frac{\gamma}{\gamma-1}}. \label{es9}
\end{equation}
These relations are valid more generally for any system with
$s/n=\lambda$ constant and $\mu=0$. This is the case for example in
the cosmology developed by Banks \& Fischler (2001) based on the holographic
principle. Central to their discussion is a perfect fluid with equation of state $P=\epsilon$ and $T\propto s\propto \epsilon^{1/2}$.

\subsection{The general relativistic Emden equation}
\label{sec_gre}

Considering the equation of state (\ref{es3}), we introduce the dimensionless variables $\xi$, $\psi$ and $M(\xi)$ by the relations (Chandrasekhar 1972, Chavanis 2002b)
\begin{equation}
\epsilon=\epsilon_{0}e^{-\psi}, \qquad r=\left\lbrace \frac{c^{4}q}{4\pi G\epsilon_{0}(1+q)}\right\rbrace^{1/2}\xi,
\label{gre1}
\end{equation}
and
\begin{equation}
M(r)=\frac{4\pi\epsilon_{0}}{c^{2}}\left\lbrace \frac{c^{4}q}{4\pi G\epsilon_{0}(1+q)}\right\rbrace^{3/2}M(\xi).
\label{gre2}
\end{equation}
In terms of these variables, the Oppenheimer-Volkoff equations
(\ref{ege1})-(\ref{ege2}) can be reduced to the following
dimensionless forms
\begin{eqnarray}
\left\lbrace 1-\frac{2q}{1+q}\frac{M(\xi)}{\xi}\right\rbrace
\frac{d\psi}{d\xi}=\frac{M(\xi)}{\xi^{2}}+q\xi e^{-\psi},
\label{gre3}
\end{eqnarray}
\begin{equation}
\frac{dM}{d\xi}=\xi^{2}e^{-\psi}.
\label{gre4}
\end{equation}
In the Newtonian limit $q\rightarrow 0$, these equations reduce to the
Emden equation (Chandrasekhar 1942). Therefore,
Eqs. (\ref{gre3})-(\ref{gre4}) represent the general relativistic
equivalent of the Emden equation.  It is in this sense that
relativistic stars described by a linear equation of state resemble
classical isothermal spheres. Just as in Newtonian gravity, there
exists a singular solution
\begin{equation}
e^{-\psi_{s}}=\frac{Q}{\xi^{2}}, \qquad {\rm where}\qquad Q=\frac{2(1+q)}{(1+q)^{2}+4q}.
\label{gre5}
\end{equation}
The singular energy density profile is
\begin{equation}
\epsilon_{s}=\frac{qQc^{4}}{4\pi G(1+q)}r^{-2}.
\label{gre6}
\end{equation}
This solution was first found by Klein (1947) and re-discovered by
numerous researchers including Misner \& Zapolsky (1964),
Chandrasekhar (1972), etc. Considering now the regular solutions of
Eqs. (\ref{gre3})-(\ref{gre4}), we can always suppose that
$\epsilon_0$ represents the energy density at the centre of the
configuration. Then, Eqs. (\ref{gre3})-(\ref{gre4}) must be solved
with the boundary conditions $\psi(0)=\psi'(0)=0$. The corresponding
solutions must be computed numerically and some density profiles are
given in Fig. 8 of Paper I for different values of $q$. It has to be
noted that the asymptotic behaviour for $\xi\rightarrow +\infty$ of the
regular solutions behave like the singular sphere (\ref{gre6}).

Since the energy density decreases as $r^{-2}$ for $r\rightarrow
+\infty$, relativistic stars with a linear equation of state have an
infinite mass like their Newtonian analogues (see, e.g., Chavanis
2002a). This reflects the tendency of the system to evaporate.  To
circumvent this difficulty, we shall enclose the system within a box
of radius $R$. Then, the solution of Eqs. (\ref{gre3})-(\ref{gre4})
must be terminated at the normalized box radius
\begin{equation}
\alpha=\left\lbrace \frac{4\pi G\epsilon_{0}(1+q)}{c^{4}q}\right\rbrace^{1/2}R.
\label{gre7}
\end{equation}
It can be noted at that point that $\alpha$ is a measure of the
central density $\epsilon_0$ for a given box radius $R$, or a
measure of the radius $R$ for a given central density $\epsilon_0$.
The energy density contrast is a monotonic function of $\alpha$
given by
\begin{equation}
{\cal R}\equiv \frac{\epsilon_{0}}{\epsilon(R)}=e^{\psi(\alpha)}.
\label{gre8}
\end{equation}
Finally, it is convenient in the analysis to introduce the Milne
variables (Chandrasekhar 1942)
\begin{equation}
u=\frac{\xi e^{-\psi}}{\psi'}, \qquad v=\xi \psi'. \label{ep3}
\end{equation}
The description of the solutions of the general relativistic Emden
equation in the Milne plane is given in Paper I. 

\subsection{Singular solution}
\label{sec_ss}

We first give the values of the thermodynamical parameters
corresponding to the singular solution of the general relativistic
Emden equation. They present exact scaling laws that share some
analogies with the scaling laws for the entropy and temperature of
black holes. We stress from the beginning, however, that the singular
solutions are unstable. The question of the stability of regular
isothermal spheres is considered in Sect. \ref{sec_me}.

The singular energy and pressure profiles are
\begin{equation}
\epsilon(r)=\frac{P(r)}{q}=\frac{qQc^{4}}{4\pi G(1+q)}r^{-2}.
\label{ss1}
\end{equation}
Using the relation (\ref{es5}), we find that the baryon number
density and the entropy profile are given by
\begin{eqnarray}
\label{ss2} n(r)=\frac{s(r)}{\lambda}=\left\lbrack\frac{q^2 Q
c^4}{4\pi G K(1+q)}\right\rbrack^{\frac{1}{q+1}}r^{-\frac{2}{q+1}}.
\end{eqnarray}
From Eq. (\ref{es9}), the temperature profile is
\begin{eqnarray}
\label{ss3} T(r)=\frac{1+q}{\lambda q}K\left\lbrack\frac{q^2 Q
c^4}{4\pi G K(1+q)}\right\rbrack^{\frac{q}{q+1}}r^{-\frac{2q}{q+1}}.
\end{eqnarray}
Finally, using Eqs. (106)-(107) of Paper I, we find that the functions
$\nu(r)$ and $\lambda(r)$ determining the metric are given by
\begin{eqnarray}
\label{ss4} e^{-\lambda(r)}=1-pQ, \qquad e^{\nu(r)}=A
r^{\frac{4q}{q+1}},
\end{eqnarray}
where $A$ is a constant and 
\begin{equation}
\label{ss6} p=\frac{2q}{1+q}.
\end{equation} 
Comparing Eqs. (\ref{ss3}) and (\ref{ss4}),
we check explicitly that the Tolman relation (see Appendix
\ref{sec_tr}) is satisfied.

Suppose now that the singular solution is terminated by a box with
radius $R$. According to Eqs. (\ref{ege2}) and (\ref{ss1}), the total
mass is
\begin{equation}
\label{ss5} M=\frac{pQc^2}{2G}R.
\end{equation}
Therefore, we find that the mass $M$ scales linearly with the radius
$R$. This is the same scaling as for the singular isothermal sphere in
Newtonian gravity (see, e.g., Chavanis 2002a,b). This is also the
scaling entering in the Schwarzschild relation (\ref{bht3}). On the other hand,
according to Eqs. (\ref{ege3}) and (\ref{ss2}), the baryon number and the total entropy
can be written
\begin{eqnarray}
\label{ss7}
N=\frac{S}{\lambda}=\frac{1+q}{3q+1}Q^{\frac{1}{1+q}}(1-pQ)^{-1/2}\nonumber\\
\times 4\pi\left\lbrack\frac{q^2c^4}{4\pi
GK(1+q)}\right\rbrack^{1/\gamma}R^{\frac{3q+1}{q+1}}.
\end{eqnarray}
Finally, using the Tolman relation (\ref{tr7}) and Eq. (\ref{ss3}),
the temperature measured by an observer at infinity is given by
\begin{eqnarray}
\label{ss8} T_0=\frac{1+q}{\lambda q}K\left\lbrack\frac{q^2 Q
c^4}{4\pi G
K(1+q)}\right\rbrack^{\frac{q}{q+1}}(1-pQ)^{1/2}R^{-\frac{2q}{q+1}}.
\end{eqnarray}
The surface temperature $T(R)$ has the same scaling with $R$ as $T_{0}$, differing only in the factor $(1-pQ)^{1/2}$ according to the Tolman relation. 

\subsection{Analogy with the black hole entropy} \label{sec_abhe}

Some interesting analogies between relativistic stars described by a
linear equation of state and black hole thermodynamics have been
discussed previously by Banks et al. (2002) and Pesci (2007). Let us
briefly develop their arguments in connection to the present
study.

When $P=\epsilon$, the velocity of sound $(dP/d\epsilon)^{1/2} c$ is
equal to the velocity of light so that Eq. (\ref{es3}) with $q=1$ is
the stiffest equation of state compatible with the principle of
causality. In that case, we see from Eqs. (\ref{ss7}) and (\ref{ss8})
that the entropy scales like the area $S\sim R^2$ and that the
temperature scales like the inverse of the radius $T\sim 1/R$. As
noticed by Banks et al. (2002), this is similar to the scaling of the
Bekenstein-Hawking entropy and temperature for black holes (see
Appendix \ref{sec_bht}). We also note from Eq. (\ref{ss2}) that the
entropy profile scales as $s(r)\sim r^{-1}$. As discussed by Pesci
(2007), this shows that the total entropy is due to the contribution
of the whole volume. Therefore, for relativistic stars with a linear
equation of state, the area scaling of the entropy is a volume effect
resulting from the inhomogeneity of the system. This differs from the
result of Oppenheim (2002) who considered self-gravitating classical
systems reaching the above scaling laws for entropy and temperature
while approaching the Schwarzschild radius. In this limit, he showed
that all entropy lies on the surface and the area scaling of the
entropy is due to external layers.

More generally, for a linear equation of state of the form
(\ref{es3}), the entropy scales as $S\sim R^{\frac{3q+1}{q+1}}$. For
$0\le q\le 1$, the exponent is always smaller than $2$ (the area
scaling law) obtained for $q=1$. This is in agreement with the
Bekenstein inequality $S\le S_{BH}=k_{B}\pi(R/L_{P})^2$ (where $L_{P}$ is the
Planck length). Finally, we note that the energy scales like $E\sim
M\sim R$ and that the temperature scales like $T\sim
R^{-\frac{2q}{q+1}}$.  This leads to $E\sim T^{-\frac{q+1}{2q}}$
implying negative specific heats $C=dE/dT<0$ (see Appendix
\ref{sec_bht} for black holes and Sect. \ref{sec_tst} for the
self-gravitating radiation). The mass-energy is positive but it
decreases with the temperature. In Newtonian gravity, isothermal
spheres also display negative specific heats but for a completely
different reason (Lynden-Bell \& Wood 1968). According to the Virial
theorem $2K+W=0$, we have $E\equiv K+W\sim -K\sim -\frac{3}{2}N k_B T$
(where $K$ is the kinetic energy and $W$ the potential energy) leading
to negative specific heats $C=dE/dT=-\frac{3}{2}N k_B<0$. In that
case, the energy becomes more and more negative as temperature
increases. Therefore, the origin of negative specific heats for
classical isothermal spheres and black holes is different.

\subsection{The mass-energy} \label{sec_me}

\begin{figure}[htbp]
\centerline{
\includegraphics[width=8cm,angle=0]{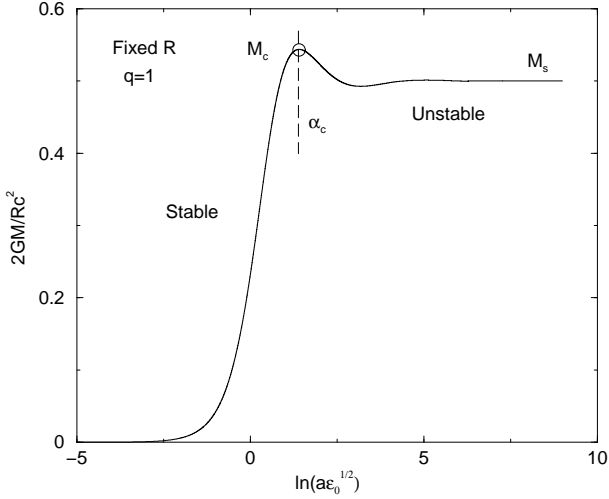}
} \caption[]{Mass as a function of the central density for a fixed box
radius. This corresponds to the curve $\chi(\alpha)$. For a system
evolving at fixed volume, the system becomes unstable at the first
mass peak.} \label{meq1}
\end{figure}

\begin{figure}[htbp]
\centerline{
\includegraphics[width=8cm,angle=0]{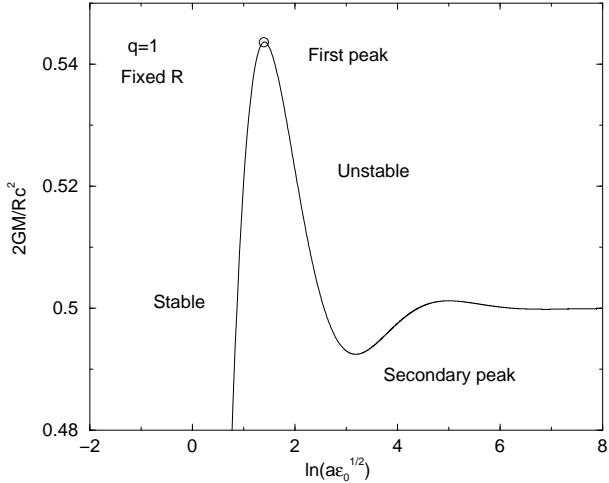}
} \caption[]{Mass as a function of the central density for a fixed box radius showing the secondary peaks. }
\label{meq1ZOOM}
\end{figure}

We now consider regular isothermal spheres and discuss the stability
of the system along the series of equilibria and the domain of
validity of the scaling laws. It is shown in Paper I that the
normalized mass-energy is related to the parameter $\alpha$ by the
relation
\begin{equation}
\chi\equiv \frac{2GM}{R c^{2}}=\frac{pv_{0}(1-qu_{0})}{1+pv_{0}},
\label{me1}
\end{equation}
where $u_{0}=u(\alpha)$ and $v_{0}=v(\alpha)$ are the values of the
Milne variables at the box radius $R$. The foregoing
relation can be rewritten
\begin{equation}
\frac{2GM}{R c^{2}}=\chi(\alpha). \label{me2}
\end{equation}
It defines the series of equilibria parametrized by $\alpha$
going from $0$ to $+\infty$. For a fixed box radius $R$, we can
write
\begin{equation}
\alpha=a\epsilon_0^{1/2}, \quad {\rm with}\quad a=\left\lbrace
\frac{4\pi G(1+q)}{c^{4}q}\right\rbrace^{1/2}R.  \label{me3}
\end{equation}
Therefore, $\alpha=a\epsilon_0^{1/2}$ is a measure of the central
density so that Eq. (\ref{me2}) determines the relation between the total mass
$M$ and the central density $\epsilon_0$. The corresponding curve
$M(\epsilon_{0})$ is plotted in Figs. \ref{meq1} and
\ref{meq1ZOOM}. It presents damped oscillations around an asymptote at
$M=M_s\equiv \chi_{s}Rc^2/(2G)$ corresponding to the singular solution
(\ref{ss5}).  For $\alpha\rightarrow 0$, the density is almost uniform
and
\begin{equation}
\chi(\alpha)\sim \frac{2q}{3(1+q)}\alpha^{2}. \label{me4}
\end{equation}
For  $\alpha\rightarrow +\infty$, the density profile tends to  the
singular sphere (\ref{gre6}) and
\begin{equation}
\chi(\alpha)\rightarrow \chi_{s}=pQ. \label{me5}
\end{equation}
The mass is maximum for a certain value of the central density
corresponding to $\alpha=\alpha_{c}$ (first peak). The maximum mass
is given by
\begin{equation}
\frac{2GM_{c}}{R c^{2}}=\chi_{c}. \label{me6}
\end{equation}
The values of $\alpha_c$, $\chi_c$, and $\chi_{s}$ depend on $q$. For
$q=1/3$, we have $\alpha_c=4.7$, $\chi_c=0.493$, $p=1/2$, $Q=6/7$,
$\chi_s=3/7$ and for $q=1$, we have $\alpha_c=4.05$, $\chi_c=0.544$,
$p=1$, $Q=1/2$, $\chi_s=1/2$. There is no equilibrium state with
$M>M_{c}$. For $M<M_{c}$, the stable configurations correspond to
$\alpha<\alpha_{c}$, i.e. to sufficiently small central densities. The
configurations with $\alpha>\alpha_{c}$ in the series of equilibria
(i.e. those located after the first mass peak) are dynamically
unstable. New modes of instability appear at each mass peak. These
results are proved analytically in Paper I where a linear dynamical
stability analysis of box-confined systems with a linear equation of
state is performed. This study is completed in Appendix \ref{sec_dsa}
where it is shown that the first mass peak is also the point at which
the steady states pass from maxima of $N[\epsilon]$ (at fixed mass) to
saddle points of $N[\epsilon]$ (at fixed mass). Therefore, the system
loses its stability at the first mass peak in agreement with the
Poincar\'e turning-point argument (Katz 2003, Chavanis 2006c) applied to
the series of equilibria $M(\epsilon_{0})$.

\begin{figure}[htbp]
\centerline{
\includegraphics[width=8cm,angle=0]{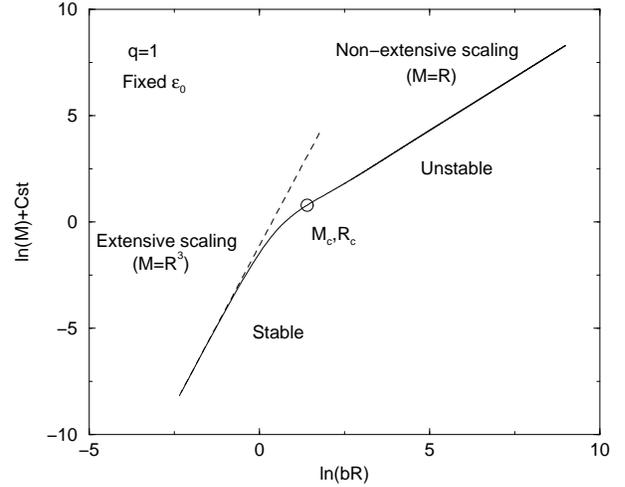}
} \caption[]{Mass as a function of the box radius for a fixed central
density. This corresponds to the curve $\alpha\chi(\alpha)$.} \label{mrq1}
\end{figure}

For a fixed central density $\epsilon_0$, we can write
\begin{equation}
\alpha=b R, \quad {\rm with}\quad b=\left\lbrace \frac{4\pi
G\epsilon_0(1+q)}{c^{4}q}\right\rbrace^{1/2}.  \label{me7}
\end{equation}
Therefore, $\alpha=bR$ is a measure of the system size $R$ so that
the equation
\begin{equation}
M=\chi(\alpha)\frac{c^{2}R}{2G}
\label{me8}
\end{equation}
determines the relation between the total mass and the  radius. In
the limit $\alpha\rightarrow 0$, the system is homogeneous and,
using Eq. (\ref{me4}),  we obtain the usual scaling for an extensive system
\begin{equation}
M=\frac{4\pi}{3}\frac{\epsilon_{0}}{c^{2}}R^{3}, \qquad
(R\rightarrow 0). \label{me9}
\end{equation}
If we now consider the  limit  $\alpha\rightarrow +\infty$
corresponding to the singular sphere, using Eq. (\ref{me5}), we get the
non-extensive scaling
\begin{equation}
M=\chi_{s}\frac{c^{2}R}{2G},\qquad (R\rightarrow +\infty).
\label{me10}
\end{equation}
More generally, the mass-radius relation for a fixed central density
is given by
\begin{equation}
\frac{2GbM}{c^{2}}=\chi(bR)bR. \label{me11}
\end{equation}
This relation is plotted in Fig. \ref{mrq1}. The scaling law $M\sim R$
is only valid in the limit $R\rightarrow +\infty$. However, as we have
seen, the system becomes unstable for $R>R_c\equiv \alpha_c/b$, i.e.
for a mass $M>M_{c}\equiv c^{2}R_{c}\chi_{c}/{2G}$. Therefore, the
solutions exhibiting a pure scaling law profile (like the singular
isothermal sphere or the regular isothermal spheres with $R\rightarrow
+\infty$) are dynamically unstable. However, for stable configurations
close to the critical radius $R_c$, we observe in Fig. \ref{mrq1} that
the linear scaling holds approximately, so that the results given in
Sect. \ref{sec_ss} are correct in that sense.

\subsection{The baryon number}
\label{sec_bn}

\begin{figure}[htbp]
\centerline{
\includegraphics[width=8cm,angle=0]{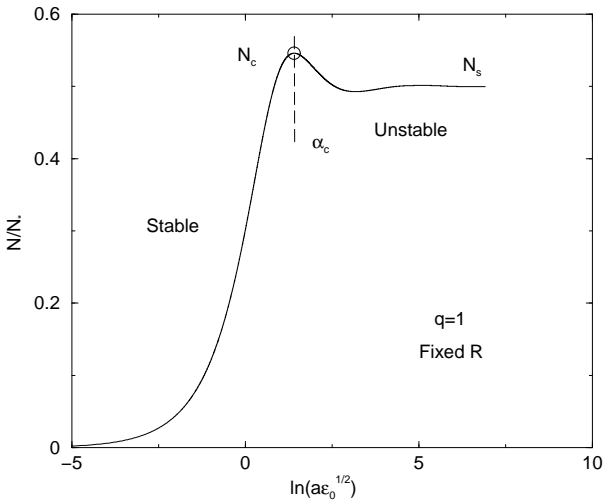}
} \caption[]{Baryon number as a function of the central
density for a fixed box radius. We have plotted $\Delta(\alpha)$.}
\label{neq1}
\end{figure}

\begin{figure}[htbp]
\centerline{
\includegraphics[width=8cm,angle=0]{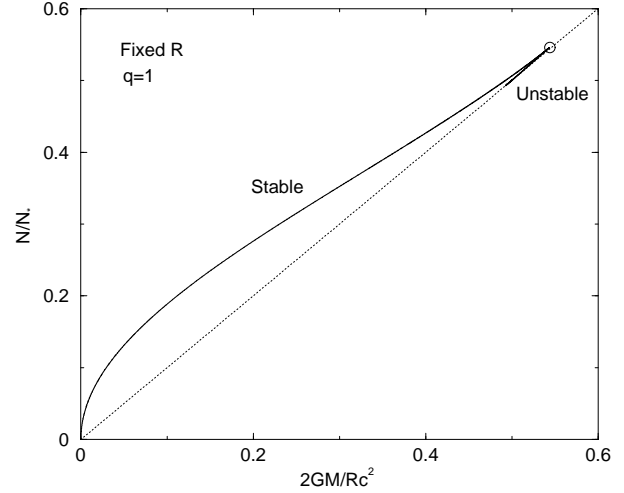}
} \caption[]{Baryon number as a function of the mass for a fixed box
radius. We have plotted $\Delta(\alpha)$ as a function of
$\chi(\alpha)$ so that the curve is parametrized by $\alpha$. For
$q=1$, we note that, asymptotically, $\Delta(\alpha)\simeq\chi(\alpha)
$ so that the relation $M\sim N$ is linear. We have drawn a line $y=x$
for comparison. This special behaviour arises only for $q=1$.}
\label{MNq1}
\end{figure}

Using Eq. (\ref{ege3})  and introducing the dimensionless variables defined
previously, we find that the baryon number is given by
\begin{equation}
\Delta\equiv \frac{N}{N_{*}}=\frac{1}{\alpha^{\frac{3q+1}{1+q}}}\int_{0}^{\alpha} e^{-\frac{\psi(\xi)}{1+q}}\left\lbrack 1-p\frac{M(\xi)}{\xi}\right\rbrack^{-1/2} \xi^{2}d\xi,
\label{bn1}
\end{equation}
with
\begin{equation}
N_{*}=4\pi R^{3}\left\lbrack \frac{q^{2}c^{4}}{4\pi GKR^{2}(1+q)}\right\rbrack^{\frac{1}{q+1}}.
\label{bn2}
\end{equation}
The foregoing relation can be rewritten
\begin{equation}
\frac{N}{N_{*}}=\Delta(\alpha).
\label{bn3}
\end{equation}
For a fixed box radius, using $\alpha=a\epsilon_0^{1/2}$, this
equation determines the relation between the baryon number and the
central density. The corresponding curve is plotted in
Fig. \ref{neq1}. It presents damped oscillations around an asymptote
at $N=N_s\equiv \Delta_{s}N_*$ corresponding to the singular solution (\ref{ss7}).  According to
Eq. (\ref{ege5}) the extrema of $N(\alpha)$ occur for the same values of
$\alpha$ as the extrema of $M(\alpha)$ in the series of
equilibria. This leads to angular points in the curve $N(M)$, as shown in
Fig. \ref{MNq1}. For $\alpha\rightarrow 0$, corresponding to an almost
uniform distribution, we have
\begin{equation}
\Delta(\alpha)\sim \frac{1}{3}\alpha^{\frac{2}{q+1}},
\label{bn4}
\end{equation}
and for $\alpha\rightarrow +\infty$ corresponding to the singular
sphere, we have
\begin{equation}
\Delta(\alpha)\rightarrow \Delta_{s}=\frac{1+q}{3q+1}Q^{\frac{1}{1+q}}(1-\chi_{s})^{-1/2}.
\label{bn5}
\end{equation}
The baryon number is maximum for $\alpha=\alpha_{c}$ and its value
is
\begin{equation}
\frac{N_c}{N_{*}}=\Delta_{c}. \label{bn6}
\end{equation}
The values of $\alpha_c$, $\Delta_c$ and $\Delta_{s}$ depend on
$q$. For $q=1/3$, we have $\alpha_c=4.7$, $\Delta_c=0.925$,
$\Delta_s=(8/21)^{1/4}=0.7856...$ and for $q=1$, we have
$\alpha_c=4.05$, $\Delta_c=0.546$, $\Delta_s=1/2$. There is no
equilibrium state with $N>N_{c}$. For $N<N_{c}$, the stable
configurations correspond to $\alpha<\alpha_{c}$, i.e. to sufficiently
small central densities. Configurations with $\alpha>\alpha_c$ in the
series of equilibria are dynamically unstable. We can note that, for
$q=1$, the asymptotic values $\chi_{s}=1/2$ and $\Delta_{s}=1/2$
coincide. More precisely, the curves $\chi(\alpha)$ and
$\Delta(\alpha)$ turn out to superimpose for $\alpha\gg 1$. This
implies that the curve $N(M)$ becomes linear in the region where
$\alpha\gg 1$, as can be seen in Fig. \ref{MNq1}. This behaviour is,
however, not general and, for $q\neq 1$, the curves $\chi(\alpha)$ and
$\Delta(\alpha)$ are different (see, e.g., Figs. 18 and 19 of Paper I
for $q=1/3$).

\begin{figure}[htbp]
\centerline{
\includegraphics[width=8cm,angle=0]{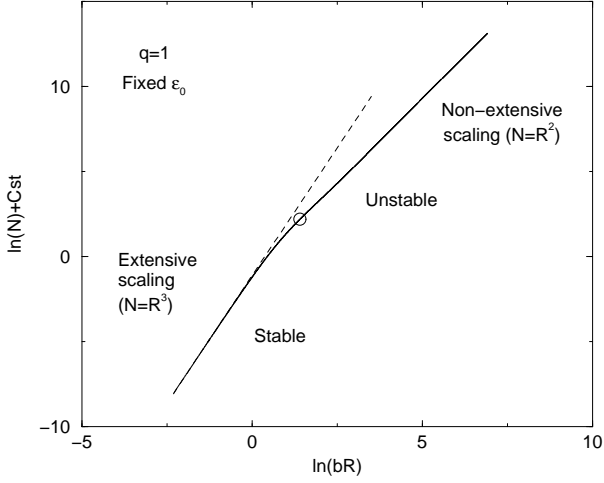}
} \caption[]{Baryon number as a function of the box radius for a fixed
central density. We have plotted
$\alpha^{\frac{3q+1}{q+1}}\Delta(\alpha)$. } \label{nrq1}
\end{figure}

\begin{figure}[htbp]
\centerline{
\includegraphics[width=8cm,angle=0]{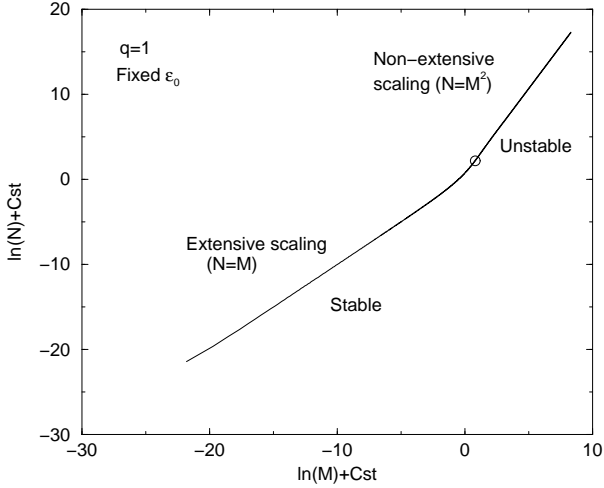}
} \caption[]{Baryon number as a function of the mass for a fixed
central density. We have plotted
$\alpha^{\frac{3q+1}{q+1}}\Delta(\alpha)$ as a function of
$\alpha\chi(\alpha)$ so that the curve is parametrized by $\alpha$. } 
\label{MNq1fixede0}
\end{figure}

For a fixed central density, using $\alpha=bR$, the equation
\begin{equation}
N=\Delta(\alpha) 4\pi \left\lbrack \frac{q^{2}c^{4}}{4\pi GK(1+q)}\right\rbrack^{\frac{1}{q+1}}R^{\frac{3q+1}{q+1}},
\label{bn7}
\end{equation}
determines the relation between the baryon number and the radius. In the
limit $\alpha\rightarrow 0$, the system is homogeneous and, using
Eq. (\ref{bn4}), we have the usual scaling for an extensive system
\begin{equation}
N=\frac{4\pi}{3}n_{0}R^{3}, \qquad (R\rightarrow 0). \label{bn8}
\end{equation}
If we now consider the  limit  $\alpha\rightarrow +\infty$,
corresponding to the singular sphere, using Eq. (\ref{bn5}) we find the
non-extensive scaling
\begin{eqnarray}
N=\Delta_{s} 4\pi \left\lbrack \frac{q^{2}c^{4}}{4\pi
GK(1+q)}\right\rbrack^{\frac{1}{q+1}}R^{\frac{3q+1}{q+1}}, \quad (R\rightarrow
+\infty). \nonumber\\
\label{bn9}
\end{eqnarray}
More generally, for a fixed central density, the baryon number is
expressed  as a function of the radius by
\begin{equation}
\frac{1}{4\pi}\left\lbrack \frac{4\pi GK(1+q)}{q^{2}c^{4}}
\right\rbrack^{\frac{1}{q+1}}b^{\frac{3q+1}{q+1}}N=(bR)^{\frac{3q+1}{q+1}}\Delta(bR).
\label{bn10}
\end{equation}
This relation is plotted in Fig. \ref{nrq1}. The scaling law $N\sim
R^{\frac{3q+1}{q+1}}$ is only valid in the limit $R\rightarrow +\infty$.
However, as we have seen, the system becomes unstable for
$R>R_c\equiv \alpha_c/b$, i.e. for a baryon number $N>N_{c}\equiv
\Delta_{c} N_*(R_{c})$. Therefore, the solutions exhibiting a pure scaling law profile (like the singular
isothermal sphere or the regular isothermal spheres with $R\rightarrow
+\infty$) are dynamically unstable. However, for 
the stable configurations close to the critical radius $R_c$, we note
that the scaling $N\sim R^{\frac{3q+1}{q+1}}$ holds approximately, so
that the results given in Sect. \ref{sec_ss} are correct in that
sense. Finally, in Fig. \ref{MNq1fixede0}, we have plotted the curve
$N(M)$ for a fixed central density. For small mass, we have the
scaling $N\sim M$ and for large mass, we have $N\sim
M^{\frac{3q+1}{q+1}}$.

As a final remark, we note that the density contrast (\ref{gre8}) is a
monotonic function of $\alpha$ (behaving like the inverse of the
density profile reported in Fig. 8 of Paper I) that could be used to
parametrize the series of equilibria $M(\alpha)$ and $N(\alpha)$ in
place of $\alpha$. In particular, the critical energy density contrast
corresponding to $\alpha_c$ is ${\cal R}_c=30.1$ for $q=1$ and ${\cal
R}_c=22.4$ for $q=1/3$. Configurations with a density contrast greater
than these values are dynamically unstable.

\section{Self-gravitating radiation: photon stars}
\label{sec_sgr}

In this section, we apply the preceding results to a radiation that is
so intense that self-gravity (in the sense of general relativity) must
be taken into account. This is sometimes referred to as ``photon
stars'' (Schmidt \& Homann 2000). This problem was first
considered by Sorkin et al. (1981). Our work confirms and extends the
results of these authors.

\subsection{Equation of state for ultra-relativistic particles}
\label{sec_u}

Let us consider a perfect gas of non-interacting particles that can be
relativistic. We call $f({\bf p})$ the numerical density of particles
with impulse ${\bf p}$. The particle number, the kinetic energy, and
the pressure are
\begin{equation}
N=V\int f d{\bf p},
\label{u1}
\end{equation}
\begin{equation}
E_{kin}=V\int f \epsilon(p) d{\bf p},
\label{u2}
\end{equation}
\begin{equation}
 P=\frac{1}{3}\int f p \frac{d\epsilon}{dp} d{\bf p},
\label{u3}
\end{equation}
where the kinetic energy of a particle is
\begin{equation}
\epsilon(p)=mc^{2}\left\lbrace \left (1+\frac{p^{2}}{m^{2}c^{2}}\right )^{1/2}-1\right \rbrace.
\label{u4}
\end{equation}
In the ultra-relativistic limit, the kinetic energy of a particle
with impulse $p$ reduces to
\begin{equation}
\epsilon(p)=p c.
\label{u5}
\end{equation}
Therefore, the pressure is related to the kinetic energy by
\begin{equation}
P=\frac{1}{3}\frac{E_{kin}}{V}.
\label{u6}
\end{equation}
In the local form, this leads to the linear equation of state
\begin{equation}
P=q\epsilon \qquad {\rm with}\quad q=\frac{1}{3}. \label{u7}
\end{equation}
Note that, in a $d$-dimensional universe (see Sect. \ref{sec_dimd}), the
corresponding value of this parameter would be
\begin{equation}
q=\frac{1}{d}. \label{u8}
\end{equation}

\subsection{The Stefan-Boltzmann law}
\label{sec_sbl}

We first recall some basic elements of thermodynamics that apply to a
pure radiation modeled as a gas of photons. We first ignore
gravitational effects. The distribution function of a perfect gas of
bosons without interaction is given by the Bose-Einstein statistics
\begin{equation}
f=\frac{1}{h^{3}}\frac{1}{e^{\beta\epsilon-\mu}-1}, \label{sbl1}
\end{equation}
where $\mu$ is the chemical potential. For ultra-relativistic
particles, we have $\epsilon=pc$. On the other hand, for particles
without rest mass, like photons, the chemical potential
$\mu=0$. Therefore, the distribution function of a gas of photons is
simply
\begin{equation}
f=\frac{1}{h^{3}}\frac{1}{e^{\beta p c}-1}.
\label{sbl2}
\end{equation}
The particle number, the kinetic energy, and the pressure can be
computed from Eqs. (\ref{u1})-(\ref{u3}). Writing these relations
locally and using the properties of the Bose integrals, we obtain the
standard results
\begin{equation}
P=\frac{1}{3}\epsilon=\frac{8\pi
}{h^{3}c^{3}}(k_{B}T)^{4}\frac{\pi^{4}}{90}, \label{sbl3}
\end{equation}
\begin{equation}
n=\frac{8\pi}{h^{3}c^{3}}(k_{B}T)^{3}\zeta(3). \label{sbl4}
\end{equation}
The relation (\ref{sbl3})  between the energy and the temperature is the
Stefan-Boltzmann law. Using the Gibbs-Duhem relation (\ref{es6})
with $\mu=0$, we directly obtain the entropy
\begin{equation}
s=k_{B}\frac{32\pi^{5}}{90 h^{3}c^{3}}(k_{B}T)^{3}.
\label{sbl5}
\end{equation}
For a pure radiation, the pressure is related to the photon
density by
\begin{equation}
P=K n^{\gamma},
\label{sbl6}
\end{equation}
with
\begin{equation}
\gamma=\frac{4}{3},  \qquad
K=\frac{\pi^{4}}{90}\frac{hc}{\left\lbrack 8\pi
\zeta(3)^{4}\right\rbrack^{1/3}}. \label{sbl7}
\end{equation}On the other hand, the entropy is related to the photon density by
\begin{equation}
s=\lambda n, \qquad {\rm with} \quad \lambda=\frac{4\pi^{4}}{90
\zeta(3)}k_{B}. \label{sbl8}
\end{equation}

If we now consider a gas of self-gravitating photons in general
relativity (photon stars), the preceding relations still hold
locally. They have the form considered in Sect. \ref{sec_es} with
$q=1/3$.  Therefore, a self-gravitating radiation can be studied with
the theory developed in Sect. \ref{sec_rs}.  We note that the singular
profiles of Sect. \ref{sec_ss} scale like
\begin{equation}
P\propto\epsilon\propto r^{-2}, \qquad s\propto n\propto r^{-3/2},
\qquad T\propto r^{-1/2}. \label{sbl9}
\end{equation}
These are also the asymptotic behaviours of the regular profiles for
$r\rightarrow +\infty$. The numerical constants in
Eqs. (\ref{ss1})-(\ref{ss3}) can be obtained explicitly by using the
values of $K$, $\gamma$, and $\lambda$ given above.

\subsection{The mass-energy of self-gravitating photons}
\label{sec_mep}

\begin{figure}[htbp]
\centerline{
\includegraphics[width=8cm,angle=0]{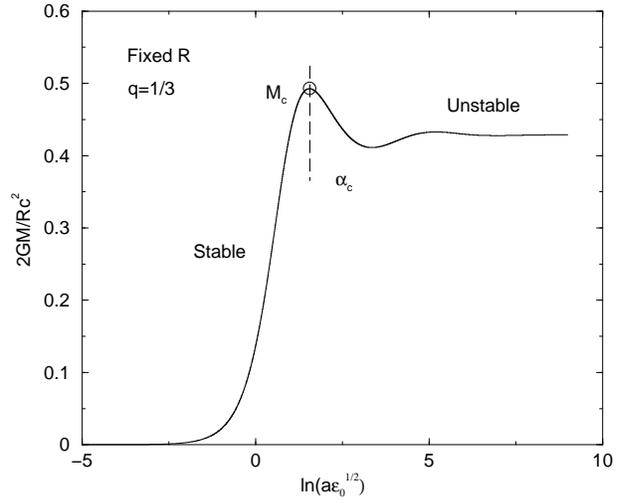}
} \caption[]{Mass as a function of the central density for a fixed box
radius. }
\label{meq13}
\end{figure}

We now apply the general results of Sect. \ref{sec_me} to the specific
context of a self-gravitating radiation. The consideration of an
explicit physical example allows us to determine the numerical value
of the multiplicative constants appearing in the scaling laws. It is
convenient to introduce the Planck length and the Planck mass,
\begin{equation}
L_{P}=\left (\frac{G\hbar}{c^{3}}\right )^{1/2}, \quad M_{P}=\left (\frac{\hbar c}{G}\right )^{1/2}.\label{mep1}
\end{equation}
In terms of these parameters, the relation (\ref{me2}) can be rewritten
\begin{equation}
\frac{M}{M_{P}}=\frac{1}{2}\chi(\alpha)\frac{R}{L_{P}}, \label{mep2}
\end{equation}
with
\begin{equation}
\alpha=\left (\frac{16\pi G\epsilon_{0}}{c^{4}}\right )^{1/2}R.
\label{mep3}
\end{equation}
For a given box radius, the mass-central density relation
$M(\epsilon_0)$ presents damped oscillations around an asymptote
\begin{equation}
M_s=\frac{1}{2}\chi_s M_P \frac{R}{L_{P}}, \label{mep4fxgdr}
\end{equation}
with $\chi_s=3/7$ corresponding to the singular solution (see
Fig. \ref{meq13}). There is no equilibrium state with a mass larger
than
\begin{equation}
M_c=\frac{1}{2}\chi_c M_P \frac{R}{L_{P}}, \label{mep4}
\end{equation}
where $\chi_c=0.493$ corresponding to $\alpha_c=4.7$. On the other
hand, the configurations with central density
\begin{equation}
\epsilon_0>\epsilon_{0,c}\equiv \frac{c^4}{16\pi G}\left
(\frac{\alpha_c}{R}\right )^2,\label{mep5}
\end{equation}
corresponding to $\alpha>\alpha_c$ are dynamically and
thermodynamically unstable (see Sect. \ref{sec_tst}).

\begin{figure}[htbp]
\centerline{
\includegraphics[width=8cm,angle=0]{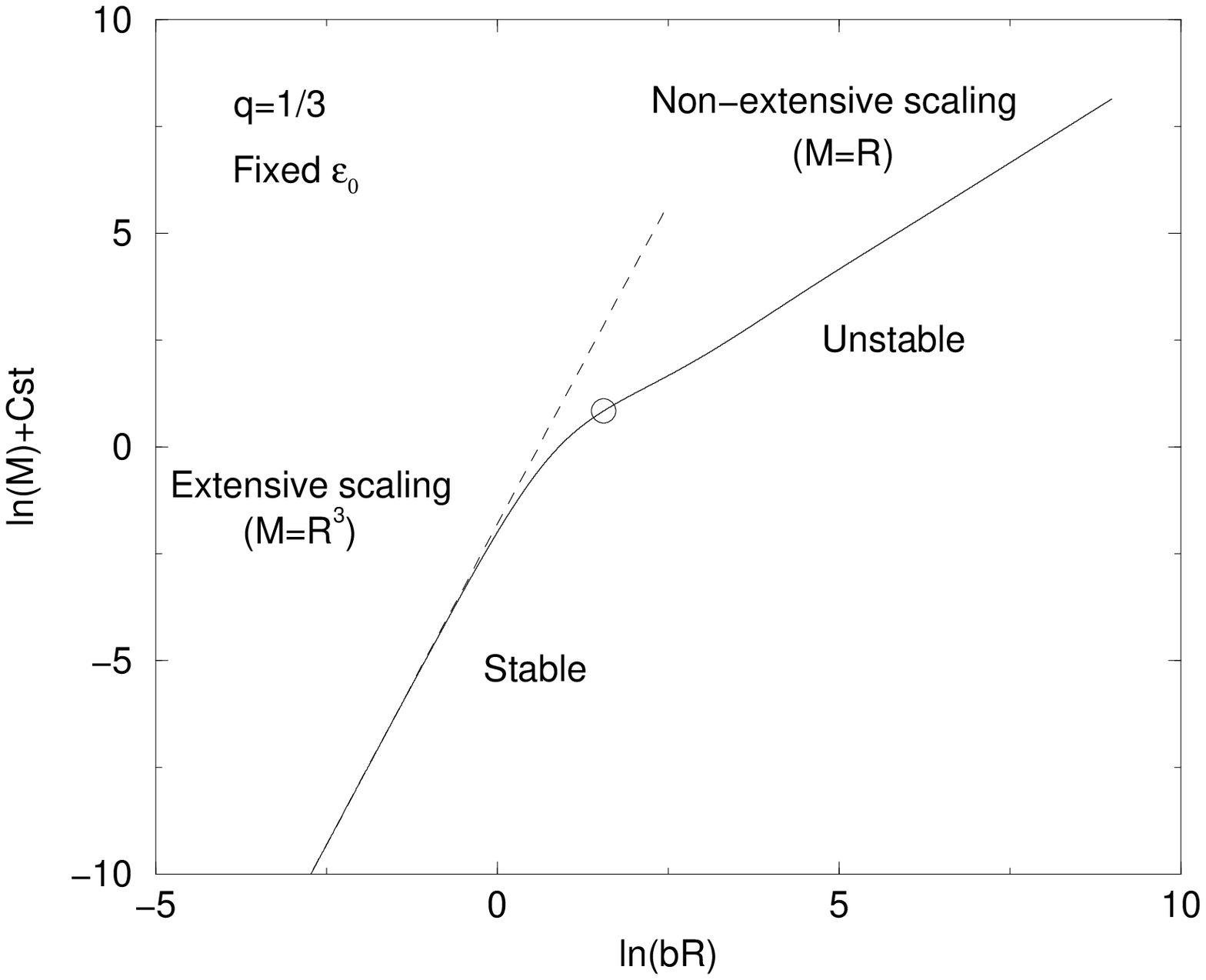}
} \caption[]{Mass as a function of the box
radius for a fixed central density. }
\label{mrq13}
\end{figure}

For a given central density, the mass-radius relation $M(R)$ is
represented in Fig. \ref{mrq13}. For small radii, we have the extensive
scaling $M\sim R^3$ and for large radii, we get the scaling law
\begin{equation}
\frac{M}{M_{P}}=\frac{1}{2}\chi_{s}\frac{R}{L_{P}}, \qquad (R\rightarrow +\infty). \label{mep6}
\end{equation}
However, configurations with
\begin{equation}
R>R_c\equiv \alpha_c \left (\frac{c^{4}}{16\pi G\epsilon_{0}}\right
)^{1/2}, \label{mep7}
\end{equation}
corresponding to $\alpha>\alpha_c=4.7$, are dynamically and
thermodynamically unstable. This corresponds to
\begin{equation}
 M>M_c\equiv \frac{1}{2}\chi_c \alpha_c \left (\frac{c^{8}}{16\pi G^{3}\epsilon_{0}}\right
)^{1/2}.\label{mep7bis}
\end{equation}

\subsection{The entropy of self-gravitating photons}
\label{sec_esgp}

\begin{figure}[htbp]
\centerline{
\includegraphics[width=8cm,angle=0]{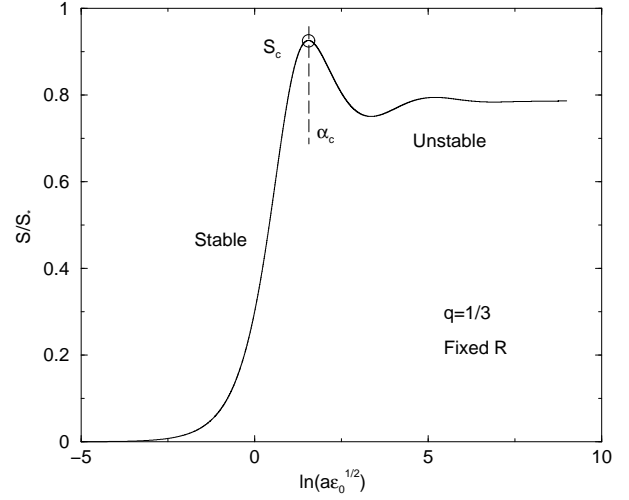}
} \caption[]{Entropy as a function of the central density for a
fixed box radius. } \label{seq13}
\end{figure}

\begin{figure}[htbp]
\centerline{
\includegraphics[width=8cm,angle=0]{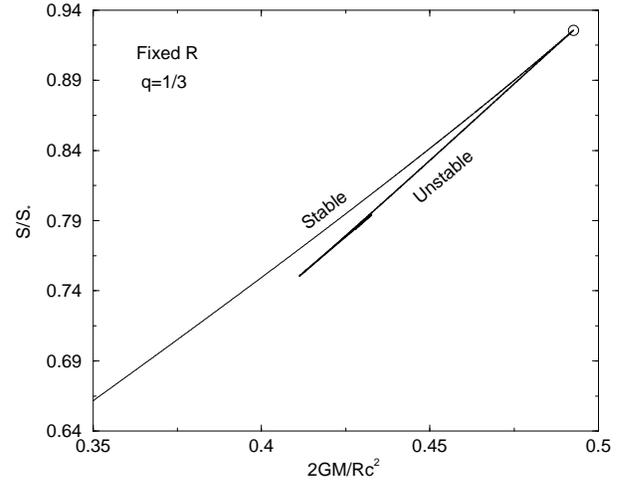}
} \caption[]{Entropy as a function of mass for a fixed box radius. }
\label{SMrfixedq13}
\end{figure}

The total entropy of the gas of photons is
\begin{equation}
S=\int_{0}^{R}s(r)\left\lbrack
1-\frac{2GM(r)}{rc^{2}}\right\rbrack^{-1/2}4\pi r^{2}dr. \label{esgp0}
\end{equation}
Using Eq. (\ref{sbl8}), we find that
\begin{equation}
S=\lambda N,
\label{esgp1}
\end{equation}
where $N$ is the number of photons (\ref{ege3}). Using the  results of Sect.
\ref{sec_bn}, we obtain
\begin{equation}
S=A\Delta(\alpha)k_{B}\left (\frac{R}{L_{P}}\right )^{3/2},
\label{esgp2}
\end{equation}
with the numerical  constant
\begin{equation}
A=\frac{1}{3}\left (\frac{8\pi^{3}}{15}\right )^{1/4}=0.672188...
\label{esgp3}
\end{equation}
For a given box radius, the entropy vs central density $S(\epsilon_0)$
presents damped oscillations around an asymptote 
\begin{equation}
S_s=A\Delta_s k_{B}\left (\frac{R}{L_{P}}\right )^{3/2} \label{esgp4fghfgy}
\end{equation}
with $\Delta_{s}=(8/21)^{1/4}$, corresponding to the singular solution
(see Fig. \ref{seq13}). There is no equilibrium state with an entropy
greater than
\begin{equation}
S_c=A\Delta_c k_{B}\left (\frac{R}{L_{P}}\right )^{3/2}, \label{esgp4}
\end{equation}
where $\Delta_c=0.925$ corresponding to $\alpha_c=4.7$. The
configurations with central density $\epsilon_0>\epsilon_{0,c}$ are
dynamically and thermodynamically unstable. Eliminating $\epsilon_0$
between Eqs. (\ref{mep2}) and (\ref{esgp2}), we obtain the entropy vs
mass curve $S(M)$. This curve is represented in
Fig. \ref{SMrfixedq13}. Since the peaks of entropy coincide with the
mass peaks in Figs. \ref{meq13} and \ref{seq13}, this implies that the
entropy vs mass curve presents angular points at each peak
\footnote{This is similar to the entropy vs energy curve $S(E)$ in
Newtonian gravity (see Fig. 4 of Chavanis 2002d).}.  However, only the
upper part of the curve, corresponding to $\alpha<\alpha_c$, is stable.

\begin{figure}[htbp]
\centerline{
\includegraphics[width=8cm,angle=0]{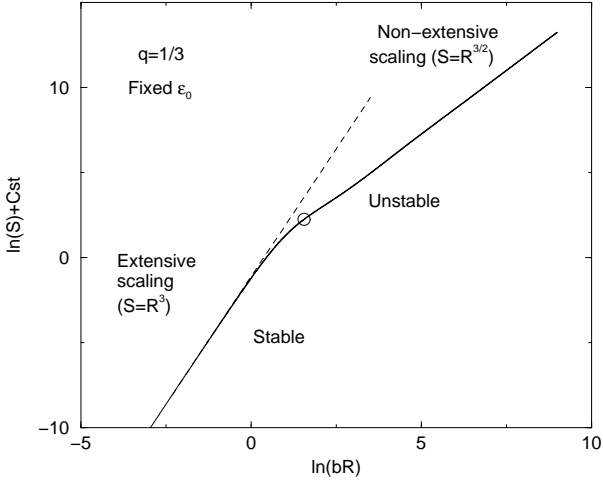}
} \caption[]{Entropy as a function of the box
radius for a fixed central density. }
\label{srq13}
\end{figure}

\begin{figure}[htbp]
\centerline{
\includegraphics[width=8cm,angle=0]{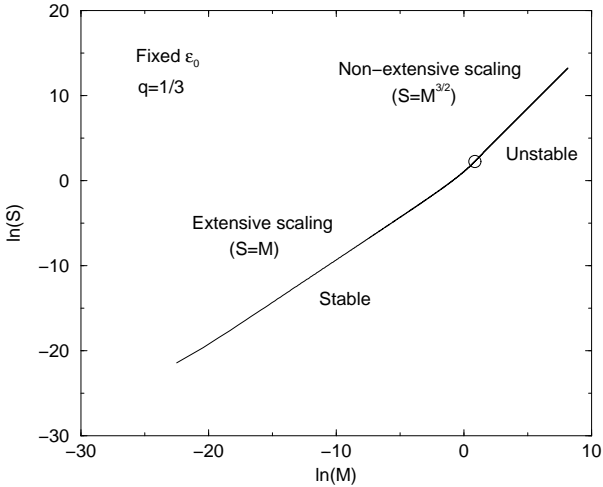}
} \caption[]{Entropy as a function of mass for a fixed central density. }
\label{SMefixedq13}
\end{figure}

For a given central density, the entropy vs radius relation $S(R)$
is represented in Fig. \ref{srq13}. For small radii, we have the extensive
scaling $S\sim R^3$ and, for large radii, we get the scaling law
\begin{equation}
S=A\Delta_{s}k_{B}\left (\frac{R}{L_{P}}\right )^{3/2}, \qquad (R\rightarrow +\infty). \label{esgp5}
\end{equation}
However, configurations with $R>R_c$
are dynamically and thermodynamically unstable. This corresponds to 
\begin{equation}
S>S_{c}=A\Delta_{c}\alpha_{c}^{3/2}k_{B}\left (\frac{c^{7}}{16\pi G^{2}\epsilon_{0}\hbar}\right )^{3/4}. \label{esgp5b}
\end{equation}
Finally, eliminating $R$
between Eqs. (\ref{mep2}) and (\ref{esgp2}), we obtain the entropy vs
mass curve $S(M)$ represented in Fig. \ref{SMefixedq13}. For small
mass, we have the extensive scaling $S\sim M$ and for large mass, we
get
\begin{equation}
S=2^{3/2}A\frac{\Delta_s}{\chi_s^{3/2}}k_B \left
(\frac{M}{M_P}\right )^{3/2}. \label{esgp6}
\end{equation}

\subsection{Thermodynamical stability and temperature}
\label{sec_tst}

We have seen in Sect. \ref{sec_ege} that the maximization of the
particles number $N$ at fixed mass-energy $M$ provides a criterion of
formal nonlinear dynamical stability.  Since the entropy of the
self-gravitating radiation is proportional to the particle number
($S=\lambda N$) and since the energy is proportional to mass
($E=Mc^2$), we conclude that the criterion of formal nonlinear dynamical
stability is equivalent to the maximization of the entropy at fixed
energy:
\begin{equation}
{\rm Max}\quad \lbrace S[\epsilon]\quad |\quad E[\epsilon]=E \quad {\rm
fixed}\rbrace, \label{tst1}
\end{equation}
that is to say, to the criterion of thermodynamical stability in the
microcanonical ensemble.  This proves very simply that the criteria of
dynamical and thermodynamical stability coincide \footnote{For
isothermal spheres in Newtonian gravity (Chavanis 2006a), we have found
that the minimization of the energy functional ${\cal W}$ at fixed
mass (formal nonlinear dynamical stability) is equivalent to the minimization
of the Boltzmann free energy $F_B$ at fixed mass (thermodynamical
stability in the canonical ensemble).}. Now, using $S=\lambda N$ and
$E=Mc^2$, the variational principle (\ref{ege5}) determining the critical
points, can be rewritten as
\begin{equation}
\delta S=\frac{1}{T}\delta E, \label{tst2}
\end{equation}
with $T\equiv c^{2}/(\lambda\mu)$. This can be interpreted as the
first principle of thermodynamics where $T$ is identified with the
thermodynamical temperature. Now, using Eqs. (\ref{f7}) and (\ref{sbl3}), we find that
\begin{equation}
T={T(R)}\left (1-\frac{2GM}{Rc^{2}}\right )^{1/2}, \label{tst3}
\end{equation}
where $T(R)$ is the surface temperature. Comparing Eq. (\ref{tst3}) with the
Tolman relation (\ref{tr7}), we observe that $T$ is equal to the temperature
measured by an observer at infinity: 
\begin{equation}
T=T_{0}. \label{tst4}
\end{equation}

\begin{figure}[htbp]
\centerline{
\includegraphics[width=8cm,angle=0]{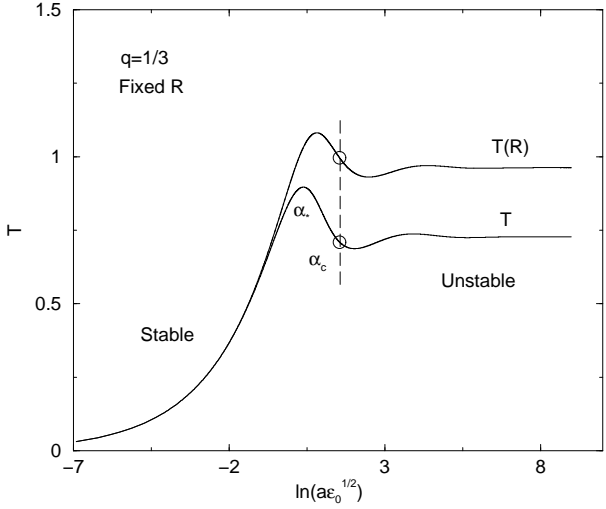}
} \caption[]{Temperature as a function of the central density for a
fixed box radius. This curve represents actually the function
$\theta(\alpha)$. }
\label{Teq13}
\end{figure}

\begin{figure}[htbp]
\centerline{
\includegraphics[width=8cm,angle=0]{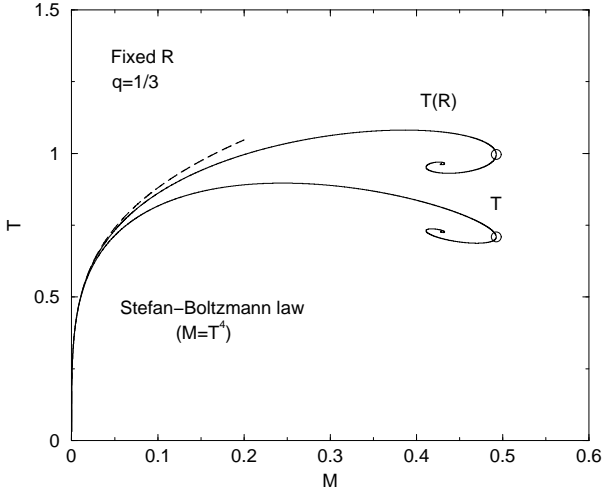}
} \caption[]{Temperature as a function of the energy for a fixed box
radius. We have plotted $\theta(\alpha)$ as a function of
$\chi(\alpha)$. 
}\label{TMrfixedq13}
\end{figure}

\begin{figure}[htbp]
\centerline{
\includegraphics[width=8cm,angle=0]{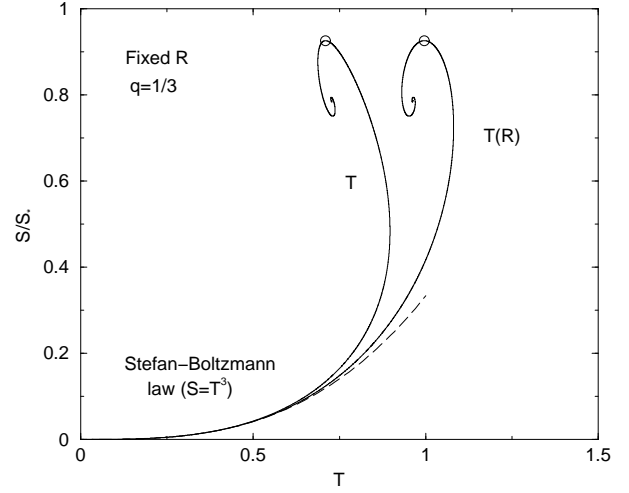}
} \caption[]{Entropy as a function of the temperature for a fixed box
radius. We have plotted $\Delta(\alpha)$ as a function of $\theta(\alpha)$.  } \label{STrfixedq13}
\end{figure}

Using Eqs (\ref{sbl3}), (\ref{gre8}), (\ref{gre7}) and (\ref{me2}), we find that the temperature $T$ is related
to the parameter $\alpha$ by the relation
\begin{equation}
k_{B}T=\frac{1}{3A}\theta(\alpha)M_{p}c^{2}\left(\frac{L_{P}}{R}\right )^{1/2}, \label{tst5}
\end{equation}
where
\begin{equation}
\theta(\alpha)=\frac{\alpha^{1/2}}{{\cal R}(\alpha)^{1/4}}\lbrack
1-\chi(\alpha)\rbrack^{1/2}. \label{tst6}
\end{equation}
For a given box radius, Eq. (\ref{tst5}) determines the relation
$T(\epsilon_0)$ between the temperature and the central density. This
curve (see Fig. \ref{Teq13}) presents damped oscillations around the
temperature  of the singular solution corresponding to $\theta_{s}=Q^{1/4}(1-\chi_{s})^{1/2}=(96/343)^{1/4}$. There is no
equilibrium state above a critical temperature
\begin{equation}
k_{B}T_*=\frac{1}{3A}\theta_* M_{p}c^{2}\left(\frac{L_{P}}{R}\right
)^{1/2}, \label{tst7}
\end{equation}
with $\theta_*=0.897$ corresponding to the first peak at
$\alpha_*=1.47$.  Since $\alpha_*<\alpha_c$, the dynamical and
thermodynamical instability occurs {\it after} the first peak of temperature
\footnote{We can wonder whether the turning point of mass-energy
$M(\epsilon_{0})$ is associated with a loss of microcanonical
stability and the turning point of temperature $T(\epsilon_{0})$ is
associated with a loss of canonical stability as for Newtonian
isothermal spheres; see Antonov (1962), Lynden-Bell \& Wood (1968),
Katz (1978), Padmanabhan (1990) and Chavanis (2002a). Note, however,
that the mass-energy $M[\epsilon]$ is a {\it linear} functional of the
density $\epsilon$, like the mass $M[\rho]$ in Newtonian gravity and
contrary to the energy $E[\rho]$ in Newtonian gravity.  Furthermore,
we note that a self-gravitating radiation becomes unstable {\it above}
a critical temperature or mass-energy, while a classical
self-gravitating isothermal gas becomes unstable {\it below} a
critical temperature or energy. These are important differences that
need to be discussed further.  }. Eliminating the central density
between Eqs. (\ref{mep2}), (\ref{esgp2}) and (\ref{tst5}), we obtain
the temperature vs energy curve $T(M)$ and the entropy vs temperature
curve $S(T)$. For small values of mass or temperature, self-gravity is
negligible and we recover the Stefan-Boltzmann law.  For larger
values, general relativistic effects must be taken into account and
the curves form a spiral. Figures
\ref{TMrfixedq13} and
\ref{STrfixedq13} show the deviation from the Stefan-Boltzmann law due
to general relativity. For a system evolving at fixed radius, the
series of equilibria is stable (dynamically and thermodynamically in
the microcanonical ensemble) until the first turning point of mass or
entropy. The caloric curve of Fig. \ref{TMrfixedq13} is strikingly
similar to the caloric curves $T(E)$ of Newtonian isothermal spheres
(see, e.g., Chavanis 2002a).

\begin{figure}[htbp]
\centerline{
\includegraphics[width=8cm,angle=0]{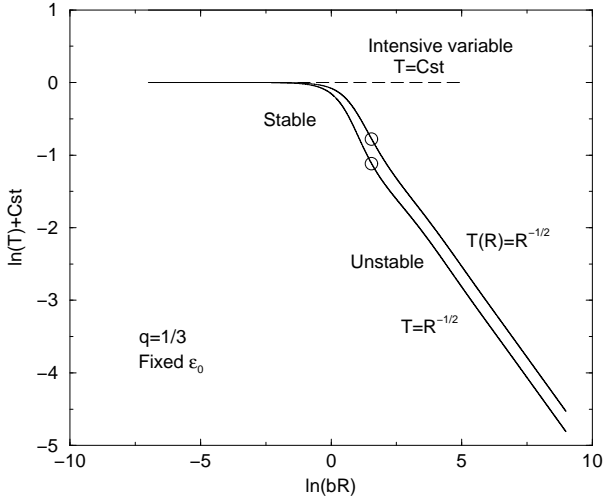}
} \caption[]{Temperature as a function of the box radius for a fixed
central density. This corresponds to the function
$\theta(\alpha)/\alpha^{1/2}$. } \label{Trq13}
\end{figure}

\begin{figure}[htbp]
\centerline{
\includegraphics[width=8cm,angle=0]{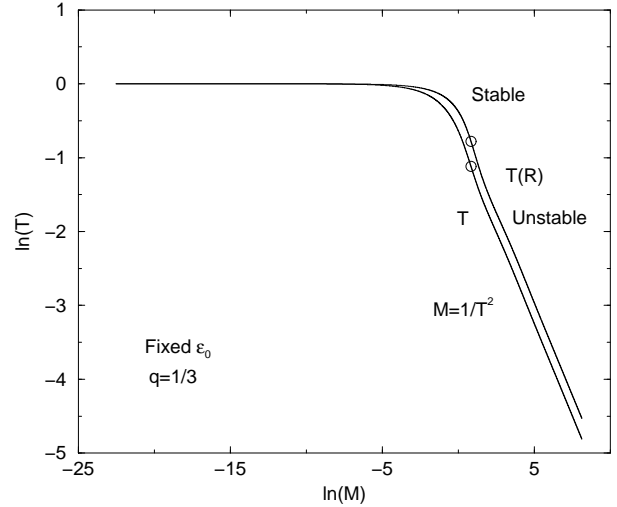}
} \caption[]{Temperature as a function of the energy for a fixed
central density. We have plotted $\theta(\alpha)/\alpha^{1/2}$ as a
function of $\alpha\chi(\alpha)$. } \label{TMefixedq13}
\end{figure}

\begin{figure}[htbp]
\centerline{
\includegraphics[width=8cm,angle=0]{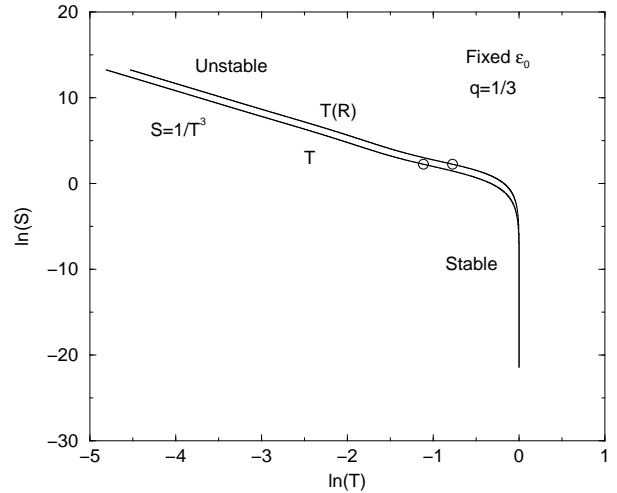}
} \caption[]{Entropy as a function of the temperature for a fixed
central density. We have plotted  $\Delta(\alpha)\alpha^{3/2}$ as a
function of $\theta(\alpha)/\alpha^{1/2}$.  }
\label{STefixedq13}
\end{figure}

For a given central density, the temperature vs radius plot $T(R)$ is
represented in Fig. \ref{Trq13}. For small radii, the temperature is
an intensive variable $T={\rm cst}$, and for large radii, we have the
scaling law
\begin{equation}
k_{B}T=\frac{\theta_{s}}{3A}
 M_{P}c^{2}\left
(\frac{L_{P}}{R}\right )^{1/2},
\label{tst8}
\end{equation}
with $\theta_{s}=(96/343)^{1/4}$. However, configurations with
$R>R_c$ are dynamically and thermodynamically unstable (in the
microcanonical ensemble). This corresponds to
\begin{equation}
T<T_c=\frac{1}{3A}(16\pi)^{3/4}\frac{\theta_{c}}{\alpha_{c}^{3/2}}\frac{G\hbar^{5/4}\epsilon_{0}^{3/4}}{c^{11/4}}. \label{tst9}
\end{equation}
Eliminating the radius between Eqs. (\ref{mep2}) and (\ref{tst5}), we
can express the temperature as a function of the mass-energy. For
small masses we have $T={\rm cst}$ and for large masses we obtain a
$T^{-2}$ law:
\begin{equation}
\frac{M}{M_P}=\frac{1}{18A^2}\chi_s \theta_s^2 \left (\frac{M_P
c^2}{k_B T}\right )^2. \label{tst10}
\end{equation}
However, the configurations are unstable for $M>M_c$ or $T<T_c$. Using
$E=Mc^{2}$, the specific heat of the self-gravitating radiation (for
sufficiently large masses or sufficiently low temperatures) is given by
\begin{equation}
C=\frac{dE}{dT}=-\frac{\chi_s \theta_s^2}{9A^2} k_{B}\frac{(M_P
c^2)^{3}}{(k_B T)^{3}}<0. \label{tst11}
\end{equation}
As for black holes (see Appendix \ref{sec_bht}), the specific heat
is negative. Finally, eliminating the radius between Eqs. (\ref{esgp2}) and
(\ref{tst5}), we can express the entropy as a function of
the temperature. For small entropies we have $T={\rm cst}$, and for
large entropies, we obtain a $T^{-3}$ law:
\begin{equation}
S= \frac{\Delta_{s}\theta_{s}^{3}}{27A^{2}} k_{B}\left
(\frac{M_{P}c^{2}}{k_{B}T}\right )^{3}. \label{tst12}
\end{equation}
However, the configurations are unstable for $S>S_c$ or $T<T_c$.  The
curves $T(M)$ and $S(T)$ for a fixed central density $\epsilon_{0}$
are plotted in Figs. \ref{TMefixedq13} and \ref{STefixedq13}.

\subsection{Comparison with the black hole entropy}
\label{sec_dbh}

The scaling laws (\ref{mep6}), (\ref{esgp5}), and (\ref{tst8}) that we
obtained for the mass, the entropy, and the temperature of a
self-gravitating radiation can be compared with the corresponding
scaling laws (\ref{bht3}), (\ref{bht6}), and (\ref{bht7}) for black
holes. For the self-gravitating radiation, we have
\begin{eqnarray}
M=\frac{3}{14}M_{P}\frac{R}{L_{P}}, \quad S_{rad}=\frac{2}{3}\left (\frac{4\pi^{3}}{315}\right )^{1/4} k_{B}\left (\frac{R}{L_{P}}\right )^{3/2},\nonumber\\
\label{dbh1}
\end{eqnarray}
\begin{equation}
k_{B} T_{rad}=\frac{3}{7}\left (\frac{140}{9\pi^{3}}\right )^{1/4} M_{P}c^{2}\left (\frac{L_{P}}{R}\right )^{1/2}.
\label{dbh2}
\end{equation}
They behave as $M\sim R$, $S_{rad}\sim R^{3/2}$, and $T_{rad}\sim R^{-1/2}$. For the black holes, we have
\begin{equation}
M=\frac{1}{2}M_{P}\frac{R}{L_{P}}, \quad S_{BH}=\pi k_{B}\left (\frac{R}{L_{P}}\right )^{2},
\label{dbh3}
\end{equation}
\begin{equation}
k_{B} T_{BH}=\frac{1}{4\pi} M_{P}c^{2} \frac{L_{P}}{R}.
\label{dbh4}
\end{equation}
They behave as $M\sim R$, $S_{BH}\sim R^{2}$, and $T_{BH}\sim
R^{-1}$. We have calculated the exact values of the numerical
constants appearing in the expressions of the mass, entropy, and
temperature of a self-gravitating radiation.  Comparing with black
holes, the form of the laws are similar but the exponents are
different. In particular, we note that $S_{rad}\propto S_{BH}^{3/4}$. For
 large enough radii $R$ where this scaling law applies we have
$S_{rad}\ll S_{BH}$ in agreement with the Bekenstein (1973) inequality.

\section{Other explicit examples}
\label{sec_gas}

In this section, we give other physical examples where the 
general theory developed in Sect. \ref{sec_rs} applies.

\subsection{Neutron stars}
\label{sec_ns}

In the simplest models, a neutron star can be viewed as an ideal gas of
self-gravitating fermions. Their density is so large that gravitation
must be described in terms of general relativity. The distribution
function of the neutrons is given by the Fermi-Dirac statistics
\begin{equation}
f=\frac{2}{h^{3}}\frac{1}{e^{\beta\epsilon-\mu}+1}, \label{ns1}
\end{equation}
where $\mu$ is the chemical potential and $\epsilon(p)$ the energy per
particle. In the completely degenerate limit ($T=0$), the neutrons
have momenta less than a threshold value $p_0$ (Fermi momentum) and
their distribution is $f=2/h^3$. There can only be $2$ neutrons in a
phase space element of size $h^3$ on account of the Pauli exclusion
principle. Furthermore, in the core of neutron stars, the neutrons are
ultra-relativistic so that $\epsilon(p)=pc$. According to
Sect. \ref{sec_u}, they are described by a linear equation of state of
the form (\ref{es3}) with $q=1/3$. On the other hand, using
Eqs. (\ref{u1}) and (\ref{u3}), the density and the pressure are given
by $n={8\pi p_0^3}/({3h^3})$ and $P={2\pi c p_0^4}/({3h^3})$.
Eliminating $p_0$ between these two expressions, we obtain
\begin{equation}
P=K n^{\gamma}, \label{ns2}
\end{equation}
with
\begin{equation}
\gamma=\frac{4}{3},  \qquad K=\frac{1}{8}\left (\frac{3}{\pi}\right
)^{1/3}hc. \label{ns3}
\end{equation}
Therefore, the core of neutron stars can be studied with the theory
developed in Sect. \ref{sec_rs} with $q=1/3$. The numerical values of $K$
and $\gamma$ are explicitly given by Eq. (\ref{ns3}). The singular
profiles are
\begin{equation}
\epsilon=\frac{3c^4}{56\pi G r^{2}}, \qquad P=\frac{c^4}{56\pi G
r^{2}}, \label{ns4}
\end{equation}
\begin{equation}
n=\left (\frac{c^9}{1029 G^3 h^3}\right )^{1/4} r^{-3/2}.
\label{ns5}
\end{equation}
These are also the asymptotic behaviours of the regular profiles for
$r\rightarrow +\infty$. On the other hand, for the singular
solution, the total mass and the total baryon number within a sphere
of radius $R$ are
\begin{equation}
\frac{M}{M_{P}}=\frac{1}{2}\chi_{s}\frac{R}{L_{P}}, \label{ns6}
\end{equation}
\begin{equation}
N=B\Delta_{s}\left (\frac{R}{L_{P}}\right )^{3/2}, \label{ns7}
\end{equation}
with $\chi_{s}=3/7$, $\Delta_{s}=(8/21)^{1/4}$ and 
\begin{equation}
B=\frac{1}{3}\left (\frac{4}{\pi}\right )^{1/4}=0.35408...  \label{ns8}
\end{equation}
The description of an ultra-relativistic and completely degenerate
gas of neutrons in a box (modelling the core of a neutron star) is
similar to the description of a self-gravitating radiation given in
Sect. \ref{sec_sgr}. We just need to replace the entropy (proportional to the
photon number) by the neutron number.

\subsection{Baryons interacting via a vector meson field}
\label{sec_bvm}

Finally, we consider a model introduced by Zel'dovich (1962) that
provides the stiffest equation of state $P=\epsilon$ compatible with
the requirements of relativity theory. It consists in a gas of
baryons interacting through a vector meson field (in addition to the
gravitational force). In the case of neutron stars, the gravitational
contraction is balanced by the quantum pressure $P=K n^{4/3}$ due to
the Pauli exclusion principle. In the Zel'dovich model, the
gravitational contraction is balanced by the pressure $P=Kn^2$ due
to the electrostatic repulsion of the baryons. For dense objects
($n\gg 1$), the form of pressure considered by Zel'dovich is expected
to prevail over the quantum pressure.

Let us first consider  a gas of charged baryons without gravitational
interaction. In the Zel'dovich model, the (repulsive) potential of interaction
between two charges is given by
\begin{equation}
u(\xi)=\frac{g^2}{\xi}e^{-\mu\xi}, \label{bvm1}
\end{equation}
where $\xi=|{\bf r}-{\bf r}'|$, $g$ is the charge of the baryons and
$\mu$ the quantum of mass of the meson (the mass of the meson is $m=\mu\hbar/c$). Since  $T=0$, the energy simply represents the rest mass energy of
the baryons $nm_b c^2$  plus the energy of interaction
\begin{eqnarray}
\epsilon=nm_bc^2+\frac{1}{2}n^2\int_0^{+\infty}u(\xi)4\pi\xi^2d\xi
=nm_bc^2+\frac{2\pi n^2 g^2}{\mu^2}. \nonumber\\
\label{bvm2}
\end{eqnarray}
The pressure is then given by
\begin{equation}
P=-\frac{\partial(\epsilon/n)}{\partial(1/n)}=\frac{2\pi n^2 g^2}{\mu^2}. \label{bvm3}
\end{equation}
It can be seen from Eqs. (\ref{bvm2}) and (\ref{bvm3}) that, in the
limit of large $n$, we have
\begin{equation}
P=\epsilon, \qquad (q=1).\label{bvm4}
\end{equation}
This represents the most rigid equation of state compatible with
relativity theory  since the velocity of sound
$(dP/d\epsilon)^{1/2}c$ is equal to the velocity of light for this
value of $q$. The value $q=1$ is clearly an upper bound
\footnote{Note that the above formulae are valid only for $\mu>1/R$
where $R$ is the system size. For $\mu=0$ (Coulomb interaction), the
potential is not screened and the baryons repel each other at large
distances so that a pressure must be imposed to retain the charges.
Now, using the Virial theorem $2K+W=3PV$ and considering $K=T=0$, we
obtain $P=\epsilon/3$ for the electromagnetic field. If we take the
contribution of the rest mass in $\epsilon$ into account, we get $P\le
\epsilon/3$ so that the upper bound is $q=1/3$ as advocated by Landau \&
Lifshitz (1960). The model of Zel'dovich (1962) shows that this bound
can be exceeded for a classical vector field with a mass $\mu$.}. On
the other hand, Eq. (\ref{bvm3}) can be rewritten
\begin{equation}
P=K n^{\gamma}, \label{bvm5}
\end{equation}
with
\begin{equation}
\gamma=2,  \qquad K=\frac{2\pi g^2\hbar^2}{m^2 c^2}. \label{bvm6}
\end{equation}
Therefore, a gas of baryons interacting through a vector meson field
can be studied with the theory developed in Sect. \ref{sec_rs} with
$q=1$. The numerical values of $K$ and $\gamma$ are explicitly given
by Eq. (\ref{bvm6}). The singular profiles are
\begin{equation}
\epsilon=P=\frac{c^4}{16\pi G}r^{-2}, \label{bvm7}
\end{equation}
\begin{equation}
n=\left (\frac{m^2 c^6}{32\pi^2 g^2 G\hbar^2}\right )^{1/2}r^{-1}. \label{bvm8}
\end{equation}
These are also the asymptotic behaviours of the regular profiles for
$r\rightarrow +\infty$. On the other hand, for the singular
solution, the total mass and total baryon number within a sphere of
radius $R$ can be written as
\begin{equation}
\frac{M}{M_{P}}=\frac{1}{2}\chi_{s}\frac{R}{L_{P}}, \label{bvm9}
\end{equation}
\begin{equation}
N=\Delta_{s}\frac{m^2 c^3}{g^2 \hbar}\left (\frac{R}{L_{P}}\right )^{2}, \label{bvm10}
\end{equation}
with $\chi_s=1/2$ and $\Delta_s=1/2$. Interestingly, these are the
same scalings as for the mass and the entropy of a black hole (see
Appendix \ref{sec_bht}). We have obtained the exact value of the
prefactor.

A more complete description of this system is provided by the analysis
of Sect. \ref{sec_rs}. The different curves have been plotted for $q=1$
so that they are directly relevant to the Zel'dovich model provided
that the numerical values  Eq. (\ref{bvm6}) are incorporated in the
discussion. The relation  (\ref{me2}) can be written
\begin{equation}
\frac{M}{M_{P}}=\frac{1}{2}\chi(\alpha)\frac{R}{L_{P}}, \label{bvm11}
\end{equation}
with
\begin{equation}
\alpha=\left (\frac{8\pi G\epsilon_{0}}{c^{4}}\right )^{1/2}R.
\label{bvm12}
\end{equation}
For a given box radius, the mass-central density relation
$M(\epsilon_0)$ presents damped oscillations around the mass $M_s$,
given by Eq. (\ref{bvm9}), corresponding to the singular solution (see
Figs. \ref{meq1} and \ref{meq1ZOOM}).  There is no equilibrium state
with a mass larger than
\begin{equation}
M_c=\frac{1}{2}\chi_c M_P \frac{R}{L_{P}}, \label{bvm13}
\end{equation}
where $\chi_c=0.544$ corresponding to $\alpha_c=4.05$. On the other
hand, the configurations with central density
\begin{equation}
\epsilon_0>\epsilon_{0,c}\equiv \frac{c^4}{8\pi G}\left
(\frac{\alpha_c}{R}\right )^2,\label{bvm14}
\end{equation}
corresponding to $\alpha>\alpha_c$ are dynamically unstable. For a
given central density, the mass-radius relation $M(R)$ is represented
in Fig. \ref{mrq1}. For small radii, we have the extensive scaling
$M\sim R^3$, and for large radii, we get the scaling law
(\ref{bvm9}). However, configurations with
\begin{equation}
R>R_c\equiv \alpha_c \left (\frac{c^{4}}{8\pi G\epsilon_{0}}\right
)^{1/2}, \label{bvm15}
\end{equation}
corresponding to $\alpha>\alpha_c$, are dynamically  unstable. This corresponds to
\begin{equation}
 M>M_c\equiv \frac{1}{2}\chi_c \alpha_c \left (\frac{c^{8}}{8\pi G^{3}\epsilon_{0}}\right
)^{1/2}.\label{bvmfg}
\end{equation}

On the other hand, the relation (\ref{bn3}) can be written
\begin{equation}
N=\Delta(\alpha)\frac{m^2 c^3}{g^2 \hbar}\left
(\frac{R}{L_{P}}\right )^{2}. \label{bvm16}
\end{equation}
For a given box radius, the baryon number vs central density
$N(\epsilon_0)$ presents damped oscillations (see Fig. \ref{neq1}). There is
no equilibrium state with $N$ larger than
\begin{equation}
N_c=\Delta_c\frac{m^2 c^3}{g^2 \hbar}\left (\frac{R}{L_{P}}\right
)^{2},\label{bvm17}
\end{equation}
where $\Delta_c=0.546$, corresponding to $\alpha_c=4.05$. The
configurations with central density $\epsilon_0>\epsilon_{0,c}$ are
dynamically unstable.  Eliminating $\epsilon_0$ between
Eqs. (\ref{bvm11}) and (\ref{bvm16}), we obtain the baryon number vs
mass curve $N(M)$. Since the peaks of baryon number coincide with the
mass peaks in Figs. \ref{meq1} and
\ref{neq1}, this implies that the baryon number vs mass curve presents
angular points at each peak $\alpha_n$ (see Fig. \ref{MNq1}). For a
given central density, the baryon number vs radius relation $N(R)$ is
represented in Fig. \ref{nrq1}. For small radii, we have the extensive
scaling $S\sim R^3$, and for large radii, we get the scaling law
(\ref{bvm10}). However, configurations with $R>R_c$ are dynamically
unstable. Eliminating $R$ between Eqs.  (\ref{bvm11}) and
(\ref{bvm16}), we obtain the baryon number vs mass curve $N(M)$
represented in Fig. \ref{MNq1fixede0}. For small mass, we have the
extensive scaling $N\sim M$ and for large mass, we get
\begin{equation}
N=4\frac{\Delta_s}{\chi_s^{2}}\frac{m^2 c^3}{g^2 \hbar} \left
(\frac{M}{M_P}\right )^{2}. \label{bvm18}
\end{equation}

\section{Relativistic stars in $d$ dimensions}
\label{sec_dimd}

In this section, we extend the theory of relativistic stars with a
linear equation of state to $d$-dimensional spheres. This study
completes our investigations (Sire \& Chavanis 2002, Chavanis \& Sire
2004, Chavanis 2004, 2006a, 2006b, 2007a) of the dependence of the
laws of physics, regarding Newtonian gravity, on the dimension of
space. It can also have applications in the theory of compact objects
where extra dimensions can appear on the microscale, an idea
originating from the Kaluza-Klein theory.

\subsection{The equations governing equilibrium}
\label{sec_eged}

Let us consider the $(d+1)$-dimensional, spherically symmetric metric
given by
\begin{eqnarray}
ds^2=e^{\nu(r)}d(ct)^2-e^{\lambda(r)}dr^2-r^2 d\Omega. \label{eged1}
\end{eqnarray}
For a perfect fluid, the Einstein equations of general relativity
governing the hydrostatic equilibrium of a spherical distribution of
matter are
\begin{eqnarray}
\frac{d}{dr}(re^{-\lambda})-(3-d)e^{-\lambda}=(d-2)-\frac{16\pi
G}{(d-1)c^4}r^2 \epsilon, \label{eged2}
\end{eqnarray}
\begin{eqnarray}
\frac{dP}{dr}=-\frac{1}{2}(P+\epsilon) \frac{d\nu}{dr}, \label{eged3}
\end{eqnarray}
\begin{eqnarray}
\frac{e^{-\lambda}}{r}\frac{d\nu}{dr}=\frac{1}{r^2}(d-2)(1-e^{-\lambda})+\frac{16\pi
G}{(d-1)c^4}P. \label{eged4}
\end{eqnarray}
Combining Eqs. (\ref{eged2})-(\ref{eged4}), we obtain the
$d$-dimensional generalization of the Oppenheimer-Volkoff equation
\begin{eqnarray}
\left\lbrace 1-\frac{aGM(r)}{c^{2}r^{d-2}}\right\rbrace \frac{dP}{dr}=\qquad\qquad\qquad\qquad\qquad\qquad\nonumber\\
-\frac{1}{c^{2}}(\epsilon+P)\left\lbrace
(d-2)\frac{aGM(r)}{2r^{d-1}}+\frac{8\pi G}{(d-1)c^{2}}P
r\right\rbrace, \label{eged5}
\end{eqnarray}
with
\begin{equation}
M(r)=\frac{S_{d}}{c^{2}}\int_{0}^{r}\epsilon r^{d-1}dr,
\label{eged6}
\end{equation}
and
\begin{equation}
S_{d}=\frac{2\pi^{d/2}}{\Gamma(d/2)}, \qquad a=\frac{16\pi}{(d-1)S_{d}}.
\label{eged7}
\end{equation}
These equations are  only defined for $d>1$. For $d=2$, we have
$S_{2}=2\pi$ and $a=8$. For $d=3$, we have $S_{3}=4\pi$ and $a=2$.  We
also recall that the value of the gravity constant $G$ and of its
dimension $G\sim R^{d}/(MT^2)$ 
changes with the dimension of space $d$. The dimension $d=2$ is
critical and will be treated separately in Sect. \ref{sec_tdg}.  In
this section, we consider that $d>2$.

Using Eqs. (\ref{eged2}) and (\ref{eged4}), we find that the metric functions
$\lambda(r)$ and $\nu(r)$ satisfy the relations
\begin{eqnarray}
e^{-\lambda(r)}=1-\frac{a G M(r)}{c^2 r^{d-2}}, \label{eged8}
\end{eqnarray}
\begin{eqnarray}
\frac{d\nu}{dr}=\frac{1+\frac{16\pi P r^d}{(d-1)(d-2)a M(r)
c^2}}{\frac{r}{d-2}\left (\frac{r^{d-2}c^2}{a G M(r)}-1\right )}.
\label{eged9}
\end{eqnarray}
In the empty space surrounding the star, $P=\epsilon=0$. Therefore, if
$M=M(R)$ denotes the total mass-energy of the star,
Eqs. (\ref{eged8})-(\ref{eged9}) become for $r>R$
\begin{eqnarray}
e^{-\lambda(r)}=1-\frac{a G M}{c^2 r^{d-2}}, \qquad
\frac{d\nu}{dr}=\frac{d-2}{r\left (\frac{r^{d-2}c^2}{a G M}-1\right
)}.\label{eged10}
\end{eqnarray}
The second equation is readily integrated in
\begin{eqnarray}
\nu(r)=\ln \left (1-\frac{aGM}{r^{d-2}c^2}\right ). \label{eged11}
\end{eqnarray}
Substituting the foregoing expressions for
$\lambda$ and $\nu$ in Eq. (\ref{eged1}), we obtain the $(d+1)$-dimensional
generalization of the Schwarzschild form of the metric outside a
star
\begin{eqnarray}
ds^2= \left (1-\frac{aGM}{r^{d-2}c^2}\right ) d(ct)^2-
\frac{dr^2}{1-\frac{a G M}{c^2 r^{d-2}}}-r^2 d\Omega. \label{eged12}
\end{eqnarray}
This metric is singular at
\begin{eqnarray}
r=\left ( \frac{a G M}{c^2}\right )^{\frac{1}{d-2}}\equiv R_S(d).
\label{eged13}
\end{eqnarray}
This does not mean that spacetime is singular at that radius but only that this particular metric is. Indeed, the singularity can be removed by a judicious change of coordinates (see, e.g,  Weinberg 1972). Furthermore, it can
be shown that, for a spherical system in hydrostatic equilibrium, the radius of the star satisfies
\begin{eqnarray}
R\ge \left \lbrack \frac{d^2}{4(d-1)} \ \frac{aGM}{c^{2}}\right\rbrack^{\frac{1}{d-2}},\label{eged14}
\end{eqnarray}
which is the generalization of the Buchdahl theorem in $d$ dimensions
(Ponce de Leon \& Cruz 2000). Therefore, for $d>2$, the points
outside the star always satisfy $r>R_S(d)$ so that the
Schwarzschild metric is never singular for these stars.

\subsection{The general relativistic Emden equation}
\label{sec_ed}

Considering the linear equation of state (\ref{es3}), we 
introduce the dimensionless variables $\xi$, $\psi$, and $M(\xi)$ by
the relations
\begin{equation}
\epsilon=\epsilon_{0}e^{-\psi}, \qquad r=\left\lbrace \frac{(d-1)c^{4}q}{8\pi G\epsilon_{0}(1+q)}\right\rbrace^{1/2}\xi,
\label{ed1}
\end{equation}
and
\begin{equation}
M(r)=\frac{S_{d}\epsilon_{0}}{c^{2}}\left\lbrace \frac{(d-1)c^{4}q}{8\pi G\epsilon_{0}(1+q)}\right\rbrace^{d/2}M(\xi).
\label{ed2}
\end{equation}
In terms of these variables, Eqs. (\ref{eged5}) and (\ref{eged6}) can be reduced to the following dimensionless forms
\begin{eqnarray}
\left\lbrace 1-\frac{2q}{1+q}\frac{M(\xi)}{\xi^{d-2}}\right\rbrace \frac{d\psi}{d\xi}=(d-2)\frac{M(\xi)}{\xi^{d-1}}+q\xi e^{-\psi},
\label{ed3}
\end{eqnarray}
\begin{equation}
\frac{dM}{d\xi}=\xi^{d-1}e^{-\psi}.
\label{ed4}
\end{equation}
The Newtonian limit corresponds to $q\rightarrow 0$ (see Appendix
\ref{sec_nl}).  Taking the derivative of Eq. (\ref{ed4}), we find that
\begin{equation}
\psi'=\frac{d-1}{\xi}-\frac{M''}{M'}.
\label{ed5}
\end{equation}
Substituting Eqs. (\ref{ed4}) and (\ref{ed5}) in Eq. (\ref{ed3}), we
obtain a differential equation for the mass profile
\begin{equation}
\frac{d-1}{\xi}-\frac{M''}{M'}=\frac{(d-2)\frac{M(\xi)}{\xi}+qM'}{\xi^{d-2}-\frac{2q}{1+q}M(\xi)}.
\label{ed6}
\end{equation}
In addition, the metric functions determined by Eqs. (\ref{eged8}) and
(\ref{eged3}) can be expressed as
\begin{equation}
e^{-\lambda}=1-\frac{2q}{1+q}\frac{M(\xi)}{\xi^{d-2}}, \qquad \nu=\frac{2q}{1+q}\psi+{\rm cst},
\label{ed7}
\end{equation}
where the constant is determined by the matching with the outer
Schwarzschild solution (\ref{eged11}) at $r=R$.

\subsection{Singular solution}
\label{sec_ssd}

For $d>2$, Eqs. (\ref{ed3}) and (\ref{ed4}) admit a singular solution
of the form
\begin{equation}
e^{-\psi_{s}}=\frac{Q}{\xi^{2}}, \quad {\rm where}\quad
Q=\frac{2(d-2)(1+q)}{(d-2)(1+q)^{2}+4q}. \label{ssd1}
\end{equation}
For this singular solution, the metric is of the form
\begin{equation}
e^{\lambda}=1+\frac{4q}{(d-2)(1+q)^{2}}, \qquad e^{\nu}=A\xi^{\frac{4q}{1+q}}. \label{ssd2}
\end{equation}
Coming back to original variables, the singular energy density profile is
\begin{equation}
\epsilon_{s}(r)=\frac{P_{s}(r)}{q}=\frac{(d-1)qQc^{4}}{8\pi G(1+q)}\ \frac{1}{r^{2}}.
\label{ssd3}
\end{equation}
Considering the thermodynamical variables of Sect. \ref{sec_ss}, we
find the scalings \footnote{In particular, the entropy scales with the
energy $E=Mc^{2}$ like $S\sim E^{b}$ with $b=(dq+d-2)/\lbrack
(q+1)(d-2)\rbrack$. Interestingly, this is the same scaling as the one
obtained by Kalyana Rama (2007) in a cosmological context once the
physical size $L_H$ of the horizon in his paper is identified with the
box size $R$ (K.R., private communication). This author studies the
phase transition, below a critical temperature, from a universe
dominated by highly excited strings to a FRW universe. He then argues
that the final spacetime configuration $(q,d)$ that emerges should
maximize the entropy at fixed energy (with the additional condition
$d\ge 3$). This leads to $q=1$ and $d=3$, providing a possible
explanation of why our universe is three-dimensional. According to
this approach, we have passed from a universe with $d=9$ dimensions
dominated by strings to a three-dimensional universe dominated by
radiation, then particles. Using completely different arguments, we
have also found that the dimension $d=3$ of our universe is
special. For example, compact objects like white dwarf stars would be
unstable in a universe with $d\ge 4$ dimensions (Chavanis 2007a). The
modifications of the laws of gravity with the dimension of space and
the special role played by the dimension $d=3$ are very intriguing.}
\begin{equation}
n(r)=\frac{s(r)}{\lambda}\propto r^{-2/(q+1)}, \qquad T(r)\propto r^{-2q/(q+1)},
\label{ssd4}
\end{equation}
\begin{equation}
M\propto R^{d-2}, \quad S\propto N\propto R^{\frac{dq+d-2}{q+1}}, \quad T\propto R^{-2q/(q+1)}.
\label{ssd5}
\end{equation}
The constants of proportionality can be easily obtained from the
expressions in Sect. \ref{sec_ss}. For $q=0$, corresponding to
Newtonian isothermal stars, we obtain the classical scalings $E\sim
N$, $S\sim N$ and $T\sim 1$ with $N\sim R^{d-2}$ (Chavanis 2004). For
$q=1$, corresponding to the stiffest stars, we find that $M\sim
R^{d-2}$ and $S\propto N\propto R^{d-1}$ so that the entropy scales
like the area in any dimension of space (Banks et al. 2002). For
$q<1$, it scales according to a power less than the area. On the other
hand, the temperature scales like $T\propto 1/R$ in any dimension. For
$q=1/d$, corresponding to a self-gravitating radiation or to a neutron
star in $d$ dimensions, we have the scaling laws
\begin{equation}
P\propto \epsilon\propto r^{-2}, \quad n\propto s\propto r^{-\frac{2d}{d+1}}, \quad  T\propto r^{-\frac{2}{d+1}},
\label{ssd6}
\end{equation}
\begin{equation}
M\propto  R^{d-2}, \quad S\propto N\propto R^{\frac{d(d-1)}{d+1}}, \quad T\propto R^{-\frac{2}{d+1}}.
\label{ssd7}
\end{equation}
Finally, for the sake of completeness, we give the expression of the
Stefan-Boltzmann law in $d$ dimensions. From the Bose-Einstein distribution 
with a chemical potential $\mu=0$, we find that the pressure, the density of photons, and the entropy density are given by
\begin{equation}
P=\frac{1}{d}\epsilon=\frac{(d-1)!S_{d}}{h^{d}c^{d}}(k_{B}T)^{d+1}\zeta(d+1),
\label{ssd8}
\end{equation}
\begin{equation}
n=\frac{(d-1)!S_{d}}{h^{d}c^{d}}(k_{B}T)^{d}\zeta(d),
\label{ssd9}
\end{equation}
\begin{equation}
s=k_{B}\frac{(1+d)(d-1)!S_{d}}{h^{d}c^{d}}(k_{B}T)^{d}\zeta(d+1).
\label{ssd10}
\end{equation}
The equivalent expressions for neutron stars in $d$ dimensions are
given in Chavanis (2007a).

\subsection{Asymptotic behaviour}
\label{sec_ab}

\begin{figure}[htbp]
\centerline{
\includegraphics[width=8cm,angle=0]{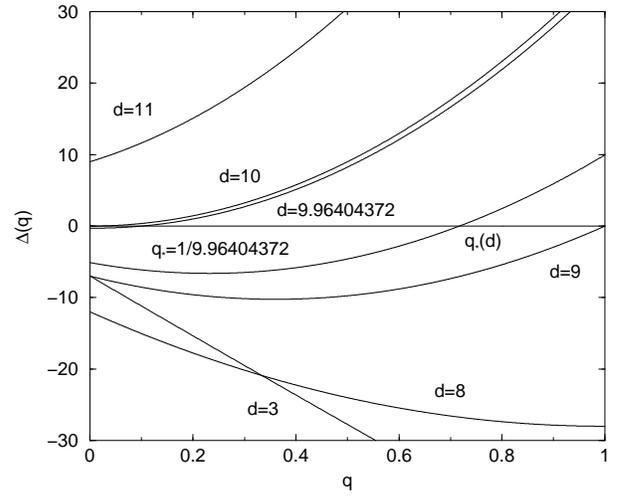}
} \caption[]{Discriminant $\Delta(q)$ as a function of $q$ for
different values of the dimension of space $d$ (characteristic
dimensions are described in the text). The mass-central density
relation $M(\epsilon_{0})$ presents damped oscillations when
$\Delta(q)<0$ and a monotonic behaviour when $\Delta(q)\ge 0$.  }
\label{deltaq}
\end{figure}

\begin{figure}[htbp]
\centerline{
\includegraphics[width=8cm,angle=0]{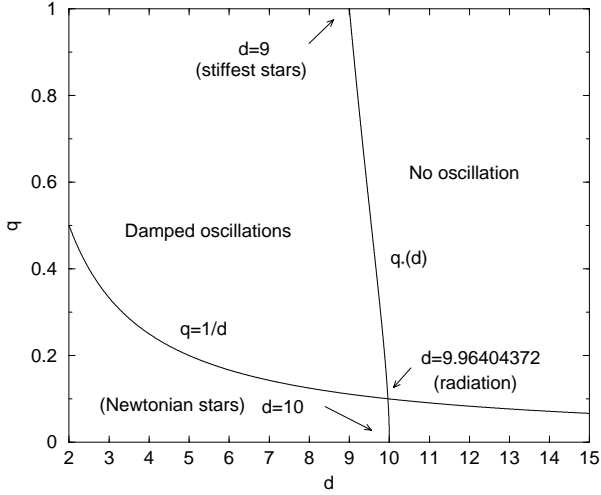}
} \caption[]{Phase diagram in the $(d,q)$ plane. The line $q_{*}(d)$
separates the region where the mass-central density relation
$M(\epsilon_{0})$ presents damped oscillations from the region where
the mass-central density relation $M(\epsilon_{0})$ is monotonic. The critical dimension corresponding to a self-gravitating radiation is obtained by taking the intersection between the line $q_{*}(d)$ and the curve $q=1/d$.  }
\label{qd}
\end{figure}

Considering now the regular solutions of Eqs. (\ref{ed3})-(\ref{ed4}),
we can always suppose that $\epsilon_0$ represents the energy density
at the centre of the configuration. Then, Eqs.
(\ref{ed3})-(\ref{ed4}) must be solved with the boundary conditions
\begin{equation}
\psi(0)=\psi'(0)=0. \label{ab1}
\end{equation}
The corresponding solutions must be computed numerically. However,
it is possible to determine the asymptotic behaviours explicitly. For
$\xi\rightarrow 0$,
\begin{equation}
\psi=a\xi^2+b\xi^4+... \label{ab2}
\end{equation}
with
\begin{equation}
a=\frac{d-2+qd}{2d},\label{ab3}
\end{equation}
and
\begin{eqnarray}
b=\frac{\lbrack d(d+2)q^2+(2d^2-4d-8)q+d(d-2)\rbrack(2-d-qd)}{8d^2
(d+2)(1+q)}.\nonumber\\
\label{ab4}
\end{eqnarray}
For $\xi\rightarrow +\infty$, the asymptotic behaviour of the solution
of Eqs. (\ref{ed3})-(\ref{ed4}) can be obtained by extending the
procedure developed by Chandrasekhar (1972) in $d=3$. We introduce a
new variable $z$ defined by
\begin{equation}
e^{-\psi}=\frac{e^z}{\xi^{2}}, \label{ab5}
\end{equation}
so that $z\rightarrow z_0=\ln Q$ for $\xi\rightarrow +\infty$. In
terms of this new variable, Eqs. (\ref{ed3})-(\ref{ed4}) become
\begin{eqnarray}
\left\lbrace 1-\frac{2q}{1+q}\frac{M(\xi)}{\xi^{d-2}}\right\rbrace
\frac{dz}{d\xi}\qquad\qquad\qquad\qquad\nonumber\\
+\frac{d-2+(d+2)q}{1+q}\frac{M(\xi)}{\xi^{d-1}}+q\frac{e^z}{\xi}-\frac{2}{\xi}=0,\label{ab6}
\end{eqnarray}
\begin{equation}
\frac{dM}{d\xi}=\xi^{d-3}e^z. \label{ab7}
\end{equation}
We set $z=z_0+f$ with $f\ll 1$ and linearize the equations. We then
find that $f$ satisfies the equation
\begin{eqnarray}
\xi^{d-1}\frac{d^2f}{d\xi^2}+\frac{(d+1)q+d-1}{q+1}\xi^{d-2}\frac{df}{d\xi}
\nonumber\\
+\frac{2(d-2+2dq+(d-2)q^2)}{(1+q)^2}\xi^{d-3}f=0. \label{ab8}
\end{eqnarray}
The further change of variables $\xi=e^t$ transforms Eq. (\ref{ab8}) to a
linear equation with constant coefficients. We find
\begin{eqnarray}
\frac{d^2f}{dt^2}+\frac{dq+d-2}{q+1}\frac{df}{dt}
+\frac{2(d-2+2dq+(d-2)q^2)}{(1+q)^2}f=0. \nonumber\\
\label{ab9}
\end{eqnarray}
Looking for solutions of the form $f\propto e^{kt}$, we obtain
\begin{eqnarray}
k=\frac{-(d-2+dq)\pm \sqrt{\Delta(q)}}{2(q+1)},
\label{ab10}
\end{eqnarray}
where $\Delta(q)$ is the discriminant 
\begin{eqnarray}
\Delta(q)=(d-4)^{2}q^{2}+2d(d-10)q+(d-2)(d-10).
\label{ab11}
\end{eqnarray}
This function is itself quadratic and its discriminant is
$\delta=-128(d-1)(d-10)$. We have $\Delta(q)\rightarrow
+\infty$ for $q\rightarrow \pm\infty$. Therefore, for $d>10$,
$\Delta(q)>0$ for all $q$. For $d<10$, $\Delta(q)=0$ has two
roots. Noting that $\Delta(0)=(d-2)(d-10)$, we conclude that one root
is negative and the other is positive. On the other hand, noting that
$\Delta(1)=4(d-9)(d-1)$, we conclude that the positive root is in the
range $]1,+\infty[$ for $d<9$ and in the range $[0,1]$ for $d\ge
9$. Summarising: (i) for $d<9$, $\Delta(q)< 0$ for all $q\in [0,1]$
(ii) for $d>10$, $\Delta(q)> 0$ for all $q\in [0,1]$ (iii) for $9\le
d\le 10$, then $\Delta(q)\le 0$ for $0\le q\le q_{*}$ and
$\Delta(q)\ge 0$ for $q_{*}\le q\le 1$ (see Figs. \ref{deltaq} and
\ref{qd}) 
where $q_{*}(d)$ is the
positive root of $\Delta(q)=0$, i.e.
\begin{eqnarray}
q_{*}(d)=\frac{d(10-d)+\sqrt{32(d-1)(10-d)}}{(d-4)^{2}}.
\label{ab11bis}
\end{eqnarray}
For $d=9$, $q_{*}=1$ and for $d=10$,
$q_{*}=0$. When $\Delta\le 0$, the asymptotic behaviour of
Eqs. (\ref{ed3})-(\ref{ed4}) for $\xi\rightarrow +\infty$ is
\begin{eqnarray}
e^{-\psi}\sim \frac{Q}{\xi^{2}}\left \lbrack 1+\frac{A}{\xi^{\frac{d-2+dq}{2(q+1)}}}\cos\left (\frac{\sqrt{-\Delta}}{2(q+1)}\ln\xi+\delta\right )\right\rbrack,
\label{ab12}
\end{eqnarray}  
and it presents damped oscillations around the singular
sphere. The curve (\ref{ab12}) intersects the
singular solution (\ref{ssd1}) infinitely often at points that
asymptotically increase geometrically in the ratio $1:{\rm exp}\lbrack
2\pi(1+q)/\sqrt{-\Delta(q)}\rbrack$. As shown in Paper I for
$d=3$ (see also Sect. \ref{sec_mdd}), this property is responsible for
the damped oscillations of the mass-central density relation
$M(\epsilon_{0})$. When $\Delta\ge 0$, the asymptotic behaviour of
Eqs. (\ref{ed3})-(\ref{ed4}) for $\xi\rightarrow +\infty$ is
\begin{eqnarray}
e^{-\psi}\sim \frac{Q}{\xi^{2}}\left \lbrack 1+\frac{A}{\xi^{\frac{d-2+dq-\sqrt{\Delta}}{2(q+1)}}}\right\rbrack,
\label{ab13}
\end{eqnarray} 
and it tends to the singular sphere without oscillating. In that case,
the mass-central density relation $M(\epsilon_{0})$ is monotonic (see
Sect. \ref{sec_mdd}). For a given value of $q$, the critical dimension above which the oscillations disappear is such that $\Delta(q)=0$ leading to
\begin{eqnarray}
d_{crit}(q)=\frac{4}{q+1}\left (q+\frac{3}{2}+\sqrt{3q+1}\right ).
\label{ab13bis}
\end{eqnarray}

Let us consider specific systems. (i) For classical isothermal spheres
($q\rightarrow 0$), the critical dimension above which the
mass-central density relation $M(\epsilon_{0})$ becomes monotonic is
such that $\Delta(0)=0$ (or $q_{*}(d)=0$) corresponding to
$d_{crit}=10$. This returns the result obtained by Sire \& Chavanis
(2002) who studied the $d$-dimensional Emden equation. As a
consequence, for $2<d<d_{crit}=10$, the caloric curve $\beta(E)$ of
Newtonian isothermal spheres presents a spiraling behaviour around the
point corresponding to the singular sphere, while for $d\ge
d_{crit}=10$, it tends to the point corresponding to the singular
sphere without spiraling (see Sect. \ref{sec_sgrd}). (ii) For
relativistic stars with the stiffest equation of state ($q=1$), the
critical dimension above which the mass-central density relation
$M(\epsilon_{0})$ becomes monotonic is such that $\Delta(1)=0$ (or
$q_{*}(d)=1$) corresponding to
$d_{crit}=9$. Interestingly, this coincides with the
critical dimension arising in superstring theory that may have some connection to the limit case $q=1$ (see, e.g., Kalyana Rama,
2006). (iii) Finally, for neutron stars or for a self-gravitating
radiation ($q=1/d$), the critical dimension above which the
mass-central density relation $M(\epsilon_{0})$ becomes monotonic is
such that $\Delta(1/d)=0$ (or $q_{*}(d)=1/d$). It is
solution of the fourth degree equation
\begin{eqnarray}
d^4-10d^3+d^2-8d+16=0,
\label{ab13tris}
\end{eqnarray} 
leading to $d_{crit}=9.96404372...$ very close to $d=10$. As a
consequence, for $2<d<d_{crit}=9.96404372...$ the mass-radius relation
of neutron stars $M(R)$ should present a spiraling behaviour around the point
corresponding to the singular sphere (see Fig. 2 of Meltzer \& Thorne
(1966) and Fig. 15 of Paper I), while for $d>d_{crit}=9.96404372...$, it
should tend to the point corresponding to the singular sphere without
spiraling. A similar property holds for the caloric curve of the
self-gravitating radiation (in Sect. \ref{sec_sgrd}).

\subsection{The Milne variables}
\label{sec_md}

\begin{figure}[htbp]
\centerline{
\includegraphics[width=8cm,angle=0]{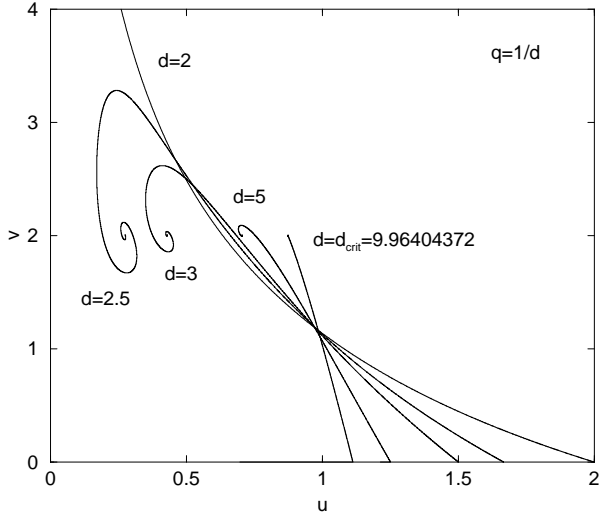}
} \caption[]{The $(u,v)$ plane for isothermal spheres in general
relativity. We have considered the case $q=1/d$ of a self-gravitating
radiation. For $2<d<d_{crit}=9.96404372$, the curve forms a spiral
around the point $(u_{s},v_{s})$ corresponding to the singular
sphere. For $d\ge d_{crit}$ it reaches that point directly, without spiraling. The case
$d=2$ is studied in Sect. \ref{sec_tdg}. }
\label{uvD10NEW}
\end{figure}

As in the Newtonian theory of isothermal spheres (Chandrasekhar 1942),
it is convenient to introduce the Milne variables
\begin{equation}
u=\frac{\xi e^{-\psi}}{\psi'}, \qquad v=\xi \psi'. \label{md1}
\end{equation}
In terms of these variables, the system of
Eqs. (\ref{ed3})-(\ref{ed4}) can be reduced to a single first-order
differential equation (see
Paper I for $d=3$). For $\xi\rightarrow 0$, one has
\begin{equation}
u=\frac{1}{2a}-\left (\frac{b}{a^2}+\frac{1}{2}\right )\xi^2+...,
\qquad v=2a\xi^2+... \label{md2}
\end{equation}
and for $\xi\rightarrow +\infty$
\begin{equation}
u\rightarrow  u_s=\frac{Q}{2}, \qquad v\rightarrow v_s=2.
 \label{md3}
\end{equation}
The solution curves in the $(u,v)$ plane in $d=3$ for different values
of $q$ are given in Fig. 9 of Paper I. According to the discussion in
Sect. \ref{sec_ab}, they tend to the point $(u_{s},v_{s})$
corresponding to the singular sphere by forming a spiral. By contrast,
for $d\ge d_{crit}(q)$ (defined in Sect. \ref{sec_ab}), they reach the
singular solution $(u_{s},v_{s})$ directly, without spiraling (see
Fig. \ref{uvD10NEW}).

\subsection{The mass-density profile}
\label{sec_mdd}

\begin{figure}[htbp]
\centerline{
\includegraphics[width=8cm,angle=0]{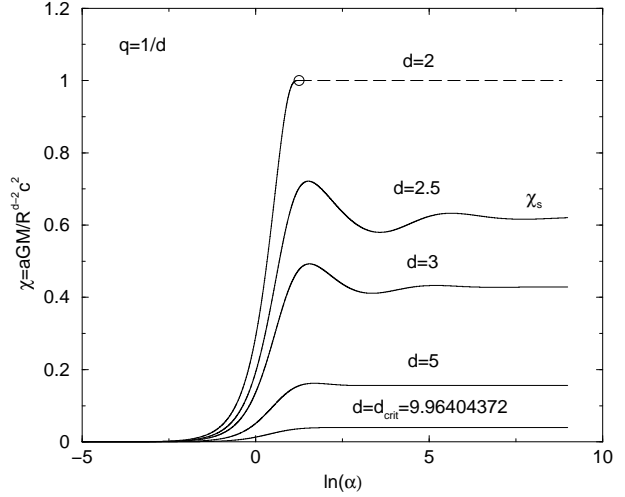}
} \caption[]{Mass vs central density for a fixed box radius. We have
considered the case $q=1/d$ of a self-gravitating radiation.  For
$2<d<d_{crit}=9.96404372$, the curve presents damped oscillations
around the mass $\chi_{s}$ corresponding to the singular sphere. The
series of equilibria becomes unstable after the first mass peak at
$(\alpha_{c},\chi_{c})$. For $d\ge d_{crit}$, the curve is
monotonic. In that case, all the configurations of the series of
equilibria (with arbitrary central density) are stable up to the
singular solution with the maximum mass $\chi=\chi_{s}$. The case
$d=2$ is studied in Sect. \ref{sec_tdg}. }
\label{achiDIM10NEW}
\end{figure}

If the system is enclosed within a box, the solution of Eqs. (\ref{ed3})-(\ref{ed4}) must be terminated at a radius $\alpha$ given by
\begin{equation}
\alpha=\left\lbrace \frac{8\pi G\epsilon_{0}(1+q)}{(d-1)c^{4}q}\right\rbrace^{1/2}R.
\label{mdd0}
\end{equation}
According to Eq. (\ref{ed2}) the relation between the total mass $M$ of the
configuration and the parameter $\alpha$ is 
\begin{equation}
M=\frac{2q}{1+q}\frac{M(\alpha)}{\alpha^{d-2}}\frac{R^{d-2}c^{2}}{aG}.
\label{mdd1}
\end{equation}
Solving for $M(\xi)$ in Eq. (\ref{ed3}) and taking $\xi=\alpha$, we get
\begin{equation}
\frac{M(\alpha)}{\alpha^{d-2}}=\frac{\alpha
\psi'(\alpha)-q\alpha^{2}e^{-\psi(\alpha)}}{d-2+p\alpha\psi'(\alpha)},
\label{mdd2}
\end{equation}
where $p$ is defined by Eq. (\ref{ss6}). Writing $u_{0}=u(\alpha)$ and
$v_{0}=v(\alpha)$, we obtain
\begin{equation}
\frac{aGM}{R^{d-2}c^{2}}\equiv \chi(\alpha)=\frac{pv_{0}(1-qu_{0})}{d-2+pv_{0}}.
\label{mdd3}
\end{equation}
The curve $\chi(\alpha)$ gives the mass $M(\epsilon_{0})$ as a
function of the central density for a fixed box radius. It starts from
$\chi=0$ for $\alpha=0$ and tends to an asymptotic value
$\chi_{s}=pQ/(d-2)=4q/\lbrack (d-2)(1+q)^{2}+4q\rbrack$, corresponding
to the singular sphere, as $\alpha\rightarrow +\infty$. The mass
associated to the singular sphere is
\begin{equation}
M_{s}=\chi_{s}\frac{R^{d-2}c^{2}}{aG}.
\label{mdd3bis}
\end{equation}
The properties of the mass-central density curve $\chi(\alpha)$ can be
obtained by a graphical construction in the Milne plane. Let us look
for the presence of oscillations by finding possible extrema of
$\chi(\alpha)$.  Taking the derivative with respect to $\alpha$ of
Eq. (\ref{mdd1}), using Eqs. (\ref{ed4}) and (\ref{mdd2}), and finally
introducing the Milne variables, we get
\begin{equation}
\frac{d\chi}{d\alpha}=\frac{p}{\alpha}\left\lbrack u_0 v_0
-(d-2)\frac{v_{0}(1-qu_{0})}{d-2+pv_{0}}\right\rbrack. \label{mdd4p}
\end{equation}
Therefore, the extrema of the curve $\chi(\alpha)$, determined by
the condition $d\chi/d\alpha=0$,  satisfy
\begin{equation}
p v_0=(d-2)\left (\frac{1}{u_0}-q-1\right ). \label{mdd4}
\end{equation}
The intersections between this curve and the solution curve $(u,v)$ in
the Milne plane determine the values of $\alpha$ for which
$\chi(\alpha)$ is extremum. We easily check that the curve
(\ref{mdd4}) passes by the point $(u_{s},v_{s})$ corresponding to the
singular solution. Therefore, when $\Delta(q)<0$, there is an infinity
of intersections, and the curve $\chi(\alpha)$ presents an infinity of
damped oscillations around the singular sphere $\chi_{s}$.  This is
the case extensively described in $d=3$ (see Paper I). When $\Delta\ge
0$, there is only one intersection (corresponding to the singular
sphere), and the curve $\chi(\alpha)$ increases monotonically up to the
singular sphere $\chi_{s}$. The corresponding curves for a
self-gravitating radiation are plotted in Fig. \ref{achiDIM10NEW}. 
If we extend the dynamical stability analysis (see Paper I and
Appendix
\ref{sec_dsa}) in a space of $d$ dimensions, we find that the point of
marginal stability in the series of equilibria precisely corresponds
to the criterion (\ref{mdd4}). Therefore, dynamical instability sets
it at the turning point of mass, as expected. When $\Delta(q)<0$, the
series of equilibria is stable until the first mass peak at
($\alpha_{c}$, $M_{c}$), and when $\Delta(q)\ge 0$, the series of
equilibria is stable for all values of the central density (including
the singular sphere that is marginally stable). This implies that, for
$d\ge d_{crit}(q)$, the pure scaling laws (\ref{ssd5}) correspond to
stable configurations contrary to the case $d=3$.

\subsection{Self-gravitating radiation} \label{sec_sgrd}

In this section, we briefly describe the behaviour of the curves
plotted in Sect. \ref{sec_sgr} as a function of the
dimension of space. For sake of generality we give the formulae for
any $q\in [0,1]$, but in the figures, we focus on the self-gravitating
radiation $q=1/d$.

\begin{figure}[htbp]
\centerline{
\includegraphics[width=8cm,angle=0]{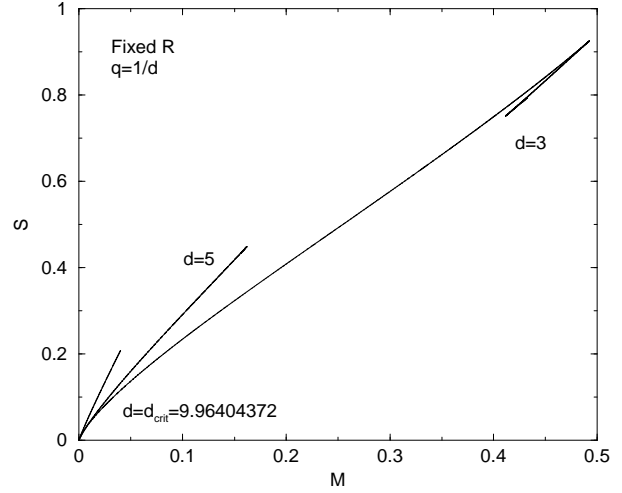}
} \caption[]{Entropy vs mass-energy for a fixed box radius. The peaks disappear for $d\ge d_{crit}$.  }
\label{smd}
\end{figure}

\begin{figure}[htbp]
\centerline{
\includegraphics[width=8cm,angle=0]{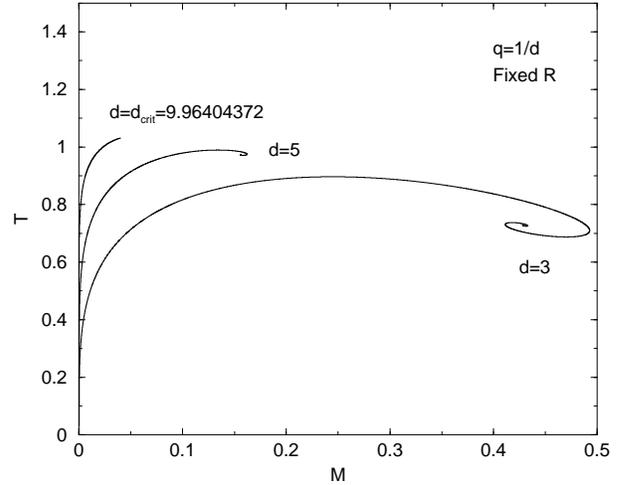}
} \caption[]{Temperature vs mass-energy for a fixed box radius. The
spiral shrinks to a point for $d\ge d_{crit}$. }
\label{tmrad}
\end{figure}

Let us first consider a fixed box radius. In that case, the parameter
$\alpha\propto \epsilon_{0}^{1/2}$ is a measure of the central density. The
mass-central density relation $M(\epsilon_{0})$ is plotted in
Fig. \ref{achiDIM10NEW}. For $2<d<d_{crit}(q)$, it presents damped
oscillations around the mass of the singular sphere, and for $d\ge
d_{crit}$ the convergence to the mass of the singular sphere is
monotonic. The entropy $S=\lambda N$, where $N$ is the number of
particles,
\begin{eqnarray}
N=\int_{0}^{R}n(r)\left\lbrack 1-\frac{aGM(r)}{c^{2}r^{d-2}}\right\rbrack^{-1/2}S_{d}r^{d-1}dr,
\label{sgrd1bds}
\end{eqnarray}
is given as a function of the central density by the relation
\begin{eqnarray}
S\propto
\Delta(\alpha)=\frac{1}{\alpha^{\frac{dq+d-2}{1+q}}}\int_{0}^{\alpha}
e^{-\frac{\psi(\xi)}{1+q}}\left\lbrack
1-p\frac{M(\xi)}{\xi^{d-2}}\right\rbrack^{-1/2}
\xi^{d-1}d\xi.\nonumber\\
\label{sgrd1}
\end{eqnarray}
For $\alpha\rightarrow +\infty$ (singular sphere), we have
\begin{equation}
\Delta(\alpha)\rightarrow \Delta_{s}=\frac{1+q}{dq+d-2}Q^{\frac{1}{1+q}}\left (1-\chi_{s}\right )^{-1/2}.
\label{sgrd2}
\end{equation}
For $d<d_{crit}$, the entropy-central density
relation $S(\epsilon_{0})$ presents damped oscillations at the same
locations as $M(\epsilon_{0})$. This implies that the curve $S(M)$
presents some peaks at these points. For $d\ge d_{crit}$, the
oscillations in $S(\epsilon_{0})$ and the peaks in $S(M)$ disappear
(see Fig. \ref{smd}). 
The temperature 
\begin{equation}
T=T(R)\sqrt{1-\frac{aGM}{R^{d-2}c^{2}}}\label{sgrd3}
\end{equation}
 is given as a function of the central density
by the relation
\begin{equation}
T\propto \theta(\alpha)=\frac{\alpha^{2q/(q+1)}}{{\cal R}(\alpha)^{q/(q+1)}}\lbrack
1-\chi(\alpha)\rbrack^{1/2}, \label{sgrd4}
\end{equation}
where ${\cal R}(\alpha)=e^{\psi(\alpha)}$ is the density contrast. For
$\alpha\rightarrow +\infty$ (singular sphere), we have
$\theta(\alpha)\rightarrow
\theta_{s}=Q^{\frac{q}{q+1}}(1-\chi_{s})^{1/2}$. For $d<d_{crit}$, the temperature-central density
relation $T(\epsilon_{0})$ presents damped oscillations at locations
different from $M(\epsilon_{0})$. This implies that the curve $T(M)$
forms a spiral. For $d\ge d_{crit}$, the oscillations in
$T(\epsilon_{0})$ and the spiral in the caloric curve $T(M)$ disappear
(see Fig. \ref{tmrad}). This is similar to what happens to the caloric
curve $\beta(E)$ of a Newtonian isothermal gas for $d\ge d_{crit}=10$
(see Sire \& Chavanis 2002).

\begin{figure}[htbp]
\centerline{
\includegraphics[width=8cm,angle=0]{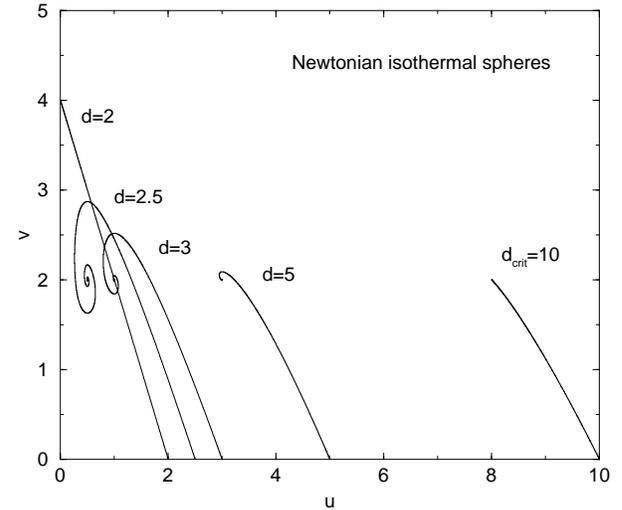}
} \caption[]{The solution of the classical Emden equation in the $(u,v)$ plane as a function of the dimension of space. }
\label{newtonuv}
\end{figure}

\begin{figure}[htbp]
\centerline{
\includegraphics[width=8cm,angle=0]{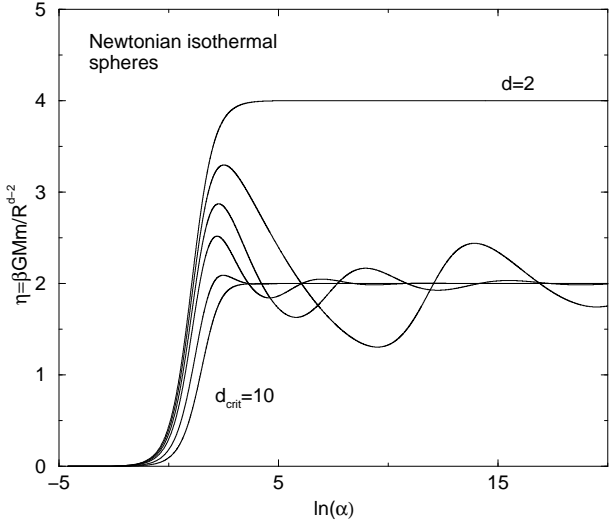}
} \caption[]{Mass-central density profile for a fixed temperature and box radius (or temperature vs density contrast for a fixed mass and box radius) as a function of the dimension of space for Newtonian isothermal spheres. }
\label{newtonah}
\end{figure}

\begin{figure}[htbp]
\centerline{
\includegraphics[width=8cm,angle=0]{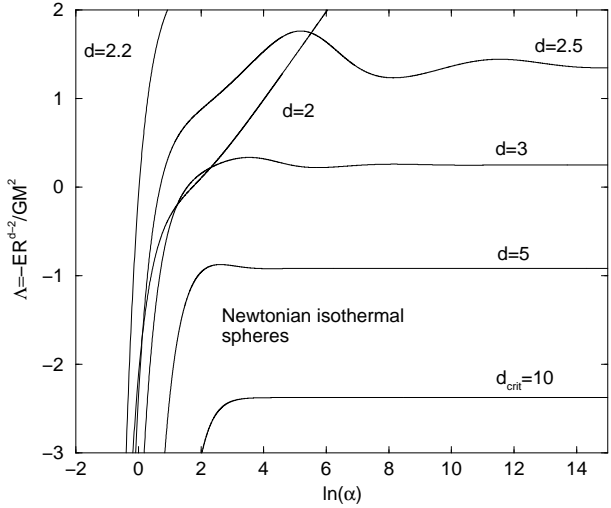}
} \caption[]{Energy vs density contrast for a fixed mass and box radius as a function of the dimension of space for Newtonian isothermal spheres. }
\label{newtonal}
\end{figure}

\begin{figure}[htbp]
\centerline{
\includegraphics[width=8cm,angle=0]{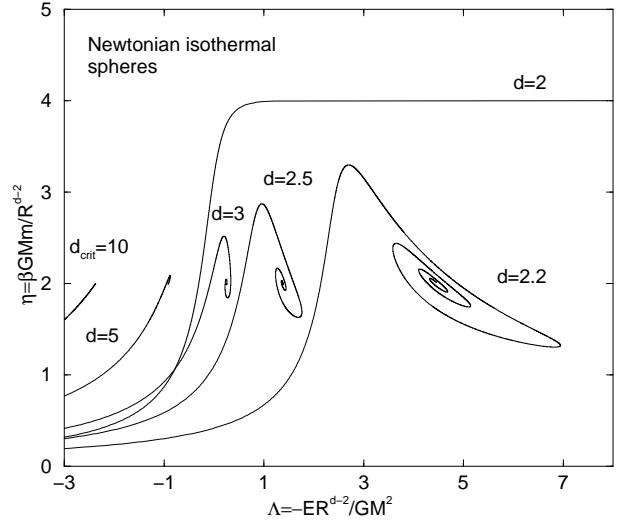}
} \caption[]{The caloric curve as a function of the dimension of space for Newtonian isothermal spheres.  }
\label{newtonlh}
\end{figure}

Alternatively, if we fix the central density, the
parameter $\alpha\propto R$ is a measure of the system size. The curves
$M(R)\propto \alpha^{d-2}\chi(\alpha)$, $S(R)\propto
\alpha^{(dq+d-2)/(q+1)}\Delta(\alpha)$, and $T(R)\propto
\alpha^{-2q/(q+1)}\theta(\alpha)$ behave like the ones
reported in Figs. \ref{mrq13}, \ref{srq13}, and \ref{Trq13} (note
that, for $d\ge d_{crit}(q)$, the small oscillations are
suppressed). For large $R$, the scaling laws are given by
Eqs. (\ref{ssd4})-(\ref{ssd5}). In particular, the mass behaves like
$M=\chi_{s}{R^{d-2}c^{2}}/({aG})$ for $R\rightarrow +\infty$. We again
emphasise that for $d\ge d_{crit}$ the solutions are stable for any
radius $R$, contrary to the case $d<d_{crit}$ where they become
unstable for $R>R_{c}$ (see Sect. \ref{sec_rs}).

Finally, we compare the results obtained previously for a
self-gravitating radiation in general relativity with the results
obtained for Newtonian isothermal spheres (Sire \& Chavanis 2002). In
that case, the critical dimension at which the oscillations disappear
is $d_{crit}=10$. In Fig. \ref{newtonuv}, we plot the solution curve
of the Emden equation in the $(u,v)$ plane. In Figs. \ref{newtonah}
and \ref{newtonal}, we plot the inverse temperature $\eta$ and the
energy $\Lambda$ as a function of the parameter $\alpha=(S_d G\beta
m\rho_0)^{1/2}R$, which is a measure of the density contrast. In
Fig. \ref{newtonlh}, we plot the caloric curve $\beta(E)$. For $d=2$,
the caloric curve tends to a plateau with temperature
$T_{c}=GMm/4k_{B}$ and energy $E\rightarrow -\infty$. This corresponds
to the formation of a Dirac peak as $T\rightarrow T_{c}$. For
$2<d<10$, the caloric curve forms a spiral. For $d\ge 10$, the spiral
shrinks to a point. These results are strikingly similar to those
obtained for a self-gravitating radiation in general relativity.

\subsection{Particular values of the parameters} \label{sec_vpre}

In this section, we regroup the values of the different parameters
defined in the text for different values of $q$ and different
dimensions of space.

For $q=1$ (stiffest stars), we have $p=1$, $Q=\theta_{s}=\frac{d-2}{d-1}$, and $\chi_{s}=\Delta_{s}=\frac{1}{d-1}$. In particular, for $d=3$, we have $p=1$, $Q=\theta_{s}=\frac{1}{2}$ and $\chi_{s}=\Delta_{s}=\frac{1}{2}$. Furthermore, $\alpha_{c}=4.05$, $\chi_{c}=0.544$, $\Delta_{c}=0.546$,  $\alpha_{*}=1.30$, and
$\theta_{*}=0.676$. On the other hand, for $d_{crit}=9$,  we have $p=1$, $Q=\theta_{s}=\frac{7}{8}$, and $\chi_{s}=\Delta_{s}=\frac{1}{8}$.

For $q=1/d$ (self-gravitating radiation), we have $p=\frac{2}{d+1}$,
$Q=2d(d-2)(d+1)/\lbrack (d-2)(d+1)^{2}+4d\rbrack$,
$\chi_{s}=pQ/(d-2)$, $\theta_{s}=Q^{1/(d+1)}(1-\chi_{s})^{1/2}$, and
$\Delta_{s}=\frac{d+1}{d(d-1)}Q^{d/(d+1)}(1-\chi_{s})^{-1/2}$. In
particular, for $d=3$, we have $p=\frac{1}{2}$, $Q=\frac{6}{7}$,
$\chi_{s}=\frac{3}{7}$, $\theta_{s}=(\frac{96}{343})^{1/4}$, and
$\Delta_{s}=(\frac{8}{21})^{1/4}$. Furthermore, $\alpha_{c}=4.7$,
$\chi_{c}=0.493$, $\Delta_{c}=0.925$, $\alpha_{*}=1.47$ and
$\theta_{*}=0.897$. On the other hand, for $P(d_{crit})=10$ (closest
upper integer value of the critical dimension), we have
$p=\frac{2}{11}$, $Q=\frac{110}{63}$, $\chi_{s}=\frac{55}{1386}$,
$\theta_{s}=(\frac{110}{63})^{1/11}(\frac{1331}{1386})^{1/2}$, and
$\Delta_{s}=\frac{11}{90}(\frac{110}{63})^{10/11}(\frac{1386}{1331})^{1/2}$.

\section{Two dimensional gravity} \label{sec_tdg}

In this section, we focus on the dimension $d=2$, which presents
peculiar features and where analytical results can be obtained.

\subsection{Self-confined solutions} \label{sec_scs}

For $d=2$, the Oppenheimer-Volkoff equations (\ref{eged5})-(\ref{eged6})
reduce to
\begin{eqnarray}
\left \lbrace 1-\frac{8GM(r)}{c^{2}}\right \rbrace\frac{dP}{dr}=
-\frac{8\pi G}{c^{4}}(\epsilon+P)Pr,
\label{scs1}
\end{eqnarray}
with
\begin{equation}
M(r)=\frac{2\pi}{c^{2}}\int_{0}^{r}\epsilon r dr.
\label{scs2}
\end{equation}
From Eqs. (\ref{eged8}) and (\ref{eged9}), we find that the functions
defining the metric are given by
\begin{equation}
e^{-\lambda}=1-\frac{8GM(r)}{c^{2}}, \quad \frac{d\nu}{dr}=\frac{\frac{16\pi GPr}{c^{4}}}{1-\frac{8GM(r)}{c^{2}}}.
\label{scs3}
\end{equation}
In the empty space surrounding the star, $P=\epsilon=0$. Therefore,
for $r>R$, we have $e^{-\lambda}=1-8GM/c^{2}$ and $\nu={\rm cst}$. The
first relation requires $M\le M_{c}\equiv c^{2}/8G$. In fact, we 
find in the following examples (valid for a linear equation of state)
that steady solutions exist only for $M=M_{c}\equiv c^{2}/8G$. We can wonder whether this result is general.

For a linear equation of state, defining
\begin{equation}
\epsilon=\epsilon_{0}e^{-\psi}, \qquad r=\left\lbrace \frac{c^{4}q}{8\pi G\epsilon_{0}(1+q)}\right\rbrace^{1/2}\xi,
\label{scs4}
\end{equation}
and
\begin{equation}
M(r)=\frac{c^{2}q}{4 G (1+q)}M(\xi),
\label{scs5}
\end{equation}
the generalized Emden equation takes the form
\begin{eqnarray}
\left\lbrace 1-\frac{2q}{1+q}M(\xi)\right\rbrace \frac{d\psi}{d\xi}=q\xi e^{-\psi},
\label{scs6}
\end{eqnarray}
\begin{equation}
\frac{dM}{d\xi}=\xi e^{-\psi}.
\label{scs7}
\end{equation}
We note that there is no Newtonian limit ($q\rightarrow 0$) in $d=2$ (see Appendix \ref{sec_nl}).
The differential equation for the mass profile is
\begin{equation}
\frac{1}{\xi}-\frac{M''}{M'}=\frac{qM'}{1-\frac{2q}{1+q}M}.
\label{scs8}
\end{equation}
This equation can be easily integrated once to yield
\begin{equation}
\frac{M'}{\xi}=A_{0}\left (1-\frac{2q}{1+q}M\right )^{\frac{1+q}{2}},
\label{scs9}
\end{equation}
where $A_{0}$ is a positive constant. This can again be integrated easily to yield
\begin{equation}
\left (1-\frac{2q}{1+q}M\right )^{\frac{1-q}{2}}=-K^{2}\xi^{2}+B,
\label{scs10}
\end{equation}
where $B$ and $K$ (related to $A_{0}$) are some constants. Here, we
assume $q\neq 1$ (the case $q=1$ will be treated separately). Because $M=0$ at $\xi=0$, we find that $B=1$. Therefore, the mass
profile is given by
\begin{equation}
M(\xi)=\frac{1+q}{2q}\left\lbrack 1-(1-K^{2}\xi^{2})^{\frac{2}{1-q}}\right\rbrack.
\label{scs11}
\end{equation}
Using Eq. (\ref{scs7}), we find that the normalized density profile is
given by
\begin{equation}
e^{-\psi}=2K^{2}\frac{1+q}{q(1-q)}(1-K^{2}\xi^{2})^{\frac{1+q}{1-q}}.
\label{scs12}
\end{equation}
Thus, we find that the density vanishes at a finite distance $\xi_{0}=1/K$. Using furthermore the fact that $\psi(0)=0$, we obtain
\begin{equation}
\xi_{0}=\left \lbrack \frac{2(1+q)}{q(1-q)}\right\rbrack^{1/2}.
\label{scs13}
\end{equation}
Therefore, the mass and density profiles can be written \footnote{It
is amusing to note that the form of the density profile is similar to
a ``Tsallis distribution'' with index $p=2q/(1+q)$. For $q=p=1$, we
obtain a ``Boltzmann distribution'' (\ref{scs25}). These analogies
with generalized thermodynamics (Tsallis 1988) are, of course, purely
coincidental. They show that the ``Tsallis distribution'' can arise in
very different contexts that are not necessarily related to
thermodynamics. }
\begin{equation}
M(\xi)=\frac{1+q}{2q}\left\lbrack 1-\left (1-\left ({\xi}/{\xi_{0}}\right )^{2}\right )^{\frac{2}{1-q}}\right\rbrack,
\label{scs14}
\end{equation}
\begin{equation}
e^{-\psi}=\left\lbrack 1-(\xi/\xi_{0})^{2}\right\rbrack^{\frac{1+q}{1-q}}.
\label{scs15}
\end{equation}
Returning to original variables, we find that
\begin{equation}
M(r)=\frac{c^{2}}{8G}\left\lbrack 1-\left (1-\left ({r}/{r_{0}}\right )^{2}\right )^{\frac{2}{1-q}}\right\rbrack,
\label{scs16}
\end{equation}
\begin{equation}
\epsilon(r)=\epsilon_{0}\left\lbrack 1-(r/r_{0})^{2}\right\rbrack^{\frac{1+q}{1-q}},
\label{scs17}
\end{equation}
\begin{equation}
r_{0}=\left \lbrack \frac{c^{4}}{4\pi G\epsilon_{0}(1-q)}\right\rbrack^{1/2}.
\label{scs18}
\end{equation}
The metric is explicitly given by
\begin{equation}
e^{-\lambda}=\left\lbrack 1-(r/r_{0})^{2}\right \rbrack^{\frac{2}{1-q}},
\label{scs20}
\end{equation}
\begin{equation}
\nu=-\frac{2q}{1-q}\ln\left\lbrack 1-(r/r_{0})^{2}\right\rbrack. 
\label{scs21}
\end{equation}
This defines a family of solutions parametrized by the central density. These solutions can have different radii $r_{0}$, but they all have the same mass $M=M(r_{0})=M_{c}$ given by
\begin{equation}
M_{c}=\frac{c^{2}}{8G}.
\label{scs19}
\end{equation}
For $r_{0}\rightarrow 0$, the density profile tends to a Dirac peak.
In two dimensions, the case of photon stars (self-gravitating
radiation) and neutron stars corresponds to $q=1/2$. On the other
hand, for the stiffest equation of state corresponding to $q=1$ ,
Eq. (\ref{scs9}) simplifies in
\begin{equation}
\frac{M'}{\xi}=A_{0} (1-M).
\label{scs22}
\end{equation}
This can be  solved easily to yield
\begin{equation}
M(\xi)=1-e^{-K^{2}\xi^{2}},
\label{scs23}
\end{equation}
where we have used $M(0)=0$. The density profile is the Gaussian
\begin{equation}
e^{-\psi}=2K^{2}e^{-K^{2}\xi^{2}}.
\label{scs24}
\end{equation}
Using $\psi(0)=0$, we get $K^{2}=1/2$ so that 
\begin{equation}
e^{-\psi}=e^{-\xi^{2}/2},
\label{scs25}
\end{equation}
\begin{equation}
M(\xi)=1-e^{-\xi^{2}/2}.
\label{scs26}
\end{equation}
Returning to original variables, we find that
\begin{equation}
\epsilon(r)=\epsilon_{0}e^{-(r/L)^{2}},
\label{scs27}
\end{equation}
\begin{equation}
M(r)=\frac{c^{2}}{8G}\left (1-e^{-(r/L)^{2}}\right ),
\label{scs28}
\end{equation}
where
\begin{equation}
L=\left (\frac{c^{4}}{8\pi G\epsilon_{0}}\right )^{1/2}
\label{scs29}
\end{equation}
is a typical lengthscale.  The metric is explicitly given by
\begin{equation}
e^{-\lambda}=e^{-(r/L)^{2}}, \qquad e^{-\nu}=e^{-(r/L)^{2}}.
\label{scs30}
\end{equation}
Again, we find a family of solutions parametrized by the central
density. It is noteworthy that the total mass $M=M(+\infty)=M_c$ is
the same for all these configurations and is again given by
Eq. (\ref{scs19}). For $\epsilon_{0}\rightarrow +\infty$, the density
profile tends to a Dirac peak.

These results have to be contrasted from their Newtonian counterpart
where the density decreases as $\xi^{-4}$ for $\xi\rightarrow
+\infty$ (see, e.g., Sire \& Chavanis 2002). However, the universal
mass $M_{c}=c^2/(8G)$ seems to be the general relativistic equivalent
of the critical temperature $T_{c}=GMm/(4k_{B})$ or the critical mass
$M_c=4k_{B}T/(Gm)$ in 2D Newtonian gravity for isothermal
spheres. Indeed, for a fixed temperature, unbounded two-dimensional,
self-gravitating isothermal spheres have a unique mass $M_c$ (see,
e.g., Chavanis 2007b).

Using Eq. (\ref{es5}), the baryon number is given by
\begin{eqnarray}
N=\int_{0}^{+\infty}n(r)e^{\lambda(r)/2} 2\pi rdr,
\label{scs31}
\end{eqnarray}
with $n(r)=(q/K)^{1/(q+1)}\epsilon(r)^{1/(q+1)}$. From the analytical
expressions Eqs. (\ref{scs17}) (\ref{scs20}) or (\ref{scs27})
(\ref{scs30}), we note that $n(r)e^{\lambda(r)/2}=n_{0}$ in the
star. For $q\neq 1$, using Eq. (\ref{scs18}) we explicitly
obtain
\begin{eqnarray}
N=\left (\frac{q}{K}\right)^{\frac{1}{q+1}}\frac{c^{4}}{4G (1-q)}\epsilon_{0}^{-\frac{q}{q+1}}.
\label{scs32}
\end{eqnarray}
The baryon number diverges for $\epsilon_{0}\rightarrow 0$,
i.e. $r_{0}\rightarrow +\infty$, suggesting that the system tends to
evaporate (recall that stable states tend to {\it maximize} $N$ at
fixed mass $M$)
\footnote{This result is different from the case of 2D Newtonian gravity
where the Boltzmann free energy $F_B[\rho]$ happens to be independent of the
central density $\rho_{0}$ parametrizing the family of isothermal solutions
at $M=M_{c}$ or $T=T_c$ (see Chavanis 2007b).}.  For $q=1$, the baryon number
diverges, whatever the central density.

\subsection{Box-confined solutions} \label{sec_bcs}

\begin{figure}[htbp]
\centerline{
\includegraphics[width=8cm,angle=0]{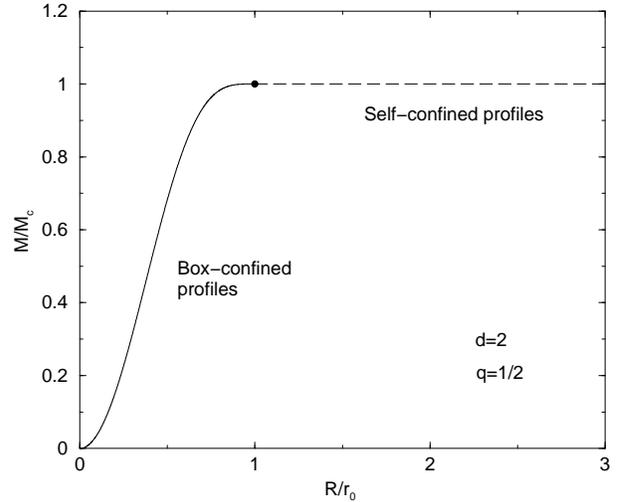}
} \caption[]{For box-confined systems, this figure represents the mass
as a function of the central density (proportional to $(R/r_{0})^{2}$)
for $q=1/2$.  }
\label{Mq12d2}
\end{figure}

\begin{figure}[htbp]
\centerline{
\includegraphics[width=8cm,angle=0]{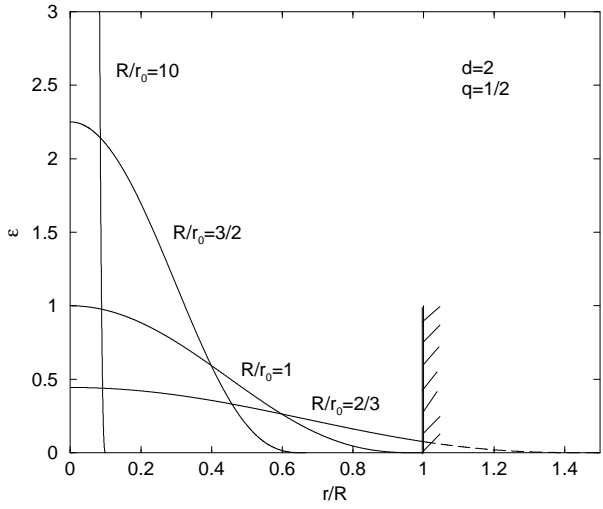}
} \caption[]{Density energy profile (in units of $c^{4}/2\pi GR^{2}$)
for different values of the central density (proportional to
$(R/r_{0})^{2}$) for $q=1/2$. The profile is self-confined if
$r_{0}\le R$ corresponding to $\epsilon_{0}>c^{4}/\lbrack 4\pi
G(1-q)R^{2}\rbrack$. For $r_{0}\rightarrow 0$ or
$\epsilon_{0}\rightarrow +\infty$, the density profile tends to a
Dirac peak.  }
\label{epsDim2}
\end{figure}

\begin{figure}[htbp]
\centerline{
\includegraphics[width=8cm,angle=0]{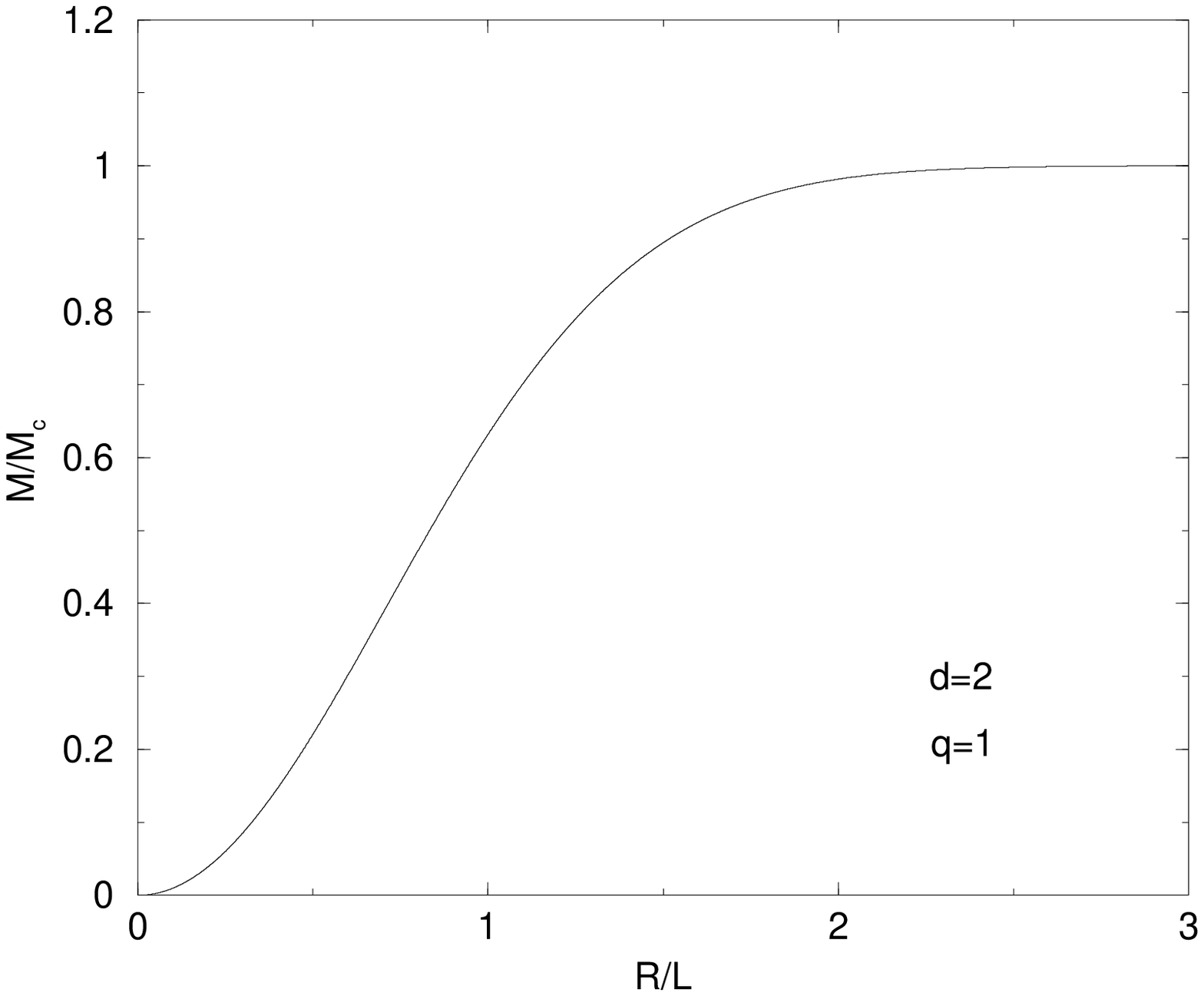}
} \caption[]{For box-confined systems, this figure represents the mass as a function of the central density (proportional to $(R/L)^{2}$) for $q=1$.   }
\label{Mq1D2}
\end{figure}

\begin{figure}[htbp]
\centerline{
\includegraphics[width=8cm,angle=0]{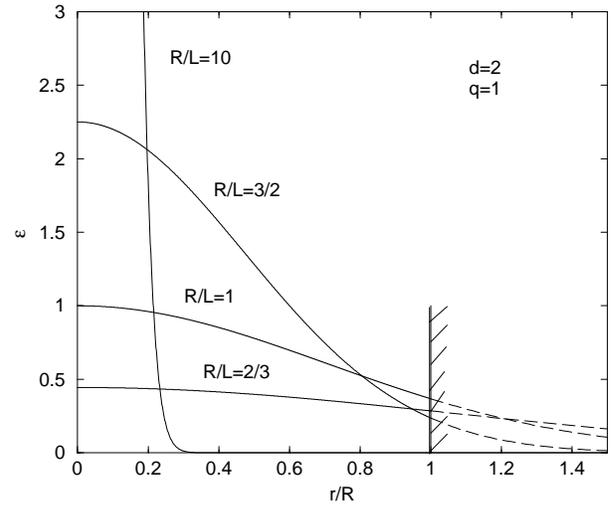}
} \caption[]{Density energy profile (in units of $c^{4}/8\pi GR^{2}$)
for different values of the central density (proportional to
$(R/L)^{2}$) for $q=1$.  For
$L\rightarrow 0$ or $\epsilon_{0}\rightarrow +\infty$, the
density profile tends to a Dirac peak.  }
\label{epsq1Dim2}
\end{figure}

If the system is confined within a box of radius $R$, the previous
results remain valid for $r\le R$. If $r_0>R$, the system is
confined by the wall ($\rho(R)\neq 0$), and if $r_0<R$, the system is
self-confined ($\rho(r_0)=0$). Let us consider here box-confined
configurations ($r_0>R)$. For $q\neq 1$, the mass-central energy
relation for fixed $R$ is given by
\begin{equation}
M=\frac{c^{2}}{8G}\left\lbrack 1-\left (1-\left ({R}/{r_{0}}\right )^{2}\right )^{\frac{2}{1-q}}\right\rbrack,
\label{bcs1}
\end{equation}
\begin{equation}
r_{0}=\left \lbrack \frac{c^{4}}{4\pi G\epsilon_{0}(1-q)}\right\rbrack^{1/2}.
\label{bcs2}
\end{equation}
The mass-central density (for a fixed box radius) is plotted in
Fig. \ref{Mq12d2} and the density profile is plotted in
Fig. \ref{epsDim2}. For $q=1$, we have
\begin{equation}
M=\frac{c^{2}}{8G}\left (1-e^{-(R/L)^{2}}\right ),
\label{bcs3}
\end{equation}
where
\begin{equation}
L=\left (\frac{c^{4}}{8\pi G\epsilon_{0}}\right )^{1/2}.
\label{bcs4}
\end{equation}
The mass-central density (for a fixed box radius) is plotted in
Fig. \ref{Mq1D2} and the density profile is plotted in
Fig. \ref{epsq1Dim2}.

\subsection{The Milne variables} \label{sec_mv}

\begin{figure}[htbp]
\centerline{
\includegraphics[width=8cm,angle=0]{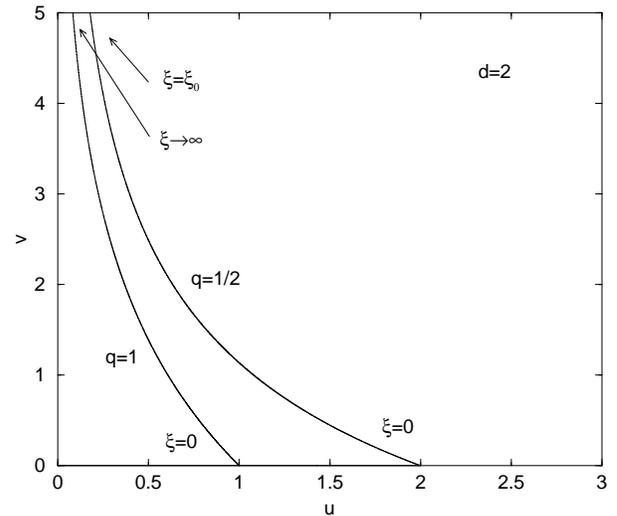}
} \caption[]{The $(u,v)$ curve of ``isothermal'' spheres in two-dimensional gravity for $q=1/2$ and $q=1$.  }
\label{uvd2}
\end{figure}

The Milne variables are defined by Eqs. (\ref{md1}). Using the
analytical solution (\ref{scs15}), we find for $q\neq 1$ that
\begin{equation}
u=\frac{1}{q}(1-(\xi/\xi_0)^{2})^{\frac{2}{1-q}}, \quad v=\frac{q\xi^{2}}{1-(\xi/\xi_{0})^{2}}.
\label{mv1}
\end{equation}
Eliminating $\xi$ between these two expressions, we obtain
\begin{equation}
v=q\xi_{0}^{2}\left\lbrack (qu)^{\frac{q-1}{2}}-1\right\rbrack.
\label{mv2}
\end{equation}
The $(u,v)$ curve is parametrized by $\xi$. For $\xi=0$, we have
$(u,v)=(1/q,0)$ and for $\xi\rightarrow \xi_{0}$, we have
$(u,v)=(0,+\infty)$.
For $q=1$,  using the
analytical solution (\ref{scs25}), we find that
\begin{equation}
u=e^{-\xi^{2}/2}, \quad v=\xi^{2}.
\label{mv3}
\end{equation}
Eliminating $\xi$ between these two expressions, we obtain
\begin{equation}
v=-2\ln u.
\label{mv4}
\end{equation}
The $(u,v)$ curve is parametrized by $\xi$. For $\xi=0$, we have
$(u,v)=(1,0)$ and for $\xi\rightarrow +\infty$, we have
$(u,v)=(0,+\infty)$. The solution curve in the $(u,v)$ plane is represented
in Fig. \ref{uvd2}. 
If the system is confined within a box of radius $R$,
the previous results remain valid for $\xi\le \alpha\le\xi_0$ where
\begin{equation}
\alpha=\left\lbrace \frac{8\pi G\epsilon_{0}(1+q)}{c^{4}q}\right\rbrace^{1/2}R.
\label{mv5}
\end{equation}
From the criterion (\ref{mdd4}), we can extrapolate that the condition
of marginal stability in $d=2$ corresponds to $u_{0}=0$. Therefore,
the solutions that are confined by a box are stable, since
$u_0=u(\alpha)>0$, while the self-confined solutions are marginally
stable since $u_0=u(\xi_0)=0$.

\section{Conclusion}
\label{sec_conclusion}

In this paper, we have carried out a thorough analysis of the
structure and stability of relativistic stars (and self-gravitating
radiation) described by a linear equation of state. In order to
prevent evaporation, we have placed these objects in a cavity. We have
studied the stability of the steady states (i) by linearizing the
Einstein equations around the stationary solution and studying the
sign of the squared pulsation $\sigma^{2}$ (see Paper I) and (ii) by
considering the sign of the second-order variations of the baryon
number (see Appendix \ref{sec_dsa}). We have found analytically that
the instability occurs precisely at the first mass peak when we vary
the central density for a fixed box radius. We have applied these
results to explicit examples: self-gravitating radiation, the core of
neutron stars, stiffest stars, Zel'dovich model etc. We have 
determined their domains of stability precisely and found some upper bounds on
the thermodynamical parameters like the entropy. We have obtained
scaling laws (with a precise determination of the prefactor) that we
compared with the scaling laws of black holes. We have stressed,
however, that pure scaling laws correspond to unstable configurations
and that they can hold only approximately close to the critical radius
$R_c$.  We
have generalized our results in $d$ dimensions and found two critical
dimensions. In $d=2$, steady state solutions exist for a unique value
of the mass $M_{c}=c^{2}/(8G)$ and they can be obtained
analytically. For this mass, we get an infinite family of solutions
parametrized by the central density. The density profiles vanish at a
finite radius for $q\neq 1$ and decrease as a Gaussian for
$q=1$. These configurations are probably marginally stable. They may
evolve dynamically either toward a completely spread profile
(evaporation) or toward a Dirac peak (black hole) with mass $M_{c}$
(collapse). On the other hand, for $d\ge d_{crit}(q)$ (where
$d_{crit}(q)$ is a nontrivial dimension depending on $q$), the mass vs
central density profile no longer displays oscillations, and all
the configurations of the series of equilibria are stable whatever
their central density. In that case, the pure scaling laws are
meaningful.  The physical implication of this result needs further
investigation. For the self-gravitating radiation, we have found that
the critical dimension has the non-integer value
$d_{crit}=9.96404372...$ This is an interesting (and intriguing)
result because it  results solely from the combination of the Einstein
equations and the Stefan-Boltzmann law, so it has a relatively
fundamental origin.

We have also shown that the structure of relativistic stars with a
linear equation of state in general relativity is strikingly similar
to the structure of isothermal stars in Newtonian gravity. Basically,
the analogy stems from the resemblence between the general
relativistic Emden equation and its Newtonian counterpart and the fact
that they coincide for $q\rightarrow 0$. On the other hand, stable
relativistic stars with a linear equation of state maximize the baryon
number $N[\epsilon]$ at fixed mass-energy $M[\epsilon]$ (formal
nonlinear dynamical stability for the Einstein equations). Similarly,
isothermal stars in Newtonian gravity minimize the energy functional
${\cal W}[\rho]$ at fixed mass $M[\rho]$ (formal nonlinear dynamical
stability for the Euler-Poisson system) or minimize the Boltzmann free
energy functional $F_B[\rho]$ at fixed mass $M[\rho]$ (thermodynamical
stability in the canonical ensemble). As a result, the curves of
Figs. \ref{meq1} and
\ref{neq1} are respectively similar to (i) the mass-versus-central
density $M(\rho_0)$ at fixed temperature and volume and (ii) the energy
functional ${\cal W}(\rho_0)$ or the Boltzmann free energy
$F_B(\rho_0)$-versus-central density at fixed temperature and volume
for Newtonian isothermal spheres (Chavanis 2002d). On the other hand,
by interpreting the mass $M[\epsilon]$ as an energy and the baryon
number $N[\epsilon]$ as an entropy, the criterion of formal nonlinear
dynamical stability in general relativity is equivalent to a criterion
of thermodynamical stability in the microcanonical
ensemble. Therefore, the curves of Figs. \ref{meq13}, \ref{seq13},
\ref{SMrfixedq13}, and \ref{TMrfixedq13} are respectively similar to
(i) the energy $E({\cal R})$ and the entropy $S({\cal
R})$-versus-density contrast at fixed mass and volume, (ii) the
entropy $S(E)$-versus-energy at fixed mass and volume, and (iii) the
caloric curve $\beta(E)$ at fixed mass and volume for Newtonian
isothermal spheres (Chavanis 2002d). Therefore, depending on the
interpretation, relativistic stars with a linear equation of state
share analogies with classical isothermal spheres in both
microcanonical and canonical ensembles. These interesting analogies
are intriguing and deserve further investigation.

\vskip0.4cm

{\bf Note added:} Coincidentally, two other authors V. Vaganov [{\tt
arXiv:0707.0864}] and J. Hammersley [{\tt arXiv:0707.0961}] have, in
two recent papers, independently carried out studies of the
self-gravitating radiation in $d$ dimensions (in an asymptotically
anti-de Sitter space $\Lambda\le 0$) related to the one developed in
Sect. \ref{sec_dimd}. These authors find that the oscillations in the
mass-central density relation disappear above a critical dimension. On
the basis of numerical calculations, J. Hammersley obtains a simple
relation between the critical density $\epsilon_{0,c}$ and the
dimension $d$ and argues that the critical dimension is $d_{crit}=10$
($+1$ if we include time). Our analytical approach (for a cosmological
constant $\Lambda=0$) is more precise and shows that the critical
dimension corresponding to the self-gravitating radiation has the {\it
noninteger} value $d_{c}=9.96404372...$ very close to $10$ (a minor
mistake in calculation was made in the first version of this
manuscript and led to a slightly different value for the critical
dimension. This mistake was pointed out to me by V. Vaganov, who
performed, in a new Appendix B of his paper [{\tt arXiv:0707.0864v4}],
a phase plane analysis of the Oppenheimer-Volkoff equation with
$\Lambda=0$. This provides an alternative derivation of the asymptotic
results obtained in our Sect. \ref{sec_ab}). Our approach also
provides a thorough description of the structure and stability of
relativistic stars with a linear equation of state $p=q\epsilon$ for
any $q\in [0,1]$ and $d\ge 2$. Note that the existence of a critical
dimension above which the oscillations of the mass-central density
relation disappear had been noted previously in Sire \& Chavanis
(2002) for Newtonian isothermal spheres (corresponding to
$q\rightarrow 0$). In that case, $d_{crit}=10$ exactly. The present
study extends this work in general relativity.

\vskip0.1cm
{\it Acknowledgements.} I am grateful to Kalyana Rama for pointing out his work after a first version of this paper was placed on arXiv:0707.2292.

\appendix

\section{Some elements of black hole thermodynamics}
\label{sec_bht}

In this appendix, we recall elementary notions of black holes
thermodynamics in order to facilitate the comparison with the results
obtained in this paper.

In the seventies, different works (Christodoulou 1970, Penrose \&
Floyd 1971, Hawking 1971) have shown that the area $A=4\pi R^2$ of a
black hole (more precisely the area of its event horizon) cannot
decrease. Noting the analogy with the second law of thermodynamics,
these results led Bekenstein (1973) to conjecture that black holes
have an entropy proportional to their area $A=4\pi R^{2}$. On the
basis of dimensional analysis, he obtained
\begin{equation}
S_{BH}=\lambda \frac{k_B A}{L_P^{2}}, \label{bht1}
\end{equation}
where $\lambda$ is a dimensionless constant. If black holes have 
entropy and  energy, they must possess a temperature. Hawking (1975)
showed that black holes emit thermal radiation at a temperature
\begin{equation}
k_B T=\frac{\hbar c^3}{8\pi GM}.\label{bht2}
\end{equation}
Considering a Schwarzschild black hole for which
\begin{equation}
M=\frac{R c^2}{2G},\label{bht3}
\end{equation}
and writing the first law of thermodynamics in the form
\begin{equation}
d(Mc^2)=T dS_{BH},\label{bht4}
\end{equation}
one gets
\begin{equation}
k_{B}T=\frac{\hbar c^{3}}{32\pi\lambda GM}.\label{bht5}
\end{equation}
Comparing this expression with the expression of the Hawking
temperature (\ref{bht2}) it is found that $\lambda=1/4$. This leads to the
following expression of the Black Hole (or Bekenstein-Hawking)
entropy \footnote{Coincidentally, the initials are the same.}
\begin{equation}
S_{BH}=\frac{k_B A}{4 L_P^{2}}. \label{bht6}
\end{equation}
On the other hand, using Eq. (\ref{bht3}), the black hole temperature (\ref{bht5}) can be
written
\begin{equation}
k_B T=\frac{c \hbar}{4\pi R},\label{bht7}
\end{equation}
emphasising the scaling $T\sim 1/R$ with the radius. On the other
hand, the energy $E=Mc^2$ of the black hole is related to its
temperature by
\begin{equation}
E=\frac{\hbar c^5}{8\pi G k_B T}.\label{bht8}
\end{equation}
Therefore, black holes have negative specific heats
\begin{equation}
C=\frac{dE}{dT}=-\frac{\hbar c^5}{8\pi G k_B T^2}<0.\label{bht9}
\end{equation}

\section{The Tolman relation}
\label{sec_tr}

Global thermodynamic equilibrium requires that the redshifted
temperature $e^{\nu(r)/2}T(r)$ is uniform throughout the medium.
This is called the Tolman (1934) relation
\begin{equation}
e^{\nu(r)/2}T(r)=T_0,\label{tr1}
\end{equation}
where $T_0$ is a constant. Since $\nu(r)\rightarrow 0$ for
$r\rightarrow +\infty$, we conclude that $T_0$ is the temperature
measured by an observer at infinity. Let us check that this relation
is satisfied by our equations. From Eq. (\ref{es9}), we find that
\begin{equation}
T(r)\propto \epsilon(r)^{q/(q+1)}.\label{tr2}
\end{equation}
On the other hand, according to Eq. (102) of Paper I, the Einstein
equations for a spherically distribution of matter give
\begin{equation}
\frac{dP}{dr}=-\frac{1}{2}(\epsilon+P)\frac{d\nu}{dr}.\label{tr3}
\end{equation}
For the linear equation of state (\ref{es3}), this relation can be easily
integrated into
\begin{equation}
\epsilon(r)=A e^{-\frac{q+1}{q}\frac{\nu(r)}{2}},\label{tr4}
\end{equation}
where $A$ is a constant. Comparing Eqs. (\ref{tr2}) and (\ref{tr4}),
we obtain the Tolman relation (\ref{tr1}). Now, applying this relation
at the boundary $r=R$ of the system, we obtain
\begin{equation}
e^{\nu(R)/2}T(R)=T_0.\label{tr5}
\end{equation}
On the other hand, according to Eq. (109) of Paper I, we have
\begin{equation}
\nu(R)=\ln (1-2GM/Rc^2).\label{tr6}
\end{equation}
Therefore, the  temperature at infinity is related to the
temperature at the surface of the star by
\begin{equation}
T_0=T(R)\sqrt{1-\frac{2GM}{R c^2}}. \label{tr7}
\end{equation}
This relation shows that the thermodynamical temperature defined by
Eq. (\ref{tst2}) coincides with the temperature measured by an observer at
infinity.

\section{The Newtonian limit in a  $d$-dimensional universe} \label{sec_nl}

In this appendix, we briefly discuss the Newtonian limit of the
Einstein equations in a $(d+1)$-dimensional space-time. Considering
first the general relativistic Emden equations
(\ref{ed3})-(\ref{ed4}), the Newtonian limit corresponds to
$q\rightarrow 0$. In that limit Eqs. (\ref{ed3})-(\ref{ed4}) reduce to
\begin{eqnarray}
\frac{d\psi}{d\xi}=(d-2)\frac{M(\xi)}{\xi^{d-1}},\quad {\rm
and}\quad \frac{dM}{d\xi}=\xi^{d-1}e^{-\psi}, \label{nl1}
\end{eqnarray}
and they combine to give
\begin{eqnarray}
\frac{1}{\xi^{d-1}}\frac{d}{d\xi}\left
(\xi^{d-1}\frac{d\psi}{d\xi}\right )=(d-2)e^{-\psi}. \label{nl2}
\end{eqnarray}
This is the familiar Emden equation (Chandrasekhar 1942) with an
additional factor $d-2$.  For $d>2$, we can rescale the parameters so
as to  recover the Emden equation exactly. For $d=2$, we see that there
is no Newtonian limit in $d=2$. This can also be seen, more
fundamentally, at the level of the Oppenheimer-Volkoff equations
(\ref{eged5})-(\ref{eged6}). The Newtonian limit corresponds to $c\rightarrow
+\infty$. In that limit, Eq. (\ref{eged5}) reduces to
\begin{equation}
\frac{dP}{dr}=-\frac{8\pi(d-2)}{(d-1)S_{d}}\rho
\frac{GM(r)}{r^{d-1}}. \label{nl3}
\end{equation}
This is to be compared with the classical condition of hydrostatic
equilibrium
\begin{equation}
\frac{dP}{dr}=-\rho \frac{G_{Newton}M(r)}{r^{d-1}}. \label{nl4}
\end{equation}
We find that the gravitational constants are related to each other
by
\begin{equation}
G_{Newton}=\frac{(d-2)8\pi}{(d-1)S_{d}}G_{Einstein}. \label{nl5}
\end{equation}
They only coincide in $d=3$.  In $d=1$, the Newtonian gravitational
constant is infinite and  it vanishes in $d=2$. Therefore, there is
apparently no Newtonian limit in one- and two-dimensional gravity.

\section{Dynamical stability analysis}
\label{sec_dsa}

In this appendix, we consider the formal nonlinear dynamical stability of a
box-confined relativistic star with a linear equation of state (\ref{es3}).
Specifically, we study the maximization problem
\begin{equation}
{\rm Max}\quad \lbrace N[\epsilon]\quad |\quad M[\epsilon]=M \quad {\rm
fixed}\rbrace, \label{dsa1}
\end{equation}
and show that it provides the same condition of stability as the
condition of linear dynamical stability studied in Paper I by
considering the growth rate of a solution of the linearized Einstein
equations.  Therefore, linear and nonlinear dynamical stability
coincide.  Stability is lost when the mass-central density profile
$M(\epsilon_0)$ presents an extremum. The solutions
on the series of equilibra $M(\epsilon_0)$ are nonlinearly dynamically
stable before the turning point of mass, and they become linearly
dynamically unstable after the turning point of mass. This is similar
to the case of barotropic stars described by the Euler-Poisson system
in Newtonian gravity (Chavanis 2002a, 2002c, 2006).

\subsection{The first-order variations}
\label{sec_f}

For a linear equation of state, using Eqs. (\ref{ege3}) and
(\ref{es5}), the baryon number can be expressed in terms of the energy
density in the form
\begin{equation}
N=\left (\frac{q}{K}\right
)^{1/\gamma}\int_{0}^{R}\epsilon^{\frac{1}{q+1}}\left\lbrack
1-\frac{2GM(r)}{rc^{2}}\right\rbrack^{-1/2}4\pi r^{2}dr. \label{f1}
\end{equation}
Therefore, the first-order variations of baryon number and mass are
\begin{eqnarray}
\delta N=\left (\frac{q}{K}\right )^{1/\gamma}\int_{0}^{R}\epsilon^{\frac{1}{q+1}}\left\lbrack 1-\frac{2GM(r)}{rc^{2}}\right\rbrack^{-1/2}\nonumber\\
\times\left
(\frac{1}{q+1}\frac{\delta\epsilon}{\epsilon}+\frac{\frac{G\delta
M(r)}{rc^{2}}}{1-\frac{2GM(r)}{rc^{2}}}\right )4\pi r^{2}dr,
\label{f2}
\end{eqnarray}
\begin{eqnarray}
\delta M=\frac{4\pi}{c^{2}}\int_{0}^{R}\delta\epsilon r^{2}dr.
\label{f3}
\end{eqnarray}
Substituting $\delta M(r)=\frac{4\pi}{c^{2}}\int_{0}^{r}\delta\epsilon
r^{'2}dr'$ in Eq.  (\ref{f2}), interchanging the order of the
integrals, and writing the first-order condition as Eq. (\ref{ege5}),
we obtain
\begin{eqnarray}
\int_{0}^{R}4\pi r^{2}dr \delta\epsilon \biggl\lbrace
\epsilon^{-\frac{q}{q+1}} \frac{1}{q+1}\left\lbrack
1-\frac{2GM(r)}{rc^{2}}\right\rbrack^{-1/2}
\nonumber\\
+\frac{G}{c^{4}}\int_{r}^{R} \epsilon(r')^{\frac{1}{q+1}}\left\lbrack 1-\frac{2GM(r')}{r'c^{2}}\right\rbrack^{-3/2} 4\pi r' dr'\biggr\rbrace \nonumber\\
-\mu \left (\frac{K}{q}\right )^{1/\gamma}
\int_{0}^{R}\frac{4\pi}{c^{2}}\delta\epsilon r^{2}dr=0. \label{f5}
\end{eqnarray}
This condition must be true for all variations, implying
\begin{eqnarray}
\mu \left (\frac{K}{q}\right )^{1/\gamma}\frac{1}{c^{2}}=
\frac{1}{q+1}\epsilon^{-\frac{q}{q+1}}\left\lbrack
1-\frac{2GM(r)}{rc^{2}}\right\rbrack^{-1/2}
\nonumber\\
+\frac{G}{c^{4}}\int_{r}^{R}
\epsilon(r')^{\frac{1}{q+1}}\left\lbrack
1-\frac{2GM(r')}{r'c^{2}}\right\rbrack^{-3/2} 4\pi r' dr'.
\label{f6}
\end{eqnarray}
If we take $r=R$, we find that the value of $\mu$ is given by
\begin{eqnarray}
\mu \left (\frac{K}{q}\right )^{1/\gamma}\frac{1}{c^{2}}=
\frac{1}{q+1}\epsilon(R)^{-\frac{q}{q+1}}\left\lbrack
1-\frac{2GM}{Rc^{2}}\right\rbrack^{-1/2}. \label{f7}
\end{eqnarray}
On the other hand, if we take the derivative of Eq. (\ref{f6}) with respect
to $r$, we obtain, after simplification,
\begin{eqnarray}
\left\lbrace 1-\frac{2GM(r)}{c^{2}r}\right\rbrace
\frac{d\epsilon}{dr}=
-\frac{1}{c^{2}}\frac{q+1}{q}\epsilon\left\lbrace \frac{GM(r)}{r^{2}}+\frac{4\pi Gq}{c^{2}}\epsilon r\right\rbrace,\nonumber\\
\label{f8}
\end{eqnarray}
which is the Oppenheimer-Volkoff equation (\ref{ege1}) with a linear
equation of state $P=q\epsilon$. This is a particular case of the
general result given by Weinberg (1972) which is valid for an
arbitrary equation of state provided that the perturbations are
adiabatic.

\subsection{The second-order variations}
\label{sec_ff}

We now turn to the more complicated second-order variations. Writing
\begin{eqnarray}
f=\delta M(r), \qquad \frac{df}{dr}=\frac{4\pi}{c^{2}}\delta\epsilon
r^{2}, \label{ff1}
\end{eqnarray}
the second variations of the baryon number are given by
\begin{eqnarray}
\delta^{2}N=\frac{G}{q+1}\int_{0}^{R}dr \frac{1}{r}\left\lbrack 1-\frac{2GM(r)}{rc^{2}}\right\rbrack^{-3/2}\epsilon^{-\frac{q}{q+1}}f\frac{df}{dr}\nonumber\\
+\frac{3 G^{2}}{2 c^{4}}\int_{0}^{R}dr 4\pi \left\lbrack 1-\frac{2GM(r)}{rc^{2}}\right\rbrack^{-5/2} \epsilon^{\frac{1}{q+1}}f^{2}\nonumber\\
-\int_{0}^{R}dr\frac{c^{4}}{4\pi r^{2}}\frac{q}{2(q+1)^{2}}\left\lbrack 1-\frac{2GM(r)}{rc^{2}}\right\rbrack^{-1/2}\epsilon^{-\frac{2q+1}{q+1}}\left (\frac{df}{dr}\right )^{2},\nonumber\\
\label{ff2}
\end{eqnarray}
where we have omitted, for brevity, the positive term
$(q/K)^{1/\gamma}$ in factor on the r.h.s. This is the sum of three
integrals that will be denoted $I_{1}$, $I_{2}$, and
$I_{3}$. Integrating by parts the first integral, we get
\begin{eqnarray}
I_{1}+I_{2}=\int_{0}^{R}\biggl\lbrace \frac{3 G^{2}}{2 c^{4}} 4\pi \left\lbrack 1-\frac{2GM(r)}{rc^{2}}\right\rbrack^{-5/2} \epsilon^{\frac{1}{q+1}}\nonumber\\
-\frac{G}{2(q+1)}\frac{d}{dr} \left (\frac{1}{r}\left\lbrack
1-\frac{2GM(r)}{rc^{2}}\right\rbrack^{-3/2}\epsilon^{-\frac{q}{q+1}}\right
)\biggr\rbrace  f^{2}dr. \nonumber\\
\label{ff3}
\end{eqnarray}
Expanding the derivative, we obtain
\begin{eqnarray}
I_{1}+I_{2}=\frac{G}{4(q+1)}\int_{0}^{R}\frac{1}{r^{2}} \left\lbrack 1-\frac{2GM(r)}{rc^{2}}\right\rbrack^{-3/2} \epsilon^{-\frac{q}{q+1}}\nonumber\\
\times \biggl\lbrace \frac{24\pi  G}{c^{4}}\left\lbrack 1-\frac{2GM(r)}{rc^{2}}\right\rbrack^{-1}q \epsilon r^{2}+\frac{2qr}{q+1}\frac{1}{\epsilon}\frac{d\epsilon}{dr}\nonumber\\
+3\left\lbrack
1-\frac{2GM(r)}{rc^{2}}\right\rbrack^{-1}-1\biggr\rbrace
f^{2}dr.\qquad  \label{ff4}
\end{eqnarray}
Using the Oppenheimer-Volkoff equation (\ref{f8}), we find after simplification
\begin{eqnarray}
I_{1}+I_{2}=\frac{G}{4(q+1)}\int_{0}^{R}\frac{1}{r^{2}} \left\lbrack 1-\frac{2GM(r)}{rc^{2}}\right\rbrack^{-3/2} \epsilon^{-\frac{q}{q+1}}\nonumber\\
\times \biggl (
2-\frac{4qr}{q+1}\frac{1}{\epsilon}\frac{d\epsilon}{dr}\biggr )
f^{2}dr. \qquad\label{ff5}
\end{eqnarray}
On the other hand, integrating by parts the third integral in Eq. (\ref{ff2})
and using the Oppenheimer-Volkoff equation (\ref{f8}), we obtain after
simplification
\begin{eqnarray}
I_{3}=\frac{c^{4}}{8\pi}\frac{q}{(q+1)^{2}}\int_{0}^{R}dr f\frac{1}{r^{2}}\left\lbrack 1-\frac{2GM(r)}{rc^{2}}\right\rbrack^{-1/2}\nonumber\\
\times \epsilon^{-\frac{2q+1}{q+1}}
\biggl\lbrack \frac{d^{2}f}{dr^{2}}+ \frac{df}{dr}\biggl\lbrace -\frac{2}{r}-\frac{1}{\epsilon}\frac{d\epsilon}{dr}\nonumber\\
+\biggl\lbrack 1-\frac{2GM(r)}{rc^{2}}\biggr\rbrack^{-1}\frac{4\pi G}{c^{4}}(q+1)\epsilon r\biggr\rbrace\biggr\rbrack. 
\label{ff6}
\end{eqnarray}
In conclusion,
\begin{eqnarray}
\delta^{2}N=\frac{G}{2(q+1)}\int_{0}^{R}\frac{1}{r^{2}}e^{3\lambda/2} \epsilon^{-\frac{q}{q+1}} \biggl (1-\frac{2qr}{q+1}\frac{1}{\epsilon}\frac{d\epsilon}{dr}\biggr )  f^{2}dr\nonumber\\
+  \frac{c^{4}}{8\pi}\frac{q}{(q+1)^{2}}\int_{0}^{R}dr f\frac{1}{r^{2}}e^{\lambda/2}\epsilon^{-\frac{2q+1}{q+1}}\nonumber\\
\times\biggl\lbrack \frac{d^{2}f}{dr^{2}}+ \frac{df}{dr}\left\lbrace -\frac{2}{r}-\frac{1}{\epsilon}\frac{d\epsilon}{dr}+e^{\lambda}\frac{4\pi G}{c^{4}}(q+1)\epsilon r\right\rbrace\biggr\rbrack.  \nonumber\\
\label{ff7}
\end{eqnarray}
To determine the sign of $\delta^2 N$, we are led to
consider the eigenvalue equation
\begin{eqnarray}
\frac{c^{4}}{8\pi}\frac{q}{(q+1)^{2}} \frac{1}{r^{2}}e^{\lambda/2}\epsilon^{-\frac{2q+1}{q+1}}\nonumber\\
\times\biggl\lbrack \frac{d^{2}f}{dr^{2}}+ \frac{df}{dr}\left\lbrace -\frac{2}{r}-\frac{1}{\epsilon}\frac{d\epsilon}{dr}+e^{\lambda}\frac{4\pi G}{c^{4}}(q+1)\epsilon r\right\rbrace\biggr\rbrack\nonumber\\
+\frac{G}{2(q+1)}\frac{1}{r^{2}}e^{3\lambda/2}
\epsilon^{-\frac{q}{q+1}} \biggl
(1-\frac{2qr}{q+1}\frac{1}{\epsilon}\frac{d\epsilon}{dr}\biggr )
f=\Lambda f,\label{ff8}
\end{eqnarray}
with the boundary conditions $f(0)=f(\alpha)=0$ (see Paper
I). Introducing the variables defined in Sect. \ref{sec_gre}, the
foregoing equation can be rewritten as
\begin{eqnarray}
\frac{d^{2}f}{d\xi^{2}}+ \left\lbrace -\frac{2}{\xi}+\frac{d\psi}{d\xi}+q\xi e^{\lambda-\psi}\right\rbrace \frac{df}{d\xi}\nonumber\\
+e^{\lambda-\psi}  \biggl (1+\frac{2q\xi}{q+1}\frac{d\psi}{d\xi}\biggr ) f\nonumber\\
=\frac{q c^{4}}{2\pi G^{2}}
\epsilon_{0}^{-\frac{1}{q+1}}\xi^{2}e^{-\lambda/2}e^{-\frac{2q+1}{q+1}\psi}\Lambda
f, \label{ff9}
\end{eqnarray}
with $f(0)=f(\alpha)=0$. This equation determines a discrete set of
eigenvalues $\Lambda_1>\Lambda_2>\Lambda_3$, ... If all the
eigenvalues $\Lambda$ are negative, then $\delta^{2}N\le 0$ and the
configuration is a maximum of $N$ at fixed mass. If at least one
eigenvalue is positive, the configuration is an unstable saddle
point. We therefore  need  to determine the point of marginal stability
$\Lambda=0$ in the series of equilibria. It is obtained by solving the
differential equation
\begin{eqnarray}
\frac{d^{2}f}{d\xi^{2}}+ \left\lbrace -\frac{2}{\xi}+\frac{d\psi}{d\xi}+q\xi e^{\lambda-\psi}\right\rbrace \frac{df}{d\xi}\nonumber\\
+e^{\lambda-\psi}  \biggl (1+\frac{2q\xi}{q+1}\frac{d\psi}{d\xi}\biggr ) f=0,\nonumber\\
\label{ff10}
\end{eqnarray}
with $f(0)=f(\alpha)=0$. The same equation was obtained in Paper I
[see Eq. (167)] by determining the condition of marginal linear
dynamical stability.  Therefore, the thresholds of linear and
nonlinear dynamical stability coincide. Equation (\ref{ff10}) was
solved in Paper I and led to the relation (I-171) identical to
(I-148). Therefore, the points where $\Lambda=0$ correspond to extrema
of mass in the series of equilibria $M(\alpha)$. A first eigenvalue
$\Lambda_1$ becomes positive at the first mass peak (implying
instability), and new modes of stability are lost at subsequent
extrema. At the $i$-th extremum, we have
$\Lambda_1>\Lambda_2>...>\Lambda_i=0>\Lambda_{i+1}>\Lambda_{i+2}>...$.


\begin{thebibliography}{99}

\bibitem{antonov}  {\small Antonov, V.A.  1962, Vest. Leningr. Gos. Univ., 7, 135}

\bibitem{banksfischler}  {\small Banks, T., \& Fischler, W. 2001, [hep-th/0111142]}

\bibitem{banksetal}  {\small Banks, T., Fischler, W., Kashani-Poor, A., McNees, R., \& Paban, S. 2002, Class. Quantum Grav., 19, 4717}

\bibitem{bekenstein73}  {\small Bekenstein, J.D.  1973, PRD, 7, 2333}

\bibitem{bekenstein81}  {\small Bekenstein, J.D.  1981, PRD, 23, 287}

\bibitem{bousso}  {\small Bousso, R.  2002, Rev. Mod. Phys., 74, 825}

\bibitem{chandra}  {\small Chandrasekhar, S. 1942, 
An Introduction to the Theory of Stellar Structure (Dover)} 

\bibitem{chandra72}  {\small Chandrasekhar, S. 1972, ``A limiting case of relativistic equilibrium'' in General Relativity, papers in honour of J.L. Synge, Edited by L.O' Raifeartaigh (Oxford)}


\bibitem{aaiso}  {\small Chavanis, P.H. 2002a, A\&A, 381, 340}

\bibitem{aarelat}  {\small Chavanis, P.H. 2002b, A\&A, 381, 709 (Paper I)}

\bibitem{aapoly}  {\small Chavanis, P.H. 2002c, A\&A, 386, 732}

\bibitem{fermions}  {\small Chavanis, P.H. 2002d, PRE, 65, 056123}

\bibitem{fermiD}  {\small Chavanis, P.H. 2004, PRE, 69, 066126} 

\bibitem{aaantonov}  {\small Chavanis, P.H. 2006a, A\&A, 451, 109}

\bibitem{cras}  {\small Chavanis, P.H. 2006b, C. R. Physique, 7, 331}

\bibitem{ijmpb}  {\small Chavanis, P.H. 2006c, Int J. Mod. Phys. B , 20, 3113}

\bibitem{prd}  {\small Chavanis, P.H. 2007a, PRD, 76, 023004 }

\bibitem{mass2d}  {\small Chavanis, P.H. 2007b, Physica A, 384, 392} 

\bibitem{rieutord}  {\small Chavanis, P.H., \& Rieutord, M. 2003, A\&A, 412, 1   }

\bibitem{lang}  {\small Chavanis, P.H., \& Sire, C. 2004, PRE, 69, 016116}

\bibitem{vs}  {\small de Vega, H.J., \& Sanchez, N. 2002, Nucl. Phys. B, 625, 409}

\bibitem{christodoulou}  {\small Christodoulou, D. 1970, PRL, 25, 1596}

\bibitem{hawking}  {\small  Hawking, S. 1971, PRL {26}, 1344 }

\bibitem{hawking75}  {\small  Hawking, S. 1975, Commun. math. Phys. {43}, 199 }

\bibitem{ht}  {\small  Hertel, P., \& Thirring, W.  1971, Commun. math. Phys. {24}, 22 }

\bibitem{hooft}  {\small 't Hooft, G. 1985, Nucl. Phys. B {256}, 727}

\bibitem{rama}  {\small  Kalyana Rama, S. 2007, Phys. Lett. B  {645}, 365}

\bibitem{katz}  {\small  Katz, J. 1978, MNRAS {183}, 765}

\bibitem{katzrev}  {\small  Katz, J. 2003, Found. Phys. {33}, 223}

\bibitem{klein}  {\small  Klein, O. 1947, Ark. Mat. Astr. Fys. {34}, 19}

\bibitem{ll}  {\small  Landau, L.D., \&  Lifshitz, E.M.  1960, The classical theory of fields (Fizmatgiz)  }

\bibitem{lbw}  {\small  Lynden-Bell, D., \& Wood, R. 1968, MNRAS {138}, 495 }

\bibitem{meltzer}  {\small  Meltzer, D.W., \& Thorne, K.S. 1966,  ApJ {145}, 514 }

\bibitem{misner}  {\small  Misner, C.W., \& Zapolsky, H.S. 1964,  Phys. Rev. Lett.  {12}, 635 }

\bibitem{oppenheim}  {\small  Oppenheim, J. 2002,  Phys. Rev. D  {65}, 024020 }

\bibitem{oppenheimer}  {\small  Oppenheimer, J.R., \& Volkoff, G.M. 1939,  Phys. Rev.   {55}, 374 }

\bibitem{paddy}  {\small  Padmanabhan, T. 1990, Phys. Rep. 188,285}

\bibitem{penrose}  {\small  Penrose, R., \& Floyd, R.M. 1971, Nature,   {229}, 177}

\bibitem{pesci}  {\small  Pesci, A. 2007,  Class. Quantum Grav.   {24}, 2283 }

\bibitem{ponce}  {\small  Ponce de Leon, J., \& Cruz, N. 2000, Gen. Relativ. Gravit.     {32}, 1207 }

\bibitem{sh}  {\small  Schmidt, H.J., \& Homann, F. 2000, Gen. Relativ. Gravit.   {32}, 919}

\bibitem{sc}  {\small Sire, C., \& Chavanis, P.H. 2002, PRE, 66, 046133}

\bibitem{sorkin}  {\small  Sorkin, R.D., Wald, R.M., \&  Jiu, Z.Z. 1981, Gen. Relativ. Gravit.   {13}, 1127}

\bibitem{srednicki}  {\small  Srednicki, M. 1993, Phys. Rev. Lett.  {71}, 666}

\bibitem{tolman}  {\small Tolman, R.C. 1934, Relativity, Thermodynamics and  Cosmology (Oxford: Clarendon)}

\bibitem{tsallis}  {\small Tsallis, C. 1988, J. Stat. Phys. {52}, 479}

\bibitem{weinberg}  {\small Weinberg, S. 1972, Gravitation and Cosmology (John Wiley)}

\bibitem{zel}  {\small Zel'dovich, Ya.B. 1962, Soviet Phys. J.E.T.P.  {14}, 1143}





\end{thebibliography}
\end{document}